\documentclass[
reprint,
superscriptaddress,
amsmath, amssymb, aps,
prb,
longbibliography,
]{revtex4-2}

\usepackage[dvipdfmx]{graphicx}

\usepackage{bm}
\usepackage{hyperref}
\usepackage{xcolor}
\definecolor{LightGray}{gray}{0.91}
\usepackage{cleveref}
\usepackage{listings}
\IfFileExists{minted2.sty}{%
  \usepackage[frozencache,cachedir=.]{minted2}%
}{%
  \usepackage[frozencache,cachedir=.]{minted}%
}
\usepackage{longtable}

\newcommand{\bnoindentnl}[1]{\hfill\break\noindent\textbf{{#1}}}
\newcommand{\example}{\hfill\break\noindent\fbox{Example}}

\begin{document}

\title{Tensor Computing Application Programming Interface: An Application-Oriented, Lightweight API for Portable High-Performance Tensor Network Computations}

\begin{abstract}
Tensor networks (TNs) are a central computational tool in quantum science and artificial intelligence.
However, the lack of a unified software interface across tensor-computing frameworks severely limits the portability of TN applications, coupling algorithmic development to specific hardware and software back ends. 
To address this challenge, we introduce the Tensor Computing Application Programming Interface (TCAPI)---an application-oriented, lightweight API designed to enable framework-independent, high-performance TN applications, currently specified for \texttt{C++} and \texttt{Python}. 
TCAPI provides well-defined type-system conventions that abstract tensor objects, together with a minimal yet expressive set of core functions for essential tensor manipulations and tensor linear-algebra operations, with operation semantics shared across language specifications. 
Through numerical demonstrations of representative tensor-network applications implemented in \texttt{C++} using TCAPI, we show that code written against TCAPI can be migrated seamlessly across heterogeneous hardware and software platforms while achieving performance comparable to native framework implementations. 
We further provide open-source implementations of TCAPI based on \textit{Cytnx} and \textit{ITensor} in \texttt{C++}, and on \textit{NumPy} in \texttt{Python}, demonstrating its practicality and ease of integration with existing tensor-computing frameworks.
\end{abstract}

\author{Rong-Yang Sun}
\affiliation{Department of Physics and Astronomy, California State University Northridge, Northridge, California 91330, USA}
\affiliation{RIKEN Interdisciplinary Theoretical and Mathematical Sciences Program (iTHEMS), Wako, Saitama 351-0198, Japan}
\affiliation{Computational Materials Science Research Team, RIKEN Center for Computational Science (R-CCS), Kobe, Hyogo, 650-0047, Japan}

\author{Tomonori Shirakawa}
\affiliation{Computational Materials Science Research Team, RIKEN Center for Computational Science (R-CCS), Kobe, Hyogo, 650-0047, Japan}
\affiliation{Quantum Computational Science Research Team, RIKEN Center for Quantum Computing (RQC), Wako, Saitama, 351-0198, Japan}
\affiliation{RIKEN Interdisciplinary Theoretical and Mathematical Sciences Program (iTHEMS), Wako, Saitama 351-0198, Japan}
\affiliation{Computational Condensed Matter Physics Laboratory, RIKEN Cluster for Pioneering Research (CPR), Saitama 351-0198, Japan}
\affiliation{Quantum Computing Simulation Unit, Quantum-HPC Hybrid Platform Division, RIKEN Center for Computational Science (R-CCS), Kobe, Hyogo, 650-0047, Japan}

\author{Hidehiko Kohshiro}
\affiliation{Computational Materials Science Research Team, RIKEN Center for Computational Science (R-CCS), Kobe, Hyogo, 650-0047, Japan}
\affiliation{Quantum Computational Science Research Team, RIKEN Center for Quantum Computing (RQC), Wako, Saitama, 351-0198, Japan}
\affiliation{Next-Generation HPC Application Development Unit, Next-Generation HPC Infrastructure Development Division, RIKEN Center for Computational Science (R-CCS), Kobe, Hyogo, 650-0047, Japan}

\author{D. N. Sheng}
\affiliation{Department of Physics and Astronomy, California State University Northridge, Northridge, California 91330, USA}

\author{Seiji Yunoki}
\affiliation{Computational Materials Science Research Team, RIKEN Center for Computational Science (R-CCS), Kobe, Hyogo, 650-0047, Japan}
\affiliation{Quantum Computational Science Research Team, RIKEN Center for Quantum Computing (RQC), Wako, Saitama, 351-0198, Japan}
\affiliation{Computational Condensed Matter Physics Laboratory, RIKEN Cluster for Pioneering Research (CPR), Saitama 351-0198, Japan}
\affiliation{Computational Quantum Matter Research Team, RIKEN Center for Emergent Matter Science (CEMS), Wako, Saitama 351-0198, Japan}
\date{\today}

\maketitle

\section{Introduction}
Tensor networks (TNs) have become a central computational paradigm for representing and manipulating high-dimensional classical and quantum data.
They have been especially influential in quantum many-body physics, where tensor-network states (TNSs) provide compact variational representations of many-body wave functions whose efficiency is governed by entanglement structure.
Prominent examples include matrix product states (MPSs) and projected entangled-pair states (PEPSs), which offer a unifying language for low-entanglement quantum states \cite{cirac2021matrix}.
Tensor-network algorithms such as the density-matrix renormalization group (DMRG) \cite{white1992density}, time-evolving block decimation (TEBD) \cite{vidal2007classical}, and time-dependent variational principle (TDVP) methods \cite{haegeman2011time,haegeman2016unifying} have enabled highly accurate simulations of one-dimensional quantum lattice systems, while PEPS and related TNSs extend the framework to strongly correlated systems in two and higher spatial dimensions \cite{vlaar2021simulation}.
Tensor renormalization group methods provide a complementary approach to classical and quantum lattice models, evaluating partition functions and thermodynamic observables through successive coarse-graining of tensor networks \cite{levin2007tensor,xie2012coarse}.
These methods have become indispensable for studying quantum many-body systems in regimes that are otherwise computationally inaccessible, and they provide powerful tools for characterizing entanglement structure and topological properties.
Examples include symmetry-protected topological states \cite{chen2011classification, schuch2011classifying, pollmann2012symmetry, chen2013symmetry} and intrinsic topological orders \cite{buerschaper2009explicit, cirac2011entanglement, zaletel2012exact, zaletel2013topological, he2014modular}.

Beyond quantum many-body physics, TNs have become important tools in quantum computing.
They enable the classical simulation of many-qubit quantum circuits with moderate depth \cite{markov2008simulating, zhou2020what, pan2022solving, tindall2023efficient, anand2023classical, sun2024improved, liu2024verifying, oh2024classical}, support circuit synthesis \cite{schon2005sequentiala, zaletel2020isometric, ran2020encoding, haghshenas2022variational, wei2022sequential, anand2023holographic, malz2024preparation}, and facilitate quantum error correction and mitigation \cite{ferris2014tensor, farrelly2021tensornetwork, cao2022quantum, guo2022quantum, filippov2023scalable} (see Ref.~\cite{berezutskii2025tensor} for a comprehensive review).

TNs and tensor decompositions are also increasingly used in machine learning.
In this setting, tensorial neural networks (TNNs) \cite{wang2025tensor} can provide efficient data representations \cite{zadeh2017tensor, ben2017mutan, zhang2018quantum, hou2019deep, kossaifi2020factorized} and replace dense model parameters with tensor-factored forms \cite{novikov2015tensorizing, garipov2016ultimate, stoudenmire2016supervised, yang2017deep, ma2019tensorized, kossaifi2020factorized}, enabling more resource-efficient machine-learning models.
Recent studies have applied TNNs to reinforcement learning \cite{sozykin2022ttopt}, prompt learning \cite{qiu2025steps}, statistical learning \cite{han2022optimal, saiapin2025tensor}, and the training and fine-tuning of large language models \cite{chekalina2023efficient, chen2024quanta, xu2024tensorgpt, anjum2024tensor, yang2024comera, tao2025transformed}.
More broadly, by connecting classical and quantum descriptions, TNs provide a natural framework for quantum machine learning \cite{huggins2019quantuma, dilip2022data, dborin2022matrix, rieser2023tensor, shin2024dequantizing}.

\begin{figure}[t]
    \centering
    \includegraphics[width=\linewidth]{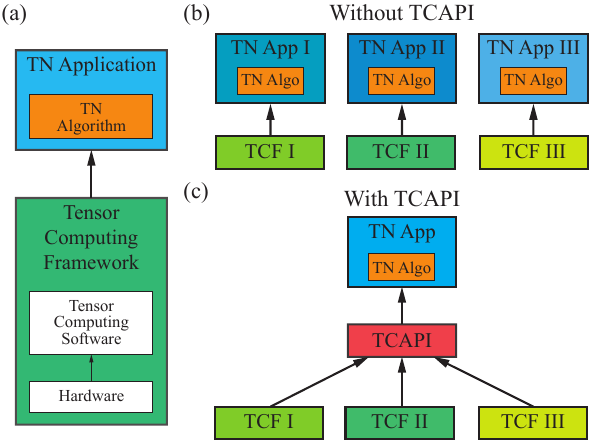}
    \caption{
        System architectures and development workflows for tensor-network (TN) applications. Arrows indicate dependency relations, with arrowheads pointing to the dependent components. (a) Typical software architecture of a TN application, in which the application logic and TN algorithms are directly tied to a specific tensor-computing framework (TCF). (b) Development workflow without TCAPI: TN applications must be reimplemented separately for each TCF, leading to duplicated development effort and limited portability. (c) Development workflow with TCAPI: application logic and TN algorithms are decoupled from back-end hardware and software through a uniform interface, enabling portable development across heterogeneous TCFs. 
    }
    \label{fig:tcapi_illus}
\end{figure}

Progress in TN algorithms and applications has, in turn, driven rapid advances in tensor computing---that is, the hardware and software stacks responsible for executing tensor operations underlying TN workloads.
Here we use the term {\it tensor computing framework} (TCF) to denote a concrete combination of a hardware platform (for example, a shared-memory CPU system) and the tensor-computing software stack designed to target it. 
In practice, TN applications are typically written directly against a specific TCF [see Fig.~\ref{fig:tcapi_illus}(a)]. As a result, application logic becomes tightly coupled to the chosen framework, since competing TCFs generally expose incompatible application programming interfaces (APIs). 
Portability across TCFs is therefore nontrivial. 
For example, an application implemented using \textit{ITensor} \cite{itensor} is not straightforward to migrate to \textit{Cytnx} \cite{wu2025the}, despite the fact that both provide \texttt{C++} APIs.
Adopting a different TCF thus commonly necessitates substantial code reimplementation, even when the underlying TN algorithm remains unchanged. This lack of portability increases development cost, complicates maintenance, and raises the risk of introducing implementation errors, as schematically illustrated in Fig.~\ref{fig:tcapi_illus}(b).

Accelerating progress in both TN algorithms and applications requires true application portability across TCFs, which in turn calls for a clear separation between application logic (i.e., TN algorithms) and the underlying execution back ends through a standardized software interface. 
A well-specified interface would enable seamless migration of applications across heterogeneous TCFs while preserving a single codebase. 
As a result, development efforts could focus on algorithmic design, whereas advances in hardware and software back ends would be inherited with little or no modification to application code. 
In the machine-learning community, where \texttt{Python} is the dominant programming language, the de facto standard interface for tensor computation is the array API introduced in \textit{NumPy} \cite{harris2020array}.
Popular frameworks such as \textit{JAX} \cite{jax2018github} and \textit{PyTorch} \cite{paszke2019pytorch} provide NumPy-like interfaces, enabling a certain degree of portability at the source-code level.  
However, this portability remains limited in practice: migrating nontrivial applications between frameworks typically requires substantial reimplementation effort. 
Moreover, although NumPy-like interfaces provide a useful convention within \texttt{Python}, they do not by themselves address portability between \texttt{Python}-based frameworks and performance-critical \texttt{C++} applications, particularly in HPC environments. Their portability within \texttt{Python} is further undermined by subtle but consequential API differences across frameworks. 
In addition, because many tensor operations are exposed as methods of the tensor object itself (for example, slicing or reshaping), it is often difficult for application developers to customize or intercept framework behavior without modifying or extending the underlying tensor classes.

Motivated by these limitations, we introduce the Tensor Computing Application Programming Interface (TCAPI), a lightweight software interface for generic tensor computations, currently specified for \texttt{C++} and \texttt{Python}.
TCAPI is designed to be application-oriented and to sit clearly between TN applications and TCFs, enabling developers to decouple application logic from execution back ends without modifying the source code of the underlying TCFs. 
At its core, TCAPI provides carefully designed type-system conventions that abstract tensor objects and their associated metadata: the \texttt{C++} specification uses compile-time tensor types and traits, whereas the \texttt{Python} specification uses runtime tensor-kind descriptors and backend-native tensor values. Built on these conventions, TCAPI defines a concise yet expressive set of functions whose operation semantics are shared across language specifications and cover essential tensor manipulations and tensor linear-algebra operations. 
As illustrated in Fig.~\ref{fig:tcapi_illus}(c), this design allows TN applications to be written against a uniform interface, enabling portable development across heterogeneous hardware and software back ends. 
As a result, application developers can focus on implementing TN algorithms rather than managing framework-specific details. 
The \texttt{C++} specification of TCAPI is formulated in modern \texttt{C++} based on the \texttt{C++17} standard.
We provide open-source \texttt{C++} implementations based on \textit{Cytnx} \cite{tcapi_cytnx} and \textit{ITensor} \cite{tcapi_itensor}, together with a \texttt{Python} implementation based on \textit{NumPy} \cite{tcapi_numpy}.
An online version of the TCAPI specification is also available \cite{tcapi_spec}.
To evaluate the practicality and performance of TCAPI, we implement two representative TN applications---ground-state simulation and real-time dynamics of quantum many-body systems---using TCAPI in \texttt{C++} and benchmark them across multiple back ends. 
Our results show that applications written against TCAPI can be migrated across TCFs without changes to their main code while achieving performance comparable to implementations based on native framework APIs.
The application-oriented scope of TCAPI also clarifies its relation to lower-level tensor-algebra standardization efforts, such as Tensor Algebra Processing Primitives (TAPP).
TAPP defines a \texttt{C}-based, BLAS-like interface for performance-critical tensor-algebra primitives, especially tensor contractions \cite{brandejs2026tapp}.
Thus, TAPP targets tensor-kernel portability, whereas TCAPI targets application-level portability of TN workloads across TCFs.

The remainder of this paper is organized as follows.
Section~\ref{sec:ten_ten_comp} reviews the fundamentals of tensor computation, introducing tensors and the basic operations that underpin TN workloads. 
Section~\ref{sec:tcapi_basic} illustrates the basic \texttt{C++} usage of TCAPI through a minimal working example, with step-by-step explanations of the core components. 
In Sec.~\ref{sec:demo}, we present numerical benchmarks that demonstrate both the portability and the performance of two representative TN applications implemented using TCAPI in \texttt{C++}.
Section~\ref{sec:summary_outlook} concludes the paper with a summary and a discussion of future directions. 
Implementation details of the \texttt{C++} TCAPI-based application for ground-state simulation are provided in Appendix~\ref{app:itebd_tcapi_impl}.
Appendix~\ref{app:bp-qc} describes the TN algorithm employed for quantum dynamics simulation.
A comprehensive specification of TCAPI for \texttt{C++} and \texttt{Python}, including type-system conventions, core functions, and supporting environment variables, is given in Appendix~\ref{app:tcapi_spec}.

\section{Tensor and tensor operations \label{sec:ten_ten_comp}}

We begin by reviewing the foundational elements of tensor computation, focusing on a computational definition of tensors and the core operations that underlie tensor workloads.
The terminology and notation introduced in this section establish a common language for the subsequent presentation and form the basis for the TCAPI specification.

\subsection{Tensor}

\begin{figure}[t]
    \centering
    \includegraphics[width=\linewidth]{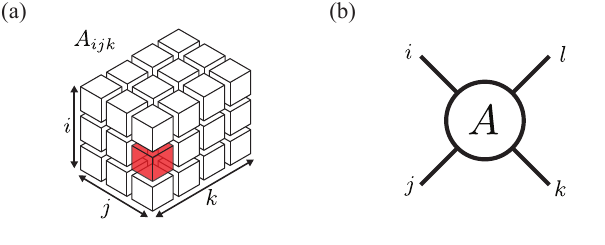}
    \caption{Graphic representations of tensors. (a) Array representation of a third-order tensor $A_{ijk}$, where the entry at coordinates $(i,j,k)$ corresponds to the tensor element $A_{i,j,k}$. (b) Penrose diagram of a fourth-order tensor $A_{ijkl}$, in which the node represents the tensor and each incident leg corresponds to a tensor mode (bond).}
    \label{fig:tensor_illustration}
\end{figure}

From a computational perspective, we regard a \textit{tensor} as an $n$-dimensional array $A_{ijk\cdots}$ that generalizes scalars $A$ (zero-dimensional), vectors $A_i$ (one-dimensional), and matrices $A_{ij}$ (two-dimensional).
The number of dimensions---equivalently, the number of indices or modes---defines the \textit{order} of the tensor. 
Accordingly, a scalar is a zeroth-order tensor, a vector is a first-order tensor, and a matrix is a second-order tensor. 
Low-order tensors (up to third order) can be conveniently depicted as multidimensional arrays [see Fig.~\ref{fig:tensor_illustration}(a) for a third-order example]. 
In contrast, higher-order tensors are more naturally represented using Penrose's graphical notation, in which a node denotes the tensor and each incident edge (or ``leg'') corresponds to a tensor dimension \cite{penrose1971applications} [see Fig.~\ref{fig:tensor_illustration}(b) for a fourth-order example].

Each mode (dimension) of the tensor array is represented by a \textit{bond}, which appears in diagrammatic notation as a ``leg'' attached to the tensor node [see Fig.~\ref{fig:tensor_illustration}(b)]. 
A bond is characterized by its \textit{bond index} $b$, which specifies its position in an explicit ordering of the tensor modes, and its \textit{bond dimension} $d_b$, which gives the size of that mode. 
For example, a $4\times 5$ matrix can be represented as a tensor $A_{ij}$. Under zero-based indexing, the bond associated with index $j$ has bond index $b = 1$ and bond dimension $d_1 = 5$. 
In an abstract TN diagram, bonds are specified by connectivity rather than by array-axis order.
In software, however, each bond must be mapped to a definite tensor mode.
This ordering fixes the bond indices used in TCAPI, determines the tensor shape, and affects the implementation of tensor operations.
For tensor networks with nontrivial exchange structure, such as fermionic or braided tensor networks, the ordering may also be tied to convention-dependent data; incorporating such structures is left for future work.
The ordered list of bond dimensions, arranged according to bond indices, defines the \textit{shape} of the tensor. For a zeroth-order tensor, the shape is an empty list. In the above example, the shape of $A$ is $(4,5)$.

We refer to the entries of a tensor array as \textit{elements}.
Each element is uniquely identified by a zero-based \textit{coordinate} tuple $\mathbf{c} = (c_0,\dots ,c_{n-1})$ and has an associated \textit{value} $A_\mathbf{c} = A_{c_0,\dots ,c_{n-1}}$. 
For instance, the highlighted entry in Fig.~\ref{fig:tensor_illustration}(a) corresponds to the coordinates $\mathbf{c}=(1,2,0)$ and has value $A_{1,2,0}$.
The \textit{size} of a tensor is defined as the total number of its elements. For a zeroth-order tensor, the size is 1. 
Throughout this paper, we denote by $\mathbb{K}$ the scalar field from which tensor elements are drawn. 
In the context of TCAPI, we restrict $\mathbb{K}$ to the real field $\mathbb{R}$ or the complex field $\mathbb{C}$, while the concrete element type and precision are determined by the underlying TCF.

\subsection{Tensor operations}

Tensor computations are performed by applying well-defined operations to tensors.
These operations generalize familiar vector- and matrix-level primitives and can be broadly classified into two categories: tensor manipulation operations and tensor linear-algebra operations.

\begin{figure}[t]
    \centering
    \includegraphics[width=\linewidth]{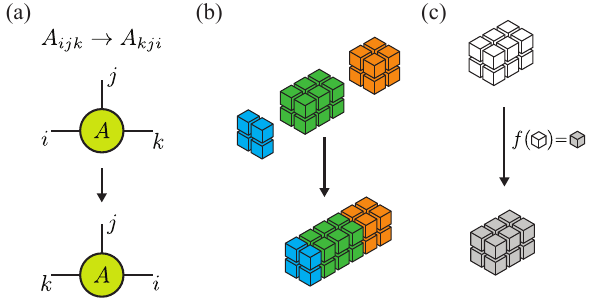}
    \caption{Typical tensor-manipulation operations. (a) Transpose, which reorders tensor bonds. (b) Concatenation, which combines multiple tensors along a specific bond. (c) Broadcasting, which applies a scalar function elementwise while preserving tensor order and shape.}
    \label{fig:tensor_manipulations}
\end{figure}

\subsubsection{Tensor manipulation operations}
This class of operations comprises tensor manipulations that act either on the structure of a tensor or independently on its individual elements. 
They include non-arithmetic transformations---such as bond reordering, reshaping, the combining or splitting of bonds (tensor modes), and slicing---that permute the ordering of elements or modify the tensor shape without changing element values, together with element-wise operations that preserve tensor order and shape. 
Below, we illustrate three representative tensor-manipulation operations shown in Fig.~\ref{fig:tensor_manipulations}.
A complete list of tensor-manipulation functions supported by TCAPI is provided in Appendix~\ref{sec:tensor_manipulation}.

\textbf{Transpose.}
Given a permutation $\pi$ of the bond indices, the transpose operation produces a tensor $B$ whose entries satisfy 
\begin{equation}
    B_{i_{\pi(0)}i_{\pi(1)}\cdots i_{\pi(n-1)}} = A_{i_0 i_1 \cdots i_{n-1}}~.
\end{equation}
This operation permutes bonds and therefore reorders tensor elements without changing their values. 
While the memory layout is typically altered, the underlying data remain unchanged. 
Figure~\ref{fig:tensor_manipulations}(a) illustrates the specific transpose $A_{ijk} \to A_{kji}$.

\textbf{Concatenation.}
Tensor concatenation combines multiple tensors of the same order along a specified bond. 
All bonds with the same bond index must have identical dimensions across the input tensors, except for the concatenated bond. 
For example, concatenating three third-order tensors along bond $k$ yields a third-order tensor whose $k$-th bond dimension equals the sum of the corresponding bond dimensions of the inputs.
Figure~\ref{fig:tensor_manipulations}(b) illustrates the concatenation of three tensors into a single tensor. 

\textbf{Broadcasting.}
Given a tensor $A\in \mathbb{K}^{d_0\times\cdots \times d_{n-1}}$ and a unary function $f:\mathbb{K} \to \mathbb{K}$, broadcasting (or entrywise mapping) defines a tensor $B = f(A)$ by
\begin{equation}
    B_{i_0i_1\cdots i_{n-1}} = f(A_{i_0i_1\cdots i_{n-1}}),\quad 0\leq i_b < d_b~.
\end{equation}
Broadcasting preserves both tensor order and shape, yielding $B\in \mathbb{K}^{d_0\times\cdots \times d_{n-1}}$.
Figure~\ref{fig:tensor_manipulations}(c) shows an example in which a scalar function $f(x)$ is applied elementwise to a third-order tensor.

\begin{figure}[t]
    \centering
    \includegraphics[width=\linewidth]{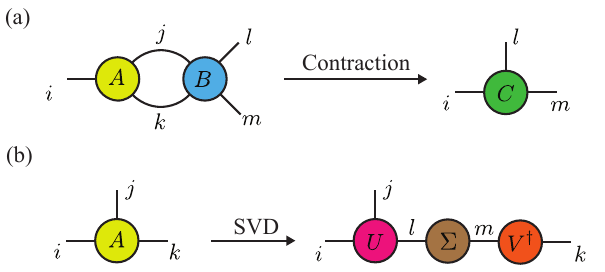}
    \caption{Representative tensor linear-algebra operations. (a) Tensor contraction over shared bonds. (b) Tensor decomposition, illustrated by the SVD as a representative example.}
    \label{fig:tensor_linalg}
\end{figure}

\subsubsection{Tensor linear-algebra operations}

This class comprises tensor linear-algebra operations, which generalize matrix-level primitives to higher-order tensors. 
Such operations typically proceed by (i) matricizing (unfolding) a tensor with respect to a chosen partition of its bonds, (ii) applying a matrix-level routine---such as multiplication, singular value decomposition, eigensolving, or linear solving---and (iii) refolding the result into tensor(s) of appropriate shape. 
In contrast to tensor manipulation operations, which act on tensor structure or on elements independently, tensor linear-algebra operations treat the tensor as a single coupled algebraic object.
In the following, we present two representative examples: contraction and decomposition.
A complete list of tensor linear-algebra operations supported by TCAPI is provided in Appendix~\ref{sec:tensor_linear_algebra}.

\textbf{Contraction.}
Tensor contraction is a fundamental primitive in TN applications.
Let tensors $A$ and $B$ share a set of bonds $S$ with matching bond dimensions.
Denoting by $I$ the ordered set of bonds $A$ not in $S$, and by $J$ the ordered set of bonds of $B$ not in $S$, contraction over the shared bonds produces a tensor $C$ with bonds $I \cup J$, whose elements are given by
\begin{equation}
    C_{IJ} = \sum_{S} A_{IS}B_{SJ}~.
\end{equation}
This expression reduces to ordinary matrix multiplication when the numbers of nonshared and shared bonds---that is, the cardinalities of the corresponding bond sets---satisfy $|I| = |J| = |S|= 1$.
As a concrete example, consider tensors $A_{ijk}$ and $B_{jklm}$ sharing bonds $j$ and $k$. Their contraction yields 
\begin{equation}
    C_{ilm} = \sum_{j=0}^{d_j-1}\sum_{k=0}^{d_k-1} A_{ijk} B_{jklm}~,
\end{equation}
where $d_j$ and $d_k$ denote the corresponding bond dimensions.
The graphical representation of this contraction is shown in Fig.~\ref{fig:tensor_linalg}(a).

\textbf{Decomposition.}
Tensor decomposition is another fundamental primitive in TN applications. 
It extends matrix factorizations to higher-order tensors by expressing a tensor---exactly, or approximately when truncation is applied---as a contracted network of lower-order factors.
As a representative example, we consider the tensor singular value decomposition (SVD). 
Let $A\in \mathbb{K}^{d_i\times d_j \times d_k}$ and matricize it with respect to the bipartition $\{i,j\}|\{k\}$, yielding 
\begin{equation}
    A' \in \mathbb{K}^{(d_i\times d_j) \times d_k},\quad A'_{pq} \equiv A_{ijk},
\end{equation}
where the composite index $p$ corresponds to $(i,j)$ and $q$ corresponds to $k$. 
Performing an SVD gives $A'=U'\Sigma V^{\dagger}$, i.e.,
\begin{equation}
    A'_{pq} = \sum_{l,m} U'_{pl}\Sigma_{lm}(V^\dagger)_{mq}~.
\end{equation}
Reshaping $U'_{pl}$ to $U_{ijl}$ and identifying $V^\dagger_{mq} = V^{\dagger}_{mk}$, we obtain the tensor decomposition 
$A = U * \Sigma * V^{\dagger}$, where $*$ denotes tensor contraction over the internal bonds $l$ and $m$. 
The corresponding graphical representation is shown in Fig.~\ref{fig:tensor_linalg}(b).

\section{Basic usage of the Tensor Computing Application Programming Interface \label{sec:tcapi_basic}}

In this section, we illustrate the basic usage of the \texttt{C++} TCAPI through a minimal working example, highlighting its concise and flexible syntax in a step-by-step manner. The \texttt{Python} interface follows the same operation semantics but uses runtime tensor-kind descriptors and backend-native tensor values; the full \texttt{C++} and \texttt{Python} specification is provided in Appendix~\ref{app:tcapi_spec}. 
The example considers generating a random many-body wavefunction, measuring the bipartite entanglement entropy (BEE) between the left and right halves of the system, and constructing a low-entanglement (low-rank) approximation of the state. 
Throughout the example, we assume that the \texttt{C++} TCAPI has been implemented on top of a TCF that provides a concrete \texttt{C++} tensor type \texttt{Ten} with element type \texttt{double}.

To use the \texttt{C++} TCAPI, we first include its header file \texttt{tcapi/tcapi.h}, along with the required \texttt{C++} standard library headers used in this example:
\begin{minted}[bgcolor=LightGray]{cpp}
#include "tcapi/tcapi.h"
#include <random>
\end{minted}
To make the code portable across different TCFs, we define a set of type aliases that abstract away TCF-dependent types used throughout the example:
\begin{minted}[bgcolor=LightGray]{cpp}
using ten = Ten;
using elem = tcapi::elem_t<ten>;
\end{minted}
TCAPI provides a collection of alias templates---such as \texttt{tcapi::elem\_t} used here---that extract concrete types associated with a given tensor type (see Appendix~\ref{sec:type_system} for details).
Note that the type \texttt{Ten} is directly accessible because the header \texttt{tcapi/tcapi.h} is supplied by the TCF exposing \texttt{Ten}.
To migrate this code to a different TCF, one simply replaces \texttt{Ten} with the corresponding tensor type provided by the new framework, without modifying the remaining code.

Before invoking any TCAPI tensor operations, a context must be created to manage resources associated with the TCF, such as GPU devices or thread pools. 
This step is typically performed at the beginning of the \texttt{main} function. 
The concrete context-handle type is obtained using the alias template \texttt{tcapi::context\_handle\_t}, and a context instance is initialized by calling \texttt{tcapi::create\_context}:
\begin{minted}[bgcolor=LightGray]{cpp}
tcapi::context_handle_t<ten> ctx;
tcapi::create_context(ctx);
\end{minted}

We next generate a random wavefunction for a qubit chain of length six using the function \texttt{tcapi::random}:
\begin{minted}[bgcolor=LightGray]{cpp}
std::mt19937 engine;
std::uniform_real_distribution<double> dis(
    -1.0, 1.0);
auto gen = [&dis, &engine]() {
    return dis(engine);
};
auto psi = tcapi::random<ten>(
    ctx, {2, 2, 2, 2, 2, 2}, gen);
\end{minted}
The wavefunction \texttt{psi} is a sixth-order tensor.
To satisfy the normalization condition, we renormalize \texttt{psi} using \texttt{tcapi::normalize}:
\begin{minted}[bgcolor=LightGray]{cpp}
tcapi::normalize(ctx, psi);
\end{minted}

To measure the BEE between the left and right halves of the chain, we perform a Schmidt decomposition of \texttt{psi} across the center of the system using the SVD function \texttt{tcapi::svd}:
\begin{minted}[bgcolor=LightGray]{cpp}
ten u, sigma, vt;
tcapi::svd(ctx, psi, 3, u, sigma, vt); 
\end{minted}
Here, the third argument, \texttt{3}, is the TCAPI parameter \texttt{num\_of\_bds\_as\_row}; it specifies that the SVD is performed across the bipartition after the first three bonds of \texttt{psi}, corresponding to the split between the left and right halves of the six-qubit chain.
The output \texttt{sigma} is a second-order diagonal tensor whose diagonal entries are the singular values; implementations may store this tensor compactly when supported by the underlying TCF.
To compute the BEE, we extract the diagonal entries of \texttt{sigma} as a first-order tensor and traverse these values using \texttt{tcapi::for\_each}:
\begin{minted}[bgcolor=LightGray]{cpp}
ten sigma_vals;
tcapi::diag(ctx, sigma, sigma_vals);

elem ee = 0.0;
auto compute_ee = [&ee](const elem sigma_val) {
    ee -= (sigma_val * sigma_val)
          * log(sigma_val * sigma_val);
};
tcapi::for_each(ctx, sigma_vals, compute_ee);
\end{minted}
The resulting BEE is stored in \texttt{ee}.

We next investigate a low-entanglement approximation of \texttt{psi} and examine the effect of truncation on the wavefunction.
Such an approximation is obtained by performing a truncated SVD using \texttt{tcapi::trunc\_svd}, retaining only the two largest singular values:
\begin{minted}[bgcolor=LightGray]{cpp}
elem trunc_err = 0.0;
tcapi::trunc_svd(
    ctx, psi, 3,
    u, sigma, vt,
    trunc_err, 2, 0.0);
\end{minted}
The approximate wavefunction \texttt{psi1} is reconstructed by contracting \texttt{u}, the diagonal tensor \texttt{sigma}, and \texttt{vt}.
The contractions are carried out using \texttt{tcapi::contract}:
\begin{minted}[bgcolor=LightGray]{cpp}
ten psi1;
tcapi::contract(
    ctx,
    u, "ijkl", sigma, "lm", psi1, "ijkm");
tcapi::contract(
    ctx,
    psi1, "ijkl", vt, "lmno", psi1, "ijkmno");
tcapi::normalize(ctx, psi1);
\end{minted}

To quantify the truncation effect, we compute the fidelity between \texttt{psi} and the normalized \texttt{psi1} by evaluating their overlap via tensor contraction:
\begin{minted}[bgcolor=LightGray]{cpp}
ten ovlp;
tcapi::contract(
    ctx,
    psi, "ijklmn", psi1, "ijklmn",
    ovlp, "");
auto ovlp_v = tcapi::get_elem(ctx, ovlp, {});
auto fide = std::norm(ovlp_v);
\end{minted}
Since this contraction eliminates all bonds, \texttt{ovlp} is a zeroth-order tensor with shape \texttt{\{\}} and size 1, whose sole element is read using \texttt{tcapi::get\_elem(ctx, ovlp, \{\})}.
The fidelity \texttt{fide} should equal $1.0 - \texttt{trunc\_err}$. 
Finally, before exiting \texttt{main}, we release the resources managed by the context using \texttt{tcapi::destroy\_context}:
\begin{minted}[bgcolor=LightGray]{cpp}
tcapi::destroy_context(ctx);
\end{minted}
The context handle \texttt{ctx} must not be passed to any TCAPI function after \texttt{tcapi::destroy\_context} has been called.

The code presented in this section can be seamlessly ported to other \texttt{C++} TCFs by modifying only the line 
\begin{minted}[bgcolor=LightGray]{cpp}
using ten = Ten;
\end{minted}
demonstrating the flexibility of TCAPI. 
For more realistic applications, readers are referred to the example code presented in Appendix~\ref{app:itebd_tcapi_impl}, which implements a TN algorithm for the imaginary-time evolution of a quantum spin chain.
For guidance on efficient use of TCAPI in large-scale production environments, including supercomputers, we refer readers to the open-source library \cite{tnbp} used for Application B in the following section.

\section{Numerical Demonstrations \label{sec:demo}}

By design, TN applications developed with TCAPI are highly portable, while the interface introduces only a minimal abstraction overhead in the \texttt{C++} case, consistent with the zero-overhead principle of modern \texttt{C++}. 
As a result, their performance is expected to be comparable to that achieved using the native APIs of the underlying TCFs. 
We substantiate these claims---namely, high portability and low abstraction overhead for the \texttt{C++} interface---by benchmarking two representative TN applications arising in quantum many-body computations. 
The first targets ground-state calculations, while the second focuses on quantum dynamics. 
Both applications are implemented using the \texttt{C++} TCAPI as well as the native APIs of one of the adopted TCFs, enabling a direct and fair performance comparison.

\begin{figure}[t]
    \centering
    \includegraphics[width=\linewidth]{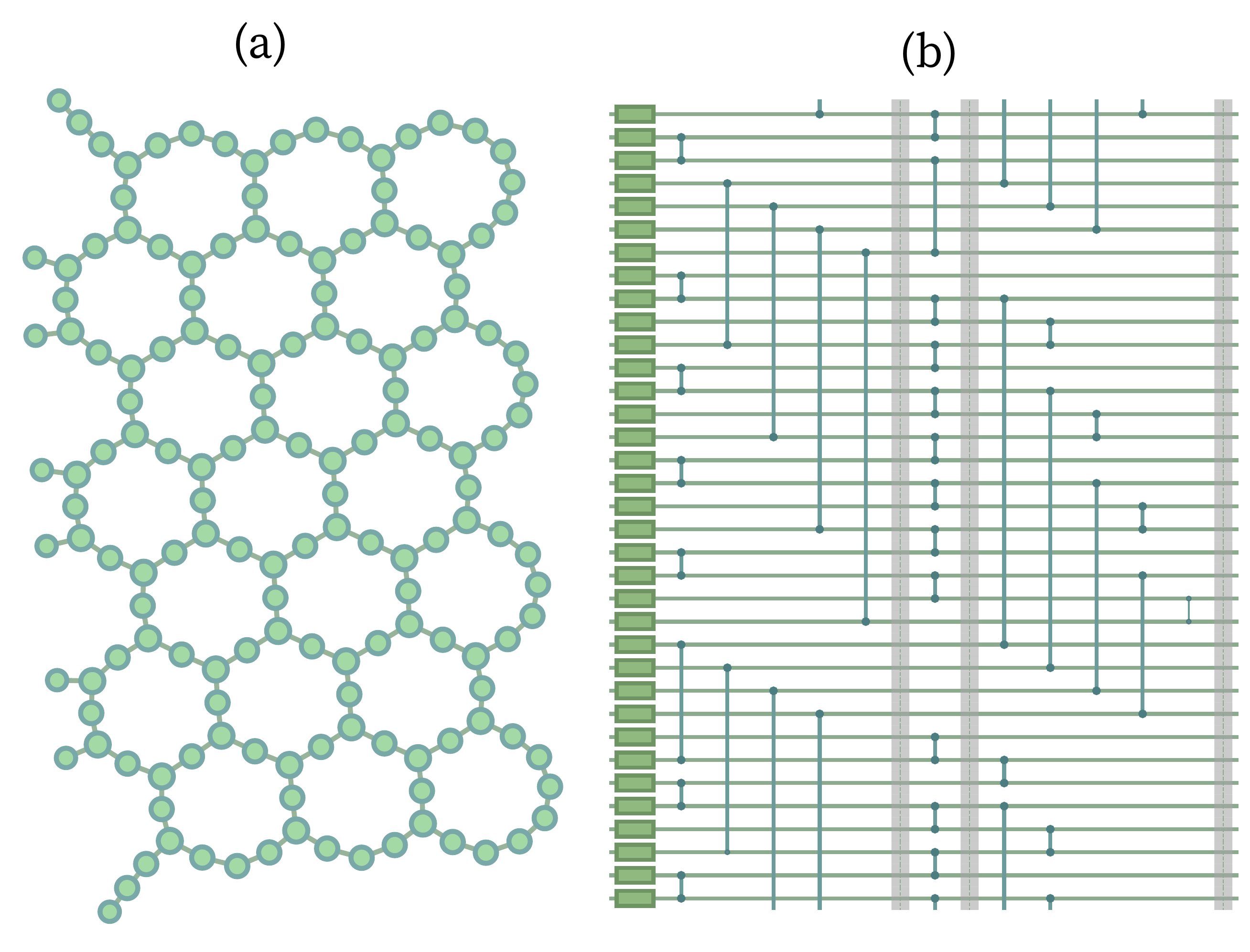}
    \caption{Configuration used in Application B. 
    (a) Heavy-hex lattice geometry corresponding to the 156-qubit superconducting chips, e.g., \texttt{ibm\_marrakesh} and \texttt{ibm\_kobe}, in IBM Quantum System Two. This connectivity is mapped directly onto the 2dTNS used in the simulation.
    (b) Representative portion of the quantum circuit implementing a single Floquet cycle of the KIM. Rectangles denote single-qubit $R^X$ gates, while wires connecting pairs of qubits represent two-qubit $R^{ZZ}$ gates. Gray vertical bands indicate groups of $R^{ZZ}$ gates executed concurrently by the parallel algorithm (see Appendix~\ref{app:bp-qc}).
    }
    \label{fig:circuit_info}
\end{figure}

\subsection{Two representative TN applications}

Our first TN application (Application A) computes the ground state of the one-dimensional (1D) transverse-field Ising model (TFIM) using imaginary-time evolution with the infinite time-evolving block decimation (iTEBD) algorithm \cite{vidal2007classical}.
The Hamiltonian is given by 
\begin{equation}
    \label{eq:1dtfim}
    H = -J\sum_{i} Z_{i}Z_{i+1} + g\sum_{i} X_{i}~,
\end{equation}
where $Z_i$ ($X_i$) denotes the Pauli-$Z$ (Pauli-$X$) operator acting on site $i$.
Here, $J$ is the nearest-neighbor Ising coupling, which we set to unity hereafter, and $g$ is the transverse-field strength along the $x$ direction.  
For $g=1$, the value used in the benchmarks below, the system undergoes a quantum phase transition \cite{pfeuty1970the}, accompanied by a closing of the excitation gap. 
In contrast to gapped one-dimensional systems, whose ground states obey an entanglement area law \cite{hastings2007area} and can therefore be efficiently approximated by finite-bond-dimension MPSs, critical one-dimensional systems exhibit logarithmic growth of bipartite entanglement with subsystem size, as described by conformal field theory \cite{calabrese2004entanglement}. 
A finite MPS bond dimension then imposes an effective entanglement and correlation-length cutoff, leading to finite-entanglement scaling \cite{tagliacozzo2008scaling,pollmann2009theory}. 
This critical setting therefore provides a controlled way to increase the bond dimension and collect performance data for increasingly demanding tensor-network workloads. 
Implementation details of the \texttt{C++} TCAPI-based version of Application A are provided in Appendix~\ref{app:itebd_tcapi_impl}.

\begin{figure}[t]
    \centering
    \includegraphics[width=\linewidth]{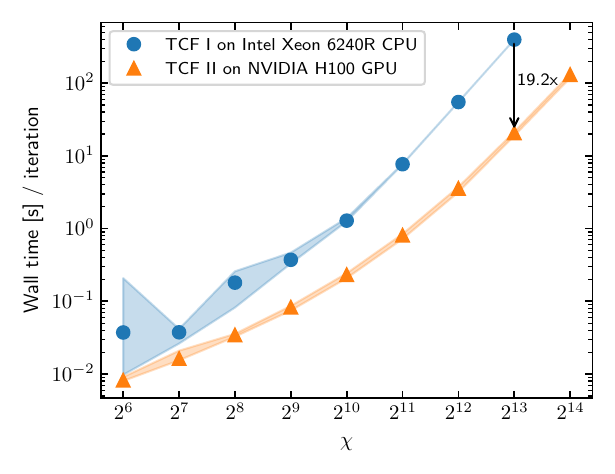}
    \caption{
    Wall-clock time per iTEBD iteration as a function of the bond dimension $\chi$ for the ground-state calculation of 1D TFIM at $g=1$ (Application A). 
    Blue dots correspond to TCF I executed on an Intel Xeon 6240R CPU, while orange triangles correspond to TCF II executed on an NVIDIA H100 GPU. 
    Marks indicate mean values computed over the final ten iterations, and shaded bands represent the corresponding minimum–maximum ranges.
    }
    \label{fig:itebd_timing}
\end{figure}

Our second TN application (Application B) simulates the dynamics of the kicked Ising model (KIM) on a heavy-hex lattice \cite{kim2023evidence, shinjo2024unveiling} using a two-dimensional tensor-network state (2dTNS) combined with belief propagation (BP). 
An overview of the 2dTNS-BP algorithm is given in Appendix~\ref{app:bp-qc}.
One Floquet cycle is implemented by the quantum circuit
\begin{equation}
    U = \prod_{\langle i,j\rangle} R^{ZZ}_{ij}(\theta_{ZZ}) \prod_{i} R^X_i(\theta_X)~,
\end{equation}
where $R^{ZZ}_{ij}(\theta_{ZZ}) = \exp(-i\theta_{ZZ} Z_i Z_j/2)$ denotes a two-qubit ZZ rotation acting on sites $i$ and $j$, and $R^X_i(\theta_X)$ is a single-qubit X rotation on site $i$. 
Here, $\langle i,j\rangle$ runs over the edges of the lattice connectivity graph. 
The lattice geometry used for benchmarking and a schematic of the corresponding circuit are shown in Figs.~\ref{fig:circuit_info}(a) and \ref{fig:circuit_info}(b), respectively.
After $M$ Floquet cycles, the state evolves to 
\begin{equation}
    \vert \psi \rangle = U^{M}\vert 00\cdots 0 \rangle~,
\end{equation}
starting from the all-zero product state.
In the benchmarks presented below, we use $\theta_X = 0.7\pi$, $\theta_{ZZ} = 0.25\pi$, and $M = 10$.

\begin{figure}[t]
    \centering
    \includegraphics[width=\linewidth]{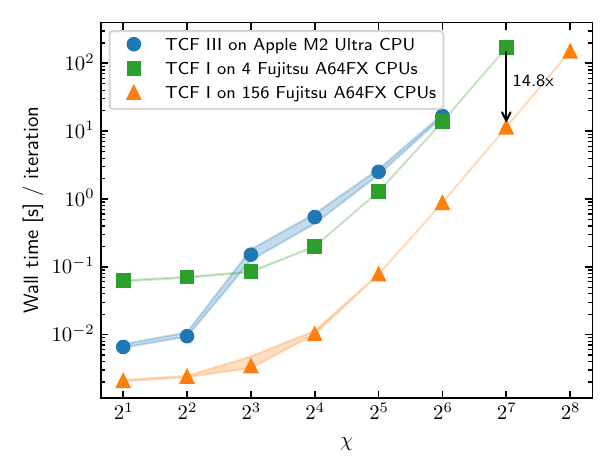}
    \caption{
        Wall-clock time per BP iteration as a function of the bond dimension $\chi$ for the KIM simulation on a heavy-hex lattice (Application B). 
        Blue dots correspond to TCF III executed on an Apple M2 Ultra CPU, while green squares and orange triangles correspond to TCF I executed on 4 and 156 nodes of supercomputer Fugaku, respectively.
        Marks indicate mean values over ten BP iterations, and shaded regions represent the corresponding minimum–maximum ranges.
    }
    \label{fig:bp_walltime_system}
\end{figure}

\subsection{Demonstration of high portability across heterogeneous platforms}

The numerical demonstrations use \texttt{C++} TCAPI implementations on top of three TCFs. TCF I employs \texttt{gqten::tensor} from the \textit{GraceQ/tensor} library \cite{gqten} for CPU execution. TCF II targets NVIDIA GPUs using \texttt{cuda\_tensor}, a thin wrapper over \textit{cuTENSOR} \cite{NVIDIA_cuTENSOR} and \textit{cuTensorNet} \cite{cuTensorNet}. TCF III is based on \textit{Cytnx}.

We first assess portability between CPU and GPU platforms by compiling the \texttt{C++} TCAPI-based implementation of Application A against two back ends: TCF I (CPU) and TCF II (GPU). 
The resulting binaries are executed on an Intel Xeon 6240R CPU and an NVIDIA H100 GPU, respectively. 
Figure~\ref{fig:itebd_timing} reports the wall-clock time as a function of the bond dimension $\chi$. 
For $\chi > 1000$, TCF II achieves speedups exceeding one order of magnitude relative to TCF I. This result demonstrates that TCAPI-based TN applications can be retargeted to accelerator-based hardware with minimal effort, while delivering substantial performance gains.

We next evaluate portability between a local development environment and a massively parallel production environment by compiling the \texttt{C++} TCAPI-based implementation of Application B against TCF I and TCF III. 
The TCF I binary is executed on the supercomputer Fugaku using 4 and 156 Fujitsu A64FX nodes (production environment), whereas the TCF III binary is executed on an Apple M2 Ultra CPU (development environment). 
Figure~\ref{fig:bp_walltime_system} shows the wall-clock time of the BP procedure applied on the tensors in the 2dTNS. 
On Fugaku, we observe favorable strong scaling: increasing the node count from 4 to 156 reduces the runtime by approximately a factor of $15$. 
Although the 4-node configuration is slightly slower than the Apple M2 Ultra, the 156-node configuration achieves substantially improved performance. 
These results indicate that TN applications developed using TCAPI can be migrated seamlessly from a local shared-memory environment to a massively parallel supercomputing architecture and can effectively exploit its resources without any modifications to the application source code.

\subsection{Demonstration of low abstraction overhead}

We next demonstrate that TN applications developed using the \texttt{C++} TCAPI achieve performance comparable to implementations written directly against the native APIs of the underlying TCFs. This confirms that TCAPI introduces only a negligible abstraction overhead.

\begin{figure}[t]
    \centering
    \includegraphics[width=\linewidth]{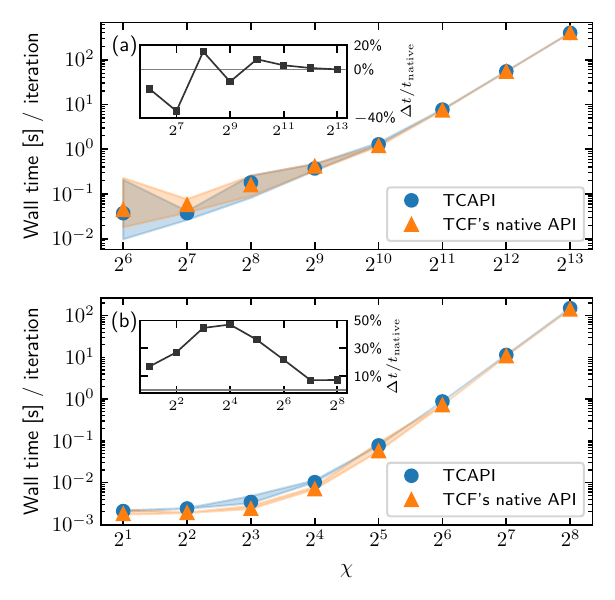}
    \caption{
        Comparison between TCAPI-based implementations and native implementations using TCF I, analogous to Fig.~\ref{fig:itebd_timing} and Fig.~\ref{fig:bp_walltime_system}.
        Blue dots correspond to TN applications developed using TCAPI, while orange triangles correspond to implementations written directly against the native API of TCF I.
        Insets show the relative timing difference, $(t_{\rm TCAPI}-t_{\rm native})/t_{\rm native}$, quantifying the abstraction overhead introduced by TCAPI.
        (a) Application A. (b) Application B.
    } 
    \label{fig:timing_tcapi_native}
\end{figure}

To quantify the overhead, we reimplemented Applications A and B using the native API of TCF I (\texttt{gqten::tensor}) and repeated the same numerical benchmarks on the corresponding hardware platforms. 
For Application A executed on an Intel Xeon 6240R CPU, the TCAPI-based implementation exhibits performance close to that of the native API implementation across all tested bond dimensions, as shown in Fig.~\ref{fig:timing_tcapi_native}(a).
The inset shows that the relative timing difference, $(t_{\rm TCAPI}-t_{\rm native})/t_{\rm native}$, remains small and does not grow systematically with the bond dimension, indicating that the abstraction overhead introduced by TCAPI is negligible compared with the overall computational cost.
For Application B executed on Fugaku using 156 nodes, the native implementation is marginally faster at small bond dimensions ($\chi = 4~\text{-}~32$), where fixed overheads are more visible because the absolute wall time is short.
However, the two implementations exhibit nearly indistinguishable performance for $\chi > 64$, as shown in Fig.~\ref{fig:timing_tcapi_native}(b), and the inset confirms that the relative difference decreases as the computational workload increases.
Overall, these results demonstrate that TCAPI does not introduce any significant performance degradation, even in large-scale supercomputing environments.

\section{Summary and outlook \label{sec:summary_outlook}}

In summary, we have introduced TCAPI---an application-oriented, lightweight API that enables portable, high-performance TN applications across diverse TCFs---and specified it for both \texttt{C++} and \texttt{Python} with shared tensor semantics. 
TCAPI comprises (i) coherent type-system conventions that model tensor objects and their associated metadata, (ii) a comprehensive set of primitives for essential tensor computations, and (iii) runtime configuration environment variables that govern execution behavior. 
We implemented the \texttt{C++} specification across TCFs spanning shared-memory CPU systems, GPU-accelerated platforms, and massively parallel supercomputing architectures, and developed representative TN applications using TCAPI. 
Our benchmark results demonstrate that applications can be migrated across TCFs with minimal code modifications while achieving performance comparable to implementations written directly against native framework APIs. 
Taken together, these results position TCAPI as a viable pathway toward standardizing an application-level interface for generic, high-performance tensor computation on current and emerging HPC architectures.

Looking ahead, several developments are required to further mature TCAPI into a fully standardized interface for generic tensor computations.
As modern computing platforms increasingly rely on CPU–accelerator architectures---most notably GPUs---exploiting concurrency has become a first-class performance objective for TN workloads. 
Accordingly, future revisions of TCAPI will introduce asynchronous APIs designed to support highly concurrent execution. 
In addition, automatic differentiation has become an essential tool for developing and optimizing TN algorithms, and we plan to incorporate a standardized automatic-differentiation API in a subsequent version of TCAPI. 
While the current specification focuses primarily on dense tensors, symmetry support is an important future direction, especially for quantum many-body applications. 
A natural TCAPI extension would expose symmetry information through API-level symmetry metadata, such as quantum-number sectors, sector dimensions, leg directions, parity or grading, and context-level fusion or sign conventions, while leaving storage layouts and backend-specific implementations to each TCF. 
This would allow contractions, permutations, and decompositions to retain the same high-level calling structure for dense and symmetry-structured tensors. 
Designing such a portable specification requires careful treatment of the differing conventions used by existing libraries and is left to future work.

The \texttt{C++} specification targets the \texttt{C++17} standard to maximize portability across leadership-class HPC systems. 
Nonetheless, the absence of modern language features such as concepts and constraints limits the expressiveness and robustness of TCAPI-based applications.
We therefore plan to migrate the \texttt{C++} specification to \texttt{C++20} or later standards in future releases. 
The \texttt{Python} specification introduced here provides a basis for integration with \texttt{Python}-based TCFs developed for machine learning, such as \textit{JAX} and \textit{PyTorch}. 
Given the widespread use of \texttt{Julia} in the tensor-computing community, a companion Julia specification remains an important future direction, particularly for quantum-physics TCFs such as \textit{ITensors.jl} \cite{ITensorsjl} and \textit{TensorKit.jl} \cite{TensorKitjl}.

\section{Acknowledgments}
We acknowledge helpful discussions with Q. Zhao, C. Li, H. Xu, and S. Niu. 
We thank K. F. Hui for developing the TCAPI implementations based on \textit{ITensor} and \textit{NumPy}.
This work was primarily supported by the U.S. Department of Energy, Office of Basic Energy Sciences, under Grant No. DE-FG02-06ER46305 (R.Y.S and D.N.S). 
This work was also supported in part by project JPNP20017, commissioned by the New Energy and Industrial Technology Development Organization (NEDO), and by JSPS KAKENHI Grant No. JP21H04446. 
Additional support was provided by JST COI-NEXT (Grant No. JPMJPF2221) and by the Program for Promoting Research of the Supercomputer Fugaku (Grant No. MXP1020230411) from MEXT, Japan. 
We further acknowledge support from the UTokyo Quantum Initiative and the RIKEN TRIP initiative (RIKEN Quantum). 
Numerical benchmarks were performed on the HOKUSAI supercomputer at RIKEN and the Fugaku supercomputer at the RIKEN Center for Computational Science (Project IDs: ra000011, ra010014), as well as through the HPCI System Research Project (Project ID: hp250069). 

The data that support the findings of this article are openly available \cite{tcapi_data}.

\clearpage
\appendix
\onecolumngrid
\section{Implementation of Application A using \texttt{C++} TCAPI 
\label{app:itebd_tcapi_impl}}

In this appendix, we provide the \texttt{C++} TCAPI-based implementation of Application A, which performs the ground-state calculation of the 1D TFIM using iTEBD with imaginary-time evolution.

\begin{minted}{cpp}
#include "tcapi/tcapi.h"
using Ten = /* The concrete tensor type, for example gqten::tensor<gqten::Real_8> */;

#include <array>
#include <vector>
#include <random>

using ContextHandle = tcapi::context_handle_t<Ten>;
using Elem = tcapi::elem_t<Ten>;
using Real = tcapi::real_t<Ten>;

int main (int argc, char *argv[]) {
  // Create context
  ContextHandle context;
  tcapi::create_context(context);

  // Define the parameters of the model and the simulation
  Real J = 1.0, g = 1.0;
  Real dt = 0.005;
  int d = 2, chi = 5;
  int N = 3000;

  // Prepare Hamiltonian and imaginary-time evolution operator
  auto g2 = - 0.5 * g;
  auto h = tcapi::zeros<Ten>(context, {4, 4});
  tcapi::set_elem(context, h, {0, 0}, J); tcapi::set_elem(context, h, {3, 3}, J);
  tcapi::set_elem(context, h, {1, 1}, -J); tcapi::set_elem(context, h, {2, 2}, -J);
  std::vector<tcapi::elem_coors_t<Ten>> coors_set {
    {0, 1}, {0, 2}, {1, 0}, {1, 3},
    {2, 0}, {2, 3}, {3, 1}, {3, 2}
  };
  for (auto &coors : coors_set) { tcapi::set_elem(context, h, coors, g2); }
  Ten u;
  tcapi::scale(context, h, -dt, u);
  tcapi::exp(context, u, 1);
  tcapi::reshape(context, u, {d, d, d, d});

  // Initialize infinite MPS
  std::mt19937 engine;
  std::uniform_real_distribution<Real> dis(-1.0, 1.0);
  auto gen = [&dis, &engine]() { return dis(engine); };
  std::array<Ten, 2> Gamma {
      tcapi::random<Ten>(context, {chi, d, chi}, gen),
      tcapi::random<Ten>(context, {chi, d, chi}, gen)
  };
  std::array<Ten, 2> Lambda {
      tcapi::random<Ten>(context, {chi}, gen),
      tcapi::random<Ten>(context, {chi}, gen)
  };
  for (auto &lambda : Lambda) { tcapi::diag(context, lambda); }

  // iTEBD iterations
  for (int i = 0; i < N; i++) {
    auto A = i % 2; 
    auto B = (i+1) % 2;
    Ten Theta;
    tcapi::contract(context, Lambda[B],"ij", Gamma[A],"jkl", Theta,"ikl");
    tcapi::contract(context, Theta, "ijk", Lambda[A],"kl", Theta,"ijl");
    tcapi::contract(context, Theta,"ijk", Gamma[B],"klm", Theta,"ijlm");
    tcapi::contract(context, Theta,"ijkl", Lambda[B],"lm", Theta,"ijkm");

    tcapi::contract(context, Theta,"ijkl", u,"jkmn", Theta,"imnl");
    Real trunc_err = 0.0;
    tcapi::trunc_svd(context,
      Theta, 2,
      Gamma[A], Lambda[A], Gamma[B],
      trunc_err,
      chi,
      1.0e-10
    );
    tcapi::normalize(context, Lambda[A]);

    auto LambdaB_inv = tcapi::copy(context, Lambda[B]);
    tcapi::diag(context, LambdaB_inv);
    tcapi::for_each(context, LambdaB_inv, [](Elem &elem) { elem = 1.0 / elem; });
    tcapi::diag(context, LambdaB_inv);
    tcapi::contract(context, LambdaB_inv,"ij", Gamma[A],"jkl", Gamma[A],"ikl");
    tcapi::contract(context, Gamma[B],"ijk", LambdaB_inv,"kl", Gamma[B],"ijl");
    if (i >= 1) {
      Elem E_iTEBD = 0.0;
      tcapi::for_each(context,
          Theta,
          [&E_iTEBD](const Elem &elem) { E_iTEBD += (elem*elem); });
      E_iTEBD = - log(E_iTEBD) / dt / 2.0;
      std::printf("Iter %5i E_iTEBD = %.15f\n", i, E_iTEBD);
    }
  }

  // Destroy context
  tcapi::destroy_context(context);
  return 0;
}
\end{minted}
\twocolumngrid

\phantom{    }
\clearpage

\section{2dTNS-BP Algorithm for Quantum Circuit Simulation \label{app:bp-qc}}

This appendix summarizes the algorithm used to simulate quantum circuits in Application B, as introduced in Sec.~\ref{sec:demo}. The overall workflow is as follows. A quantum circuit is first partitioned into smaller segments, each of which---while still representing a valid subcircuit---is transformed into a tensor-network operator (TNO). These TNOs are then applied sequentially to a tensor-network state (TNS). During this process, the bond dimensions are regulated through low-rank approximations, enabling efficient state updates while preventing exponential growth in computational cost.

The subsections below describe (i) the construction of TNOs from quantum circuits and (ii) the BP-based approximation used to contract TNOs with TNSs, which together enables scalable and efficient circuit simulation.

\subsection{Construction of TNOs from quantum circuits}

\begin{figure}[t]
\includegraphics[width=0.9\hsize]{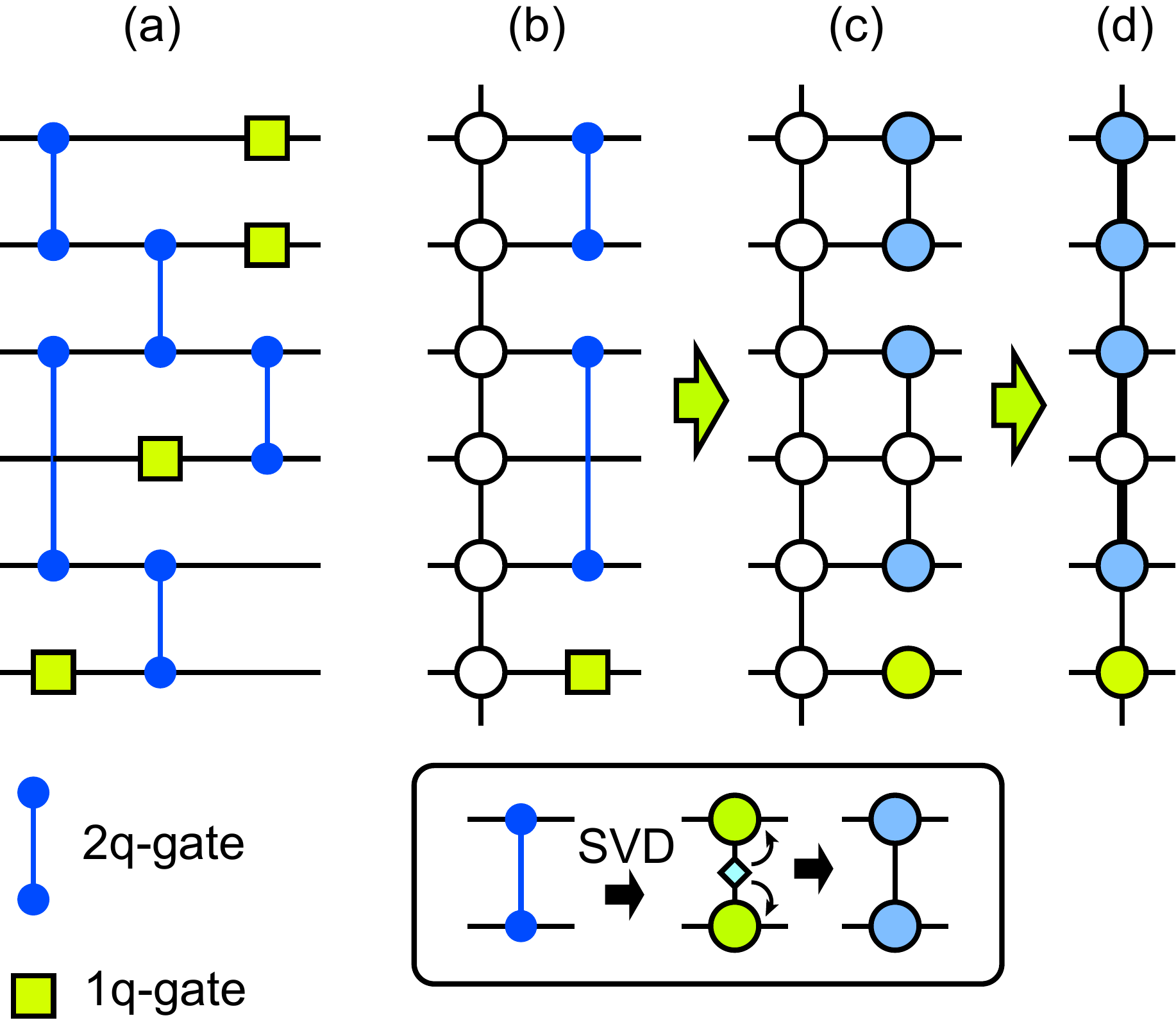}
\caption{
(a) Schematic of a quantum circuit composed of single-qubit (1q) and two-qubit (2q) gates.
(b) Construction of the base TNO representing the identity operator together with the first gate layer. 
White circles denote tensors of bond dimension one on the \textit{virtual} bonds, corresponding to the operator $1 \otimes 1 \otimes \cdots \otimes 1$.
(c) Decomposition of a two-qubit unitary gate into two site tensors---each acting on one qubit and a shared virtual bond---obtained via SVD.
Non-nearest-neighbor gates are handled by inserting identity operators along the intermediate path connecting the two target qubits. 
(d) Assembly of the full TNO by attaching the decomposed site tensors to the base network.
Tensors updated through gate contractions are highlighted in color, and thicker lines indicate enlarged bond dimensions.}
\label{fig:qasm-to-tno}
\end{figure}

In this implementation, quantum circuits specified in \texttt{OpenQASM} format~\cite{OpenQASM2} are converted into TNO representations suitable for TN processing, as outlined in Fig.~\ref{fig:qasm-to-tno}. 
Within the standard \texttt{OpenQASM} specification, one-, two-, and three-qubit gates constitute the basic instruction set. 
For clarity, we restrict the present discussion to circuits composed of single- and two-qubit gates [Fig.~\ref{fig:qasm-to-tno}(a)].

The conversion procedure starts from a bond-dimension-one TNO corresponding to the identity operator [Fig.~\ref{fig:qasm-to-tno}(b)].
Gate tensors are then incorporated sequentially. 
Single-qubit gates are applied directly to the corresponding local operator tensors, whereas each two-qubit unitary gate---represented as a fourth-order tensor---is decomposed, typically using SVD, into two local tensors connected by an auxiliary virtual bond. 
Each resulting tensor acts on one qubit and the shared virtual bond [Figs.~\ref{fig:qasm-to-tno}(b) and \ref{fig:qasm-to-tno}(c)], ensuring compatibility with the target TN topology.
For two-dimensional geometries, these identity operators are placed along the shortest path connecting the target qubits on the hardware coupling graph.

To avoid the computational explosion that would arise from attaching all gates simultaneously, barriers are inserted between layers of parallelizable gates, and separate TNOs are constructed for each layer. All \texttt{OpenQASM} circuits used in this work were generated using \textit{Qiskit} \cite{Qiskit}.

\subsection{BP-based approximate contraction scheme} 

\begin{figure*}[t]
\includegraphics[width=0.9\hsize]{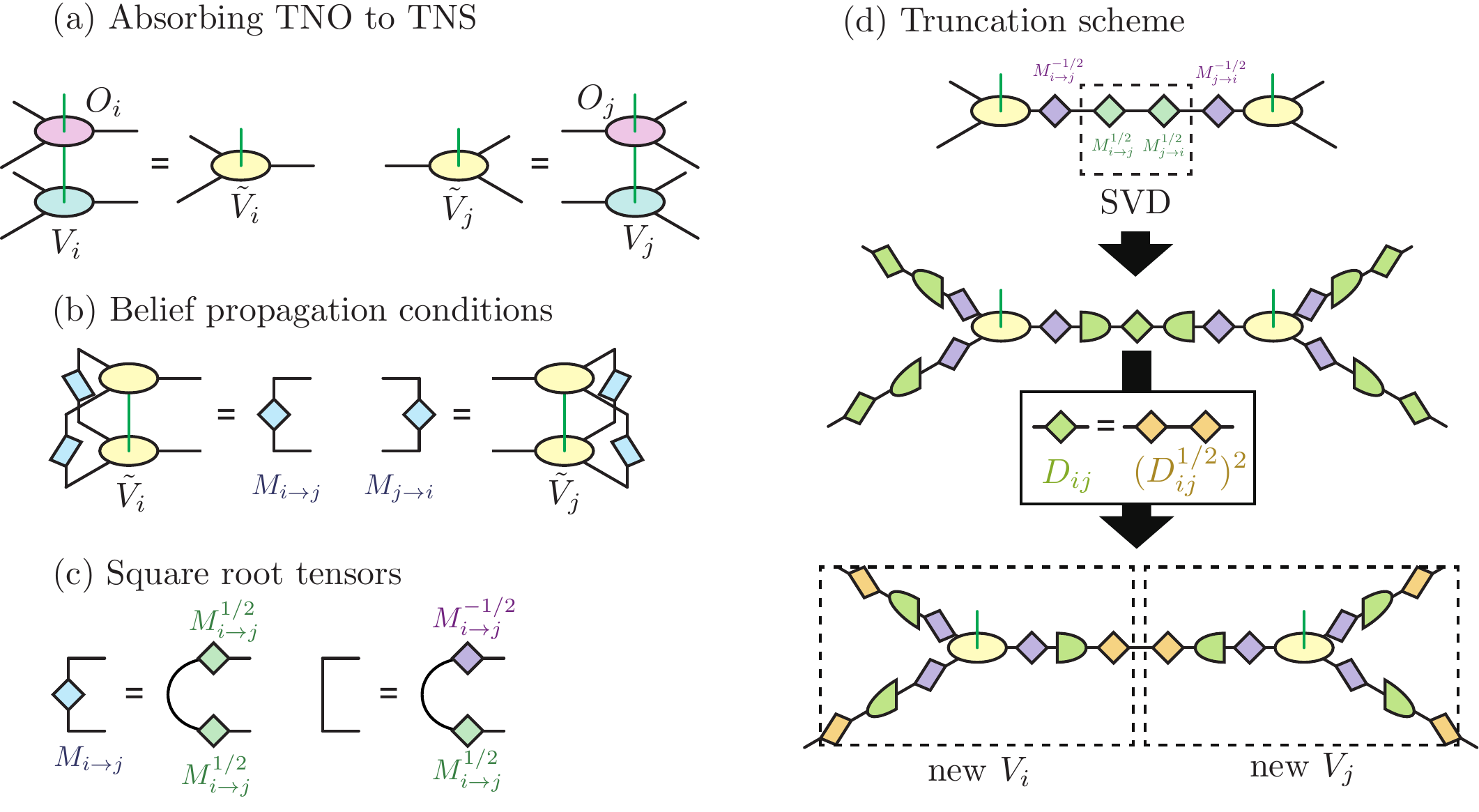}
\caption{
Workflow for updating a TNS by applying a TNO using BP, followed by low-rank approximation of the resulting tensors.
(a) Local contraction of TNO site tensors $O_i$ with TNS site tensors $V_i$,  producing intermediate tensors $\tilde V_i$. 
(b) Fixed-point relations for the BP message tensors. 
(c) Consistency constraints for the square-root and inverse square-root message tensors, $M_{i\to j}^{1/2}$ and $M_{i\to j}^{-1/2}$.
(d) SVD-based truncation of the message tensors $M_{i\to j}$. With $D_{ij}$ denoting the diagonal matrix of singular values, the largest $\chi$ singular values are retained to reconstruct the updated site tensors $V_i$ within the prescribed maximum bond dimension.
Dashed boxes highlight the subnetwork in which the updated tensors $V_i$ are assembled.
}
\label{fig:bp-tno-tns}
\end{figure*}

We employ a BP–based regauging procedure to efficiently contract a TNO with a TNS, while applying low-rank truncations to the updated tensors.
The approach follows Ref.~\cite{tindall2023belief} and is summarized schematically in Fig.~\ref{fig:bp-tno-tns}. The procedure proceeds as follows. 
First, the TNO site tensors $O_i$ are contracted with the corresponding TNS site tensors $V_i$ [Fig.~\ref{fig:bp-tno-tns}(a)]. BP is then iterated over all bonds until the message tensors $M_{i\to j}$ converge [Fig.~\ref{fig:bp-tno-tns}(b)]. Upon convergence, square-root and inverse square-root tensors, $M_{i\to j}^{1/2}$ and $M_{i\to j}^{-1/2}$, are constructed so as to satisfy the consistency relations illustrated in Fig.~\ref{fig:bp-tno-tns}(c). These tensors are inserted on each bond [Fig.~\ref{fig:bp-tno-tns}(d)], after which SVD-based low-rank truncation is applied to obtain the updated site tensors $V_i$. Although BP involves nearest-neighbor communication, all contractions are strictly local, allowing the algorithm to be parallelized naturally with only local data exchanges.

Relative to TEBD, the present approach offers greater flexibility. TEBD requires a fixed ordering in which adjacent pairs of sites are updated sequentially, whereas operating directly on the TNO enables more general update schedules. For example, gates acting on non-adjacent sites can be incorporated without explicit reordering. The main trade-off is that the computational cost increases with the bond dimension of the TNO; consequently, very large TNOs reduce overall efficiency.
An additional distinction concerns the regauging strategy. In TEBD with the simple-update scheme, regauging is performed after truncation, whereas in the present method it is carried out during TNO attachment. In principle, this ordering can provide more context-aware local environments for approximation, although our numerical tests did not reveal a significant practical advantage.
Conversely, unlike TEBD---which can omit certain regauging passes and truncate directly---the present TNO–TNS scheme requires regauging at every update step, since truncation is applied to the BP message tensors. As a result, the overall update loop is modestly more computationally demanding.

We provide an open-source implementation of the 2dTNS-BP algorithm, \textit{TNBP}~\cite{tnbp}. Beyond the benchmarks reported in this work, \textit{TNBP} is ready for simulating generic large-scale quantum circuits on massively parallel computing systems.

\clearpage

\section{Tensor Computing Application Programming Interface \label{app:tcapi_spec}}

In this appendix, we present the TCAPI specification for the \texttt{C++} and \texttt{Python} language interfaces.
The tensor terminology and operation semantics introduced above are shared by all TCAPI language interfaces.
Each language exposes these concepts through type-system conventions appropriate to that language: the \texttt{C++} interface uses compile-time tensor types and traits, while the \texttt{Python} interface uses runtime tensor-kind descriptors and raw backend tensor values.

\subsection{Type system \label{sec:type_system}}

\subsubsection{\texttt{C++} type system}

To expose a uniform interface across heterogeneous TCFs, the \texttt{C++} interface relies heavily on generic programming.
TCAPI therefore defines a self-contained type system for abstract tensor objects and their associated traits.
Building on this foundation, TCAPI provides core tensor operations and runtime configuration variables for diagnostics and performance reporting.
Unless stated otherwise, \texttt{C++} classes and functions reside in the \texttt{tcapi} namespace and are expected to be accessible through the single header \texttt{tcapi/tcapi.h}.

The \texttt{C++} TCAPI type system consists of three components: an abstract tensor type, denoted \texttt{TenT}; a set of trait types associated with \texttt{TenT}, exposed through \texttt{tcapi::tensor\_traits}; and a collection of auxiliary types.

\textbf{TenT ---}
\texttt{TenT} is the core abstraction in the \texttt{C++} TCAPI interface.
It denotes a generic, compile-time-resolved tensor type whose concrete implementation and associated types, such as element and shape types, are accessed through \texttt{tcapi::tensor\_traits}.
An object of type \texttt{TenT} represents a tensor through its logical tensor elements and metadata, and may manage backend resources associated with that tensor, including data storage and representation-specific metadata.
These logical tensor elements need not all be stored explicitly: different objects of the same \texttt{TenT} may use dense storage or implementation-defined compact storage, such as diagonal storage.
In the \texttt{C++} TCAPI interface, \texttt{TenT} is not required to be copy-constructible or copy-assignable.
TCAPI code that is intended to work across conforming backends must therefore not duplicate tensors by ordinary \texttt{C++} copying; explicit standalone deep duplication is provided through \texttt{tcapi::copy(ctx, orig)}, which receives the backend context required to perform the copy.
Ownership transfer is provided through \texttt{tcapi::move(ctx, from)}.
If a backend provides ordinary \texttt{C++} copy construction or copy assignment for \texttt{TenT}, those operations are implementation extensions and are not part of the portable TCAPI contract.

Each \texttt{C++} TCAPI implementation exposes its concrete tensor types through the namespace-level alias \texttt{tcapi::supported\_ten\_types\_t}, accessible through \texttt{tcapi/tcapi.h}:
\begin{minted}[bgcolor=LightGray]{cpp}
namespace tcapi {
using supported_ten_types_t =
  std::tuple<
    /* implementation-provided tensor types */
  >;
}
\end{minted}
This is a compile-time type catalog.
Its ordering is implementation-defined.
The catalog lists the types provided by the implementation, while their compatibility with a particular execution context remains backend-dependent.
The number of listed types, an individual tensor type, and its element type can be obtained using standard tuple traits and \texttt{tcapi::elem\_t}:
\begin{minted}[bgcolor=LightGray]{cpp}
using Types = tcapi::supported_ten_types_t;
constexpr auto count = std::tuple_size_v<Types>;
using FirstTen = std::tuple_element_t<0, Types>;
using FirstElem = tcapi::elem_t<FirstTen>;
\end{minted}

\begin{table}
    \centering
    \def\arraystretch{1.5}
    \begin{tabular}{p{0.3\linewidth} p{0.7\linewidth}}
        \hline
        \hline
        \textbf{Member type}        & \textbf{Description} \\
        \hline
        \texttt{ten\_t}          & Concrete tensor type corresponding to \texttt{TenT}. \\
        \texttt{order\_t}        & Integral type for the tensor order. \\
        \texttt{shape\_t}        & Shape type in bond-index order; \texttt{List<bond\_dim\_t>}. \\
        \texttt{bond\_dim\_t}    & Integral type for a bond dimension. \\
        \texttt{bond\_idx\_t}    & Integral type for a bond index (zero-based). \\
        \texttt{bond\_label\_t}  & Integral type for a user-defined bond label. \\
        \texttt{ten\_size\_t}    & Integral type for the number of tensor elements, i.e., the product of the dimensions in the tensor shape. \\
        \texttt{elem\_t}         & Element (value) type of \texttt{ten\_t}; a real or complex floating-point type supporting arithmetic (\texttt{+}, \texttt{-}, \texttt{*}, \texttt{/}), comparisons, and common math functions via argument-dependent lookup (ADL). \\
        \texttt{elem\_coor\_t}   & Integral type for a single coordinate along a bond (zero-based). \\
        \texttt{elem\_coors\_t}  & Coordinate tuple type for an element; \texttt{List<elem\_coor\_t>} in bond-index order. \\
        \texttt{real\_t}         & Real scalar type; equal to \texttt{elem\_t} if \texttt{elem\_t} is real, otherwise the real component type of \texttt{elem\_t}. \\
        \texttt{real\_ten\_t}    & Tensor type identical to \texttt{ten\_t} but with \texttt{elem\_t} replaced by \texttt{real\_t}. \\
        \texttt{cplx\_t}         & Complex scalar type; equal to \texttt{elem\_t} if \texttt{elem\_t} is complex, otherwise the complex counterpart of \texttt{elem\_t}. \\
        \texttt{cplx\_ten\_t}    & Tensor type identical to \texttt{ten\_t} but with \texttt{elem\_t} replaced by \texttt{cplx\_t}. \\
        \texttt{context\_handle\_t} & Handle to the backend context maintained by the underlying TCF that provides \texttt{ten\_t}. \\
        \hline
    \end{tabular}
    \caption{Member types defined in \texttt{tcapi::tensor\_traits<TenT>} and their intended roles.}
    \label{tab:tensor_traits_member_types}
\end{table}

\textbf{tcapi::tensor\_traits ---}
Properties intrinsically associated with \texttt{TenT} are exposed as member types of the \texttt{tcapi::tensor\_traits} class template:
\begin{minted}[bgcolor=LightGray]{cpp}
template <typename TenT>
struct tensor_traits;
\end{minted}
Table~\ref{tab:tensor_traits_member_types} lists the required member types and their roles.
To ensure a uniform API across heterogeneous backends, some associations are currently constrained; for example, \texttt{tcapi::tensor\_traits<TenT>::shape\_t} must be \texttt{List<tcapi::tensor\_traits<TenT>::bond\_dim\_t>}.
These constraints may be relaxed in future TCAPI revisions.

To streamline generic programming workflows, TCAPI provides alias templates that forward the associated types listed in Tab.~\ref{tab:tensor_traits_member_types} from \texttt{tcapi::tensor\_traits}:
\begin{minted}[bgcolor=LightGray]{cpp}
template <typename TenT>
using target_type =
typename tensor_traits<TenT>::target_type;
\end{minted}
Here, \texttt{target\_type} represents any required associated types (e.g., \texttt{order\_t}, \texttt{shape\_t}, or \texttt{elem\_t}); these aliases eliminate repetitive boilerplate of the form \texttt{typename tensor\_traits<TenT>::...}, so a function template that processes a single element can be written as
\begin{minted}[bgcolor=LightGray]{cpp}
template <typename TenT>
void process_element(
    const tcapi::elem_t<TenT> el
);
\end{minted}

\textbf{Auxiliary types --- \label{sec:aux_types}}
In addition to the associated types described above, TCAPI defines a set of auxiliary types.
These types are used both to construct the member types listed in Tab.~\ref{tab:tensor_traits_member_types} and as parameter or argument types for the core functions introduced in Sec.~\ref{sec:functions}.
All auxiliary types are summarized in Tab.~\ref{tab:auxiliary_types}.

For notational convenience in Sec.~\ref{sec:functions}, we introduce two implementation-level aliases: \texttt{bond\_idx\_pairs\_t<TenT>}, defined as an alias for \texttt{List<Pair<tcapi::bond\_idx\_t<TenT>, tcapi::bond\_idx\_t<TenT>>>}, and \texttt{bond\_idx\_elem\_coor\_pair\_map<TenT>}, defined as an alias for \texttt{Map<tcapi::bond\_idx\_t<TenT>, Pair<tcapi::elem\_coor\_t<TenT>, tcapi::elem\_coor\_t<TenT>>>}.
These aliases are intended solely for internal specification purposes.
They are not part of the formal TCAPI specification and are therefore not exported in the \texttt{tcapi} namespace.

\begin{table}[h]
    \centering
    \def\arraystretch{1.5}
    \begin{tabular}{p{0.3\linewidth} p{0.7\linewidth}}
        \hline
        \hline
        \textbf{Auxiliary type}        & \textbf{Description} \\
        \hline
        \texttt{List<T>} & Sequence container of values of type \texttt{T}; alias of \texttt{std::vector<T>}. \\
        \texttt{CRef<T>} & Non-owning const reference to an object of type \texttt{T}; alias of \texttt{std::reference\_wrapper<const T>}. \\
        \texttt{Pair<T, U>} & Pair of values of types \texttt{T} and \texttt{U}; alias of \texttt{std::pair<T,U>}. \\
        \texttt{Map<T, U>} & Associative container mapping \texttt{T} $\to$ \texttt{U}; alias of \texttt{std::unordered\_map<T,U>}. \\
        \hline
    \end{tabular}
    \caption{Auxiliary types defined in TCAPI and their intended roles.}
    \label{tab:auxiliary_types}
\end{table}

\subsubsection{\texttt{Python} type system}
\texttt{Python} TCAPI exposes tensor types through a runtime descriptor called \texttt{TenKind}.
A \texttt{TenKind} is the \texttt{Python} analogue of the \texttt{C++} tensor type \texttt{TenT}: it identifies a family of backend-native tensor values and carries the type information needed by TCAPI functions.
A \texttt{TenKind} is not itself a tensor object, and \texttt{Python} TCAPI does not wrap returned tensors in TCAPI-defined tensor classes.
Instead, \texttt{Python} TCAPI functions return backend-native tensors.
A \texttt{TenKind} identifies the backend tensor family and element type, not a particular storage representation.
Dense and diagonal tensors of the same backend family and dtype therefore have the same \texttt{TenKind}.
If the underlying TCF does not support compact diagonal storage, TCAPI routines that produce diagonal tensors return dense second-order backend tensors with the diagonal entries determined by the operation and all off-diagonal entries set to zero.

The required metadata fields associated with a \texttt{TenKind} mirror the \texttt{C++} traits in \texttt{tcapi::tensor\_traits<TenT>}:
\begin{table}[h]
    \centering
    \def\arraystretch{1.5}
    \begin{tabular}{p{0.3\linewidth} p{0.7\linewidth}}
        \hline
        \hline
        \textbf{Metadata} & \textbf{Description} \\
        \hline
        \texttt{ten\_type} & Raw backend tensor class returned by TCAPI functions. \\
        \texttt{dtype} & Backend-native dtype of elements; \texttt{Python} analogue of \texttt{tcapi::elem\_t<TenT>}. \\
        \texttt{real\_dtype} & Backend-native real dtype; \texttt{Python} analogue of \texttt{tcapi::real\_t<TenT>}. \\
        \texttt{cplx\_dtype} & Backend-native complex dtype; \texttt{Python} analogue of \texttt{tcapi::cplx\_t<TenT>}. \\
        \texttt{real\_ten\_kind} & Tensor kind corresponding to \texttt{tcapi::real\_ten\_t<TenT>}. \\
        \texttt{cplx\_ten\_kind} & Tensor kind corresponding to \texttt{tcapi::cplx\_ten\_t<TenT>}. \\
        \hline
    \end{tabular}
    \caption{Metadata associated with a \texttt{Python} \texttt{TenKind} and their intended roles.}
    \label{tab:python_ten_kind_metadata}
\end{table}

Concrete \texttt{TenKind} objects are provided and documented by each TCAPI implementation.
\texttt{Python} TCAPI provides the following ten-kind introspection facilities:
\begin{minted}[bgcolor=LightGray]{python}
tcapi.ten_kind_ten_type(TenKind)
tcapi.ten_kind_dtype(TenKind)
tcapi.ten_kind_real_dtype(TenKind)
tcapi.ten_kind_cplx_dtype(TenKind)

tcapi.real_ten_kind(TenKind)
tcapi.cplx_ten_kind(TenKind)

tcapi.ten_kind(ctx, tensor)

tcapi.supported_ten_kinds
\end{minted}
Here, \texttt{TenKind} denotes any \texttt{Python} tensor-kind descriptor.
The function \texttt{tcapi.ten\_kind(ctx, tensor)} returns the \texttt{TenKind} associated with a raw backend tensor in the context \texttt{ctx}.
The module-level attribute \texttt{tcapi.supported\_ten\_kinds} is an implementation-wide tuple containing the \texttt{TenKind} descriptors provided by the TCAPI implementation, in implementation-defined order.
Compatibility with a particular execution context remains backend-dependent.
The element dtype associated with each descriptor can be obtained using \texttt{tcapi.ten\_kind\_dtype}.

A \texttt{Python} TCAPI context is an implementation-defined runtime handle passed to context-dependent TCAPI operations, usually as the first argument.
Its creation and destruction are specified by \texttt{tcapi.create\_context} and \texttt{tcapi.destroy\_context}.

% \paragraph{\texttt{Python} auxiliary types and values}

\texttt{Python} TCAPI does not prescribe dedicated container classes for shapes, coordinates, bond indices, or bond labels.
Instead, it uses ordinary \texttt{Python} sequences, such as tuples, lists, or implementation-defined sequence-like objects accepted by the backend.

Tensor shapes and coordinate tuples are \texttt{Python} sequences of nonnegative integers in bond-index order.
Bond indices are \texttt{Python} integers, while bond-index lists and bond-label lists are \texttt{Python} sequences of integers.
String labels are accepted by operations that define a string-label API.

Element values and scalar return values are represented by ordinary \texttt{Python} scalar values or by backend-native scalar values consistent with the relevant \texttt{TenKind}.

\subsection{Core functions \label{sec:functions}}

Building on the type system described above, TCAPI exposes a suite of core functions that cover the most common tensor operations.
Operation semantics are shared across language interfaces, while each language exposes them through its own signatures and calling conventions.
These functions are organized into six categories: (i) \textbf{Read-only queries}, which provide access to tensor metadata and single-element values (e.g., order, shape, and element lookup); (ii) \textbf{Construction and destruction} routines, which allocate, initialize, and release tensor storage; (iii) \textbf{Input/output (I/O)} facilities for loading tensors from and saving tensors to the file system; (iv) \textbf{Tensor manipulation} operations, such as reshaping, transposition, and concatenation; (v) \textbf{Tensor linear-algebra} operations, including contractions and decompositions; and (vi) \textbf{Miscellaneous routines}, which provide auxiliary functionality that is not performance critical, such as pretty-printing for debugging.

TCAPI represents scalar-valued tensor results as backend tensor objects rather than standalone scalar values.
A zeroth-order tensor has empty shape, size 1, and a single empty coordinate sequence.
Its sole element is read with \texttt{get\_elem} and updated with \texttt{set\_elem} using that empty coordinate sequence.
In the examples below, this sequence is written as \texttt{\{\}} in \texttt{C++} and \texttt{()} in \texttt{Python}.
Operations whose definitions require selecting, reordering, inserting, removing, slicing, or partitioning existing bonds are not applicable to zeroth-order tensors unless explicitly specified by the operation.

In the \texttt{C++} interface, many TCAPI routines are available in both in-place and out-of-place variants.
In-place variants mutate the supplied tensor operand(s), whereas out-of-place variants leave all inputs unmodified and return the results either as a function return value for single-tensor outputs or via designated output parameters for multi-tensor outputs or when return-value usage would lead to name collisions.

In the \texttt{Python} interface, context-dependent tensor operations generally use the procedural form \texttt{tcapi.function(ctx, ...)}. Routines that do not require an existing context, such as tensor-kind introspection and context creation or query routines, use their individually specified signatures.
Tensor-producing operations return raw backend tensors, multi-output operations return tuples, and constructor or loader functions that require a target tensor kind take a \texttt{TenKind} argument.
\texttt{Python} TCAPI does not use \texttt{C++}-style output arguments.
Tensor operations leave their input tensors unmodified.
Unless a routine explicitly guarantees a deep copy, storage sharing between inputs and outputs is implementation-defined.
In the \texttt{Python} signature blocks below, calls are shown without assignment; the accompanying text specifies the return value of each call, such as a tensor, a tuple, a scalar, or no value, and the examples show any required assignment or rebinding.

TCAPI treats a diagonal tensor as an ordinary second-order tensor with shape $\{N,N\}$ whose off-diagonal logical tensor elements are zero.
Diagonal tensors have the same public tensor type or tensor kind as other tensors with the same backend family and element type; implementations may store them compactly or densely.
Unless stated otherwise, TCAPI operations are defined by logical tensor elements, not by storage representation.

For each function, we provide a minimal usage example following the function signature and description.
Throughout the examples, \texttt{Ten} denotes either a \texttt{C++} tensor type or a \texttt{Python} \texttt{TenKind} with 32-bit real floating-point elements, and \texttt{CplxTen} denotes its complex counterpart.
In \texttt{C++}, this means a concrete tensor type with element type \texttt{float}; in \texttt{Python}, this means an implementation-provided \texttt{TenKind} with a backend-native 32-bit real floating-point \texttt{dtype}.

\subsubsection{Read-only queries}

This subsection specifies read-only query functions provided by TCAPI, which retrieve tensor metadata and individual elements without modifying the tensor.

\bnoindentnl{order}\hfill\break

\noindent\textit{\texttt{C++}:}
\begin{minted}[bgcolor=LightGray]{cpp}
template <typename TenT>
tcapi::order_t<TenT> order(
    tcapi::context_handle_t<TenT> &ctx,
    const TenT &a
);
\end{minted}
\noindent\textit{\texttt{Python}:}
\begin{minted}[bgcolor=LightGray]{python}
tcapi.order(ctx, a)
\end{minted}

Return the order of \texttt{a}, i.e., the number of bonds (dimensions).
For a zeroth-order tensor, this function returns 0.
In the \texttt{Python} interface, the returned value is an integer-like value.

\example\hfill\break

\noindent\textit{\texttt{C++}:}
\begin{minted}[bgcolor=LightGray]{cpp}
// a is a 3rd-order tensor
auto ord = tcapi::order(ctx, a);
// ord == 3
\end{minted}
\noindent\textit{\texttt{Python}:}
\begin{minted}[bgcolor=LightGray]{python}
# a is a 3rd-order tensor
ord = tcapi.order(ctx, a)
# ord == 3
\end{minted}

\bnoindentnl{shape}\hfill\break

\noindent\textit{\texttt{C++}:}
\begin{minted}[bgcolor=LightGray]{cpp}
template <typename TenT>
tcapi::shape_t<TenT> shape(
    tcapi::context_handle_t<TenT> &ctx,
    const TenT &a
);
\end{minted}
\noindent\textit{\texttt{Python}:}
\begin{minted}[bgcolor=LightGray]{python}
tcapi.shape(ctx, a)
\end{minted}

Return the shape (the bond dimensions) of \texttt{a}, listed in bond-index order.
For a zeroth-order tensor, this function returns the empty shape, i.e., the empty sequence of bond dimensions.
In the \texttt{Python} interface, the returned value is a sequence of integer-like values.

\example\hfill\break

\noindent\textit{\texttt{C++}:}
\begin{minted}[bgcolor=LightGray]{cpp}
// a is a 3rd-order tensor
// with the shape {3, 4, 2}
auto s = tcapi::shape(ctx, a);
// s[0] == 3; s[1] == 4; s[2] == 2
\end{minted}
\noindent\textit{\texttt{Python}:}
\begin{minted}[bgcolor=LightGray]{python}
# a is a 3rd-order tensor
# with the shape (3, 4, 2)
s = tcapi.shape(ctx, a)
# s[0] == 3; s[1] == 4; s[2] == 2
\end{minted}

\bnoindentnl{size}\hfill\break

\noindent\textit{\texttt{C++}:}
\begin{minted}[bgcolor=LightGray]{cpp}
template <typename TenT>
tcapi::ten_size_t<TenT> size(
    tcapi::context_handle_t<TenT> &ctx,
    const TenT &a
);
\end{minted}
\noindent\textit{\texttt{Python}:}
\begin{minted}[bgcolor=LightGray]{python}
tcapi.size(ctx, a)
\end{minted}

Return the total number of tensor elements in \texttt{a}.
For a zeroth-order tensor, this function returns 1.
For a diagonal tensor with shape $\{N,N\}$, this value is $N^2$, independent of whether the tensor is stored compactly.
In the \texttt{C++} interface, this is consistent with the semantics of \texttt{std::size}.
In the \texttt{Python} interface, the returned value is an integer-like value.

\example\hfill\break

\noindent\textit{\texttt{C++}:}
\begin{minted}[bgcolor=LightGray]{cpp}
// a is a 3rd-order tensor
// with the shape {3, 4, 2}
auto n = tcapi::size(ctx, a);
// n == 24
\end{minted}
\noindent\textit{\texttt{Python}:}
\begin{minted}[bgcolor=LightGray]{python}
# a is a 3rd-order tensor
# with the shape (3, 4, 2)
n = tcapi.size(ctx, a)
# n == 24
\end{minted}

\bnoindentnl{size\_bytes}\hfill\break

\noindent\textit{\texttt{C++}:}
\begin{minted}[bgcolor=LightGray]{cpp}
template <typename TenT>
std::size_t size_bytes(
    tcapi::context_handle_t<TenT> &ctx,
    const TenT &a
);
\end{minted}
\noindent\textit{\texttt{Python}:}
\begin{minted}[bgcolor=LightGray]{python}
tcapi.size_bytes(ctx, a)
\end{minted}

Return the physical memory footprint of \texttt{a} in bytes.
A compactly stored diagonal tensor may use fewer bytes than a dense tensor with the same shape.
In the \texttt{Python} interface, the returned value is an integer-like value.

\example\hfill\break

\noindent\textit{\texttt{C++}:}
\begin{minted}[bgcolor=LightGray]{cpp}
// a is a 3rd-order float32 tensor
// with the shape {3, 4, 2}
auto bytes = tcapi::size_bytes(ctx, a);
// bytes == 96
\end{minted}
\noindent\textit{\texttt{Python}:}
\begin{minted}[bgcolor=LightGray]{python}
# a is a 3rd-order float32 tensor
# with the shape (3, 4, 2)
num_bytes = tcapi.size_bytes(ctx, a)
# num_bytes == 96
\end{minted}

\bnoindentnl{get\_elem}\hfill\break

\noindent\textit{\texttt{C++}:}
\begin{minted}[bgcolor=LightGray]{cpp}
template <typename TenT>
tcapi::elem_t<TenT> get_elem(
    tcapi::context_handle_t<TenT> &ctx,
    const TenT &a,
    const tcapi::elem_coors_t<TenT> &coors
);
\end{minted}
\noindent\textit{\texttt{Python}:}
\begin{minted}[bgcolor=LightGray]{python}
tcapi.get_elem(ctx, a, coors)
\end{minted}

Return the element value of \texttt{a} at the zero-based coordinate tuple \texttt{coors}, specified in bond-index order.
For a zeroth-order tensor, \texttt{coors} is the empty coordinate sequence, and the function returns the sole element.
For a diagonal tensor, off-diagonal elements return zero.
In the \texttt{Python} interface, \texttt{coors} is a \texttt{Python} sequence, and the returned value is a scalar value consistent with the relevant \texttt{TenKind}.

\example\hfill\break

\noindent\textit{\texttt{C++}:}
\begin{minted}[bgcolor=LightGray]{cpp}
// a is a 3-by-3 identity matrix
auto v = tcapi::get_elem(ctx, a, {1, 1});
// v == 1.0
v = tcapi::get_elem(ctx, a, {0, 1});
// v == 0.0
\end{minted}
\noindent\textit{\texttt{Python}:}
\begin{minted}[bgcolor=LightGray]{python}
# a is a 3-by-3 identity matrix
v = tcapi.get_elem(ctx, a, (1, 1))
# v == 1.0
v = tcapi.get_elem(ctx, a, (0, 1))
# v == 0.0
\end{minted}

\subsubsection{Construction and destruction}

This subsection specifies core routines for allocating, initializing, copying, moving, and releasing tensor objects.
Tensor element type and precision are selected through \texttt{TenT} in the \texttt{C++} interface and \texttt{TenKind} in the \texttt{Python} interface.
In the \texttt{Python} interface, constructors that create a tensor of a specified kind take a \texttt{TenKind} argument and return raw backend tensors.
\texttt{C++}-specific ownership routines that do not have a \texttt{Python} analogue are identified explicitly.

\bnoindentnl{allocate}\hfill\break

\noindent\textit{\texttt{C++}:}
\begin{minted}[bgcolor=LightGray]{cpp}
template <typename TenT>
TenT allocate(
    tcapi::context_handle_t<TenT> &ctx,
    const tcapi::shape_t<TenT> &shape
);
\end{minted}
\noindent\textit{\texttt{Python}:}
\begin{minted}[bgcolor=LightGray]{python}
tcapi.allocate(ctx, Ten, shape)
\end{minted}

Allocate and return a tensor with the specified shape \texttt{shape}.
The empty shape is valid; in that case, the function returns a zeroth-order tensor whose sole element is uninitialized.
Storage is allocated, but tensor elements are intentionally left \emph{uninitialized} to avoid initialization overhead.
Consequently, element values are indeterminate until they are explicitly assigned.
In the \texttt{Python} interface, \texttt{Ten} is a \texttt{TenKind}, \texttt{shape} is a \texttt{Python} sequence, and the returned value is a raw backend tensor.

\example\hfill\break

\noindent\textit{\texttt{C++}:}
\begin{minted}[bgcolor=LightGray]{cpp}
auto a = tcapi::allocate<Ten>(ctx, {3, 4, 2});
auto v = tcapi::get_elem(ctx, a, {0, 2, 1});
// the value of v is indeterminate
\end{minted}
\noindent\textit{\texttt{Python}:}
\begin{minted}[bgcolor=LightGray]{python}
a = tcapi.allocate(ctx, Ten, (3, 4, 2))
v = tcapi.get_elem(ctx, a, (0, 2, 1))
# the value of v is indeterminate
\end{minted}

\bnoindentnl{zeros}\hfill\break

\noindent\textit{\texttt{C++}:}
\begin{minted}[bgcolor=LightGray]{cpp}
template <typename TenT>
TenT zeros(
    tcapi::context_handle_t<TenT> &ctx,
    const tcapi::shape_t<TenT> &shape
);
\end{minted}
\noindent\textit{\texttt{Python}:}
\begin{minted}[bgcolor=LightGray]{python}
tcapi.zeros(ctx, Ten, shape)
\end{minted}

Allocate and return a tensor of shape \texttt{shape} with all elements initialized to zero.
The empty shape is valid; in that case, the function returns a zeroth-order tensor whose sole element is initialized to zero.
In the \texttt{Python} interface, \texttt{Ten} is a \texttt{TenKind}, \texttt{shape} is a \texttt{Python} sequence, and the returned value is a raw backend tensor.

\example\hfill\break

\noindent\textit{\texttt{C++}:}
\begin{minted}[bgcolor=LightGray]{cpp}
auto a = tcapi::zeros<Ten>(ctx, {3, 4, 2});
auto v = tcapi::get_elem(ctx, a, {0, 2, 1});
// v == 0.0
\end{minted}
\noindent\textit{\texttt{Python}:}
\begin{minted}[bgcolor=LightGray]{python}
a = tcapi.zeros(ctx, Ten, (3, 4, 2))
v = tcapi.get_elem(ctx, a, (0, 2, 1))
# v == 0.0
\end{minted}

\bnoindentnl{assign\_from\_range}\hfill\break

\noindent\textit{\texttt{C++}:}
\begin{minted}[bgcolor=LightGray]{cpp}
template <typename TenT,
          typename RandomIt,
          typename Func>
TenT assign_from_range(
    tcapi::context_handle_t<TenT> &ctx,
    const tcapi::shape_t<TenT> &shape,
    RandomIt first,
    Func &&coors2idx
);
\end{minted}
\noindent\textit{\texttt{Python}:}
\begin{minted}[bgcolor=LightGray]{python}
tcapi.assign_from_range(
    ctx, Ten, shape, values, coors2idx
)
\end{minted}

Construct and return a tensor of shape \texttt{shape} by assigning its elements from an input range.
In \texttt{C++}, the range is a random-access range starting at \texttt{first}; in \texttt{Python}, it is the sequence \texttt{values}.
The element at coordinate tuple \texttt{coors} is initialized as \texttt{*(first + coors2idx(coors))} in \texttt{C++} or \texttt{values[coors2idx(coors)]} in \texttt{Python}.
The empty shape is valid; in that case, the function returns a zeroth-order tensor and invokes the coordinate-to-index map once with the empty coordinate sequence.

In the \texttt{Python} interface, \texttt{Ten} is a \texttt{TenKind}, \texttt{shape} is a \texttt{Python} sequence, \texttt{values} is a \texttt{Python} sequence, \texttt{coors2idx} is a callable, and the returned value is a raw backend tensor.
In \texttt{C++}, the mapping function \texttt{coors2idx} must have the signature
\begin{minted}[bgcolor=LightGray]{cpp}
std::iterator_traits<
        RandomIt
>::difference_type coors2idx(
    const tcapi::elem_coors_t<TenT> &coors
);
\end{minted}
and must map valid coordinate tuples to zero-based indices in the range.

\example\hfill\break

\noindent\textit{\texttt{C++}:}
\begin{minted}[bgcolor=LightGray]{cpp}
using Elem = tcapi::elem_t<Ten>;
using ElemVec = std::vector<Elem>;
std::iterator_traits<
    typename ElemVec::iterator
>::difference_type coors2idx(
    const tcapi::elem_coors_t<Ten> &coors
) {
  return 3 * coors[0] + coors[1];
}
ElemVec els {1.0, 2.0, 3.0, 4.0, 5.0, 6.0};
auto a = tcapi::assign_from_range<Ten>(
    ctx, {2, 3}, els.begin(), coors2idx
);
auto el = tcapi::get_elem(ctx, a, {1, 1});
// el == 5.0
\end{minted}
\noindent\textit{\texttt{Python}:}
\begin{minted}[bgcolor=LightGray]{python}
def coors2idx(coors):
    return 3 * coors[0] + coors[1]

values = [1.0, 2.0, 3.0, 4.0, 5.0, 6.0]
a = tcapi.assign_from_range(
    ctx, Ten, (2, 3), values, coors2idx
)
el = tcapi.get_elem(ctx, a, (1, 1))
# el == 5.0
\end{minted}

\bnoindentnl{random}\hfill\break

\noindent\textit{\texttt{C++}:}
\begin{minted}[bgcolor=LightGray]{cpp}
template <typename TenT, typename RandNumGen>
TenT random(
    tcapi::context_handle_t<TenT> &ctx,
    const tcapi::shape_t<TenT> &shape,
    RandNumGen &gen
);
\end{minted}
\noindent\textit{\texttt{Python}:}
\begin{minted}[bgcolor=LightGray]{python}
tcapi.random(ctx, Ten, shape, gen)
\end{minted}

Construct and return a tensor of shape \texttt{shape} whose elements are generated by repeatedly invoking the random number generator \texttt{gen}.
The empty shape is valid; in that case, the function returns a zeroth-order tensor and invokes the random-number generator once.
Each call to \texttt{gen()} must return a value convertible to \texttt{tcapi::elem\_t<TenT>} in \texttt{C++} or consistent with \texttt{tcapi.ten\_kind\_dtype(Ten)} in the \texttt{Python} interface.
In the \texttt{Python} interface, \texttt{Ten} is a \texttt{TenKind}, \texttt{shape} is a \texttt{Python} sequence, \texttt{gen} is a callable, and the returned value is a raw backend tensor.

\example\hfill\break

\noindent\textit{\texttt{C++}:}
\begin{minted}[bgcolor=LightGray]{cpp}
std::mt19937 engine;
std::uniform_real_distribution<float> dis(
  0.0, 1.0
);
auto gen = [&dis, &engine]() {
  return dis(engine);
};
auto a = tcapi::random<Ten>(
  ctx, {3, 4, 2}, gen
);
auto el = tcapi::get_elem(ctx, a, {1, 2, 0});
// el is generated by gen and lies in
// [0.0, 1.0)
\end{minted}
\noindent\textit{\texttt{Python}:}
\begin{minted}[bgcolor=LightGray]{python}
import random

def gen():
    return random.random()

a = tcapi.random(ctx, Ten, (3, 4, 2), gen)
el = tcapi.get_elem(ctx, a, (1, 2, 0))
# el is generated by gen and lies in [0.0, 1.0)
\end{minted}

\bnoindentnl{eye}\hfill\break

\noindent\textit{\texttt{C++}:}
\begin{minted}[bgcolor=LightGray]{cpp}
template <typename TenT>
TenT eye(
    tcapi::context_handle_t<TenT> &ctx,
    const tcapi::bond_dim_t<TenT> N
);
\end{minted}
\noindent\textit{\texttt{Python}:}
\begin{minted}[bgcolor=LightGray]{python}
tcapi.eye(ctx, Ten, N)
\end{minted}

Construct and return the $N \times N$ identity tensor as a diagonal tensor, with ones on the main diagonal and zeros elsewhere.
Implementations may store the result compactly when supported by the underlying TCF.
In the \texttt{Python} interface, \texttt{Ten} is a \texttt{TenKind}, and the returned value is a raw backend tensor.

\example\hfill\break

\noindent\textit{\texttt{C++}:}
\begin{minted}[bgcolor=LightGray]{cpp}
auto a = tcapi::eye<Ten>(ctx, 3);
auto el = tcapi::get_elem(ctx, a, {1, 1});
// el == 1.0
el = tcapi::get_elem(ctx, a, {1, 2});
// el == 0.0
\end{minted}
\noindent\textit{\texttt{Python}:}
\begin{minted}[bgcolor=LightGray]{python}
a = tcapi.eye(ctx, Ten, 3)
el = tcapi.get_elem(ctx, a, (1, 1))
# el == 1.0
el = tcapi.get_elem(ctx, a, (1, 2))
# el == 0.0
\end{minted}

\bnoindentnl{fill}\hfill\break

\noindent\textit{\texttt{C++}:}
\begin{minted}[bgcolor=LightGray]{cpp}
template <typename TenT>
TenT fill(
    tcapi::context_handle_t<TenT> &ctx,
    const tcapi::shape_t<TenT> &shape,
    const tcapi::elem_t<TenT> v
);
\end{minted}
\noindent\textit{\texttt{Python}:}
\begin{minted}[bgcolor=LightGray]{python}
tcapi.fill(ctx, Ten, shape, v)
\end{minted}

Construct and return a tensor of shape \texttt{shape} with all elements initialized to \texttt{v}.
The empty shape is valid; in that case, the function returns a zeroth-order tensor whose sole element is initialized to \texttt{v}.
In the \texttt{Python} interface, \texttt{Ten} is a \texttt{TenKind}, \texttt{shape} is a \texttt{Python} sequence, \texttt{v} is a scalar value consistent with \texttt{tcapi.ten\_kind\_dtype(Ten)}, and the returned value is a raw backend tensor.

\example\hfill\break

\noindent\textit{\texttt{C++}:}
\begin{minted}[bgcolor=LightGray]{cpp}
auto a = tcapi::fill<Ten>(ctx, {3, 2, 4}, 2.0);
auto el = tcapi::get_elem(ctx, a, {0, 1, 3});
// el == 2.0
\end{minted}
\noindent\textit{\texttt{Python}:}
\begin{minted}[bgcolor=LightGray]{python}
a = tcapi.fill(ctx, Ten, (3, 2, 4), 2.0)
el = tcapi.get_elem(ctx, a, (0, 1, 3))
# el == 2.0
\end{minted}

\bnoindentnl{copy}\hfill\break

\noindent\textit{\texttt{C++}:}
\begin{minted}[bgcolor=LightGray]{cpp}
template <typename TenT>
TenT copy(
    tcapi::context_handle_t<TenT> &ctx,
    const TenT &orig
);
\end{minted}
\noindent\textit{\texttt{Python}:}
\begin{minted}[bgcolor=LightGray]{python}
tcapi.copy(ctx, orig)
\end{minted}

Construct and return a \emph{deep} copy of \texttt{orig}.
The returned tensor has the same shape and element values as \texttt{orig}, and the two tensors do not share memory.
In the \texttt{Python} interface, the returned value is a raw backend tensor.

\example\hfill\break

\noindent\textit{\texttt{C++}:}
\begin{minted}[bgcolor=LightGray]{cpp}
Ten a;
// perform operations on a
auto b = tcapi::copy(ctx, a);
// b == a
\end{minted}
\noindent\textit{\texttt{Python}:}
\begin{minted}[bgcolor=LightGray]{python}
a = tcapi.zeros(ctx, Ten, (3, 4, 2))
# perform operations on a
b = tcapi.copy(ctx, a)
# b has the same shape and element values as a
\end{minted}

\bnoindentnl{move}\hfill\break

\noindent\textit{\texttt{C++}:}
\begin{minted}[bgcolor=LightGray]{cpp}
template <typename TenT>
TenT move(
    tcapi::context_handle_t<TenT> &ctx,
    TenT &from
);
\end{minted}
\noindent\textit{\texttt{Python}:} No \texttt{Python} interface is specified for \texttt{move}.

Transfer the data storage and metadata owned by \texttt{from} to a newly returned tensor.
No deep copy is performed.
Upon return, \texttt{from} is placed in a valid, default-constructed state (\texttt{TenT\{\}}).
This is a \texttt{C++} ownership utility; \texttt{Python} tensor lifetime and ownership follow normal \texttt{Python} and backend-library semantics.

\example\hfill\break

\noindent\textit{\texttt{C++}:}
\begin{minted}[bgcolor=LightGray]{cpp}
Ten a;
// perform operations on a
auto a_cpy = tcapi::copy(ctx, a);
auto b = tcapi::move(ctx, a);
// b == a_cpy
// a == Ten{}
\end{minted}

\bnoindentnl{clear}\hfill\break

\noindent\textit{\texttt{C++}:}
\begin{minted}[bgcolor=LightGray]{cpp}
template <typename TenT>
void clear(
    tcapi::context_handle_t<TenT> &ctx,
    TenT &a
);
\end{minted}
\noindent\textit{\texttt{Python}:} No \texttt{Python} interface is specified for \texttt{clear}.

Reset \texttt{a} to an empty, default-constructed state, releasing its data storage and associated metadata.
Upon return, \texttt{a == TenT\{\}}.
This is a \texttt{C++} ownership utility; \texttt{Python} tensor lifetime and ownership follow normal \texttt{Python} and backend-library semantics.

\example\hfill\break

\noindent\textit{\texttt{C++}:}
\begin{minted}[bgcolor=LightGray]{cpp}
Ten a;
// perform operations on a
tcapi::clear(ctx, a);
// a == Ten{}
\end{minted}

\subsubsection{Input and output (I/O)}

This subsection specifies routines for serializing tensors to persistent storage and deserializing them into memory.

\bnoindentnl{load}\hfill\break

\noindent\textit{\texttt{C++}:}
\begin{minted}[bgcolor=LightGray]{cpp}
template <typename TenT, typename Storage>
TenT load(
    tcapi::context_handle_t<TenT> &ctx,
    Storage &&strg
);
\end{minted}
\noindent\textit{\texttt{Python}:}
\begin{minted}[bgcolor=LightGray]{python}
tcapi.load(ctx, Ten, strg)
\end{minted}

Deserialize a tensor from the storage location identified by \texttt{strg} and return it.
In \texttt{C++}, the type of \texttt{strg} is represented by the template parameter \texttt{Storage}; in \texttt{Python}, \texttt{strg} denotes a path-like or handle-like object that identifies a file on the host file system.
\texttt{C++} implementations are required to accept string-like inputs, such as \texttt{std::string}, \texttt{std::string\_view}, or \texttt{const char*}, and may additionally support \texttt{std::filesystem::path} or other storage abstractions.
\texttt{Python} implementations are required to accept string-like inputs, such as \texttt{str}, and may additionally support \texttt{pathlib.Path} or other storage abstractions.
In the \texttt{Python} interface, \texttt{Ten} is a \texttt{TenKind}, and the returned value is a raw backend tensor.
The on-disk format is backend-defined and must be compatible with the requested tensor kind and the execution context \texttt{ctx}.

\example\hfill\break

\noindent\textit{\texttt{C++}:}
\begin{minted}[bgcolor=LightGray]{cpp}
// ./a.ten is a file on disk
// containing the serialized representation
// of a Ten tensor
auto a = tcapi::load<Ten>(ctx, "./a.ten");
\end{minted}
\noindent\textit{\texttt{Python}:}
\begin{minted}[bgcolor=LightGray]{python}
# ./a.ten is a file on disk
# containing the serialized representation
# of a tensor compatible with Ten
a = tcapi.load(ctx, Ten, "./a.ten")
\end{minted}

\bnoindentnl{save}\hfill\break

\noindent\textit{\texttt{C++}:}
\begin{minted}[bgcolor=LightGray]{cpp}
template <typename TenT, typename Storage>
void save(
    tcapi::context_handle_t<TenT> &ctx,
    const TenT &a,
    Storage &&strg
);
\end{minted}
\noindent\textit{\texttt{Python}:}
\begin{minted}[bgcolor=LightGray]{python}
tcapi.save(ctx, a, strg)
\end{minted}

Serialize the tensor \texttt{a} and write it to the storage location identified by \texttt{strg}.
In \texttt{C++}, the type of \texttt{strg} is represented by the template parameter \texttt{Storage}; in \texttt{Python}, \texttt{strg} denotes a path-like or handle-like object that identifies a file on the host file system.
\texttt{C++} implementations are required to accept string-like inputs, such as \texttt{std::string}, \texttt{std::string\_view}, or \texttt{const char*}, and may additionally support \texttt{std::filesystem::path} or other storage abstractions.
\texttt{Python} implementations are required to accept string-like inputs, such as \texttt{str}, and may additionally support \texttt{pathlib.Path} or other storage abstractions.
This function is the counterpart to \texttt{load}.

\example\hfill\break

\noindent\textit{\texttt{C++}:}
\begin{minted}[bgcolor=LightGray]{cpp}
Ten a;
// do some operations on a
tcapi::save(ctx, a, "./a.ten");
// the file "./a.ten" is created on disk
\end{minted}
\noindent\textit{\texttt{Python}:}
\begin{minted}[bgcolor=LightGray]{python}
a = tcapi.zeros(ctx, Ten, (3, 4, 2))
# do some operations on a
tcapi.save(ctx, a, "./a.ten")
# the file "./a.ten" is created on disk
\end{minted}

\subsubsection{Tensor manipulation operations \label{sec:tensor_manipulation}}

This subsection specifies tensor-manipulation routines that operate on tensor structure or elementwise values without invoking tensor linear algebra.
In the \texttt{Python} interface, tensor-manipulation routines return their results as raw backend tensors.
Unless otherwise stated, TCAPI tensor-manipulation routines operate on tensor elements rather than stored elements.
Implementations may preserve diagonal storage when the result remains diagonal and may densify when the result cannot be represented as a diagonal tensor.

\bnoindentnl{set\_elem}\hfill\break

\noindent\textit{\texttt{C++}:}
\begin{minted}[bgcolor=LightGray]{cpp}
template <typename TenT>
void set_elem(
    tcapi::context_handle_t<TenT> &ctx,
    TenT &a,
    const tcapi::elem_coors_t<TenT> &coors,
    const tcapi::elem_t<TenT> el
);
\end{minted}
\noindent\textit{\texttt{Python}:}
\begin{minted}[bgcolor=LightGray]{python}
tcapi.set_elem(ctx, a, coors, el)
\end{minted}

Assign the value \texttt{el} to the element at coordinates \texttt{coors}.
For a zeroth-order tensor, \texttt{coors} is the empty coordinate sequence, and the function updates the sole element.
In the \texttt{C++} interface, this operation updates \texttt{a} in place.
In the \texttt{Python} interface, \texttt{coors} is a \texttt{Python} sequence, \texttt{el} is a scalar value consistent with the relevant \texttt{TenKind}, and the returned value is a raw backend tensor with the specified element updated.

\example\hfill\break

\noindent\textit{\texttt{C++}:}
\begin{minted}[bgcolor=LightGray]{cpp}
auto a = tcapi::zeros<Ten>(ctx, {3, 4, 2});
tcapi::set_elem(ctx, a, {2, 1, 0}, 1.0);
auto v = tcapi::get_elem(ctx, a, {2, 1, 0});
// v == 1.0
\end{minted}
\noindent\textit{\texttt{Python}:}
\begin{minted}[bgcolor=LightGray]{python}
a = tcapi.zeros(ctx, Ten, (3, 4, 2))
a = tcapi.set_elem(ctx, a, (2, 1, 0), 1.0)
v = tcapi.get_elem(ctx, a, (2, 1, 0))
# v == 1.0
\end{minted}

\bnoindentnl{reshape}\hfill\break

\noindent\textit{\texttt{C++}:}
\begin{minted}[bgcolor=LightGray]{cpp}
template <typename TenT>
void reshape(
    tcapi::context_handle_t<TenT> &ctx,
    TenT &inout,
    const tcapi::shape_t<TenT> &new_shape
);                                      // (1)

template <typename TenT>
void reshape(
    tcapi::context_handle_t<TenT> &ctx,
    const TenT &in,
    const tcapi::shape_t<TenT> &new_shape,
    TenT &out
);                                      // (2)
\end{minted}
\noindent\textit{\texttt{Python}:}
\begin{minted}[bgcolor=LightGray]{python}
tcapi.reshape(ctx, a, new_shape)
\end{minted}

Reshape a tensor to \texttt{new\_shape}, with the total number of logical tensor elements unchanged.
This operation also converts between the empty shape and any shape with total size 1, such as $\{1\}$ or $\{1,1\}$, because the number of logical elements is preserved.
No element reordering, transposition, or value modification occurs; the linear order of logical tensor elements is preserved.
For densely stored tensors, this operation may be implemented without copying the element data.
For diagonal tensors or other compact storage formats, the same logical-element semantics apply; implementations may preserve compact storage when possible or densify the tensor when the reshaped result is no longer representable compactly.
The operation also covers combining consecutive bonds into a composite bond or splitting a bond into several bonds, with the relevant dimensions multiplying to the same value; nonconsecutive bonds can first be made consecutive by \texttt{transpose} before reshaping.
In the \texttt{C++} interface, (1) reshapes \texttt{inout} in place and (2) reads from \texttt{in} and writes the reshaped tensor into \texttt{out}.
In the \texttt{Python} interface, \texttt{new\_shape} is a \texttt{Python} sequence, and the returned value is a raw backend tensor with the requested shape.

\example\hfill\break

\noindent\textit{\texttt{C++}:}
\begin{minted}[bgcolor=LightGray]{cpp}
auto a = tcapi::zeros<Ten>(ctx, {3, 4, 2});
tcapi::reshape(ctx, a, {4, 2, 3});
auto s = tcapi::shape(ctx, a);
// s == {4, 2, 3}

Ten b;
tcapi::reshape(ctx, a, {2, 3, 4}, b);
auto sb = tcapi::shape(ctx, b);
// sb == {2, 3, 4}
\end{minted}
\noindent\textit{\texttt{Python}:}
\begin{minted}[bgcolor=LightGray]{python}
a = tcapi.zeros(ctx, Ten, (3, 4, 2))
a = tcapi.reshape(ctx, a, (4, 2, 3))
s = tcapi.shape(ctx, a)
# s == (4, 2, 3)

b = tcapi.reshape(ctx, a, (2, 3, 4))
sb = tcapi.shape(ctx, b)
# sb == (2, 3, 4)
\end{minted}

\bnoindentnl{transpose}\hfill\break

\noindent\textit{\texttt{C++}:}
\begin{minted}[bgcolor=LightGray]{cpp}
template <typename TenT>
void transpose(
    tcapi::context_handle_t<TenT> &ctx,
    TenT &inout,
    const List<
        tcapi::bond_idx_t<TenT>
    > &new_order
);                                      // (1)

template <typename TenT>
void transpose(
    tcapi::context_handle_t<TenT> &ctx,
    const TenT &in,
    const List<
        tcapi::bond_idx_t<TenT>
    > &new_order,
    TenT &out
);                                      // (2)
\end{minted}
\noindent\textit{\texttt{Python}:}
\begin{minted}[bgcolor=LightGray]{python}
tcapi.transpose(ctx, a, new_order)
\end{minted}

Permute tensor bonds according to the permutation \texttt{new\_order}.
Element values are preserved; only their coordinates are permuted (the memory layout may also change).
In the \texttt{C++} interface, (1) applies the permutation to \texttt{inout} in place and (2) reads from \texttt{in} and writes the permuted tensor into \texttt{out}.
In the \texttt{Python} interface, \texttt{new\_order} is a \texttt{Python} sequence of bond indices, and the returned value is a raw backend tensor with the permuted bonds.

\example\hfill\break

\noindent\textit{\texttt{C++}:}
\begin{minted}[bgcolor=LightGray]{cpp}
auto a = tcapi::random<Ten>(
  ctx, {3, 2, 4}, gen
);
auto el1 = tcapi::get_elem(ctx, a, {1, 0, 0});
tcapi::transpose(ctx, a, {1, 0, 2});
auto el2 = tcapi::get_elem(ctx, a, {0, 1, 0});
// el1 == el2

Ten b;
tcapi::transpose(ctx, a, {2, 1, 0}, b);
auto sb = tcapi::shape(ctx, b);
// sb == {4, 3, 2}
\end{minted}
\noindent\textit{\texttt{Python}:}
\begin{minted}[bgcolor=LightGray]{python}
a = tcapi.random(ctx, Ten, (3, 2, 4), gen)
el1 = tcapi.get_elem(ctx, a, (1, 0, 0))
a = tcapi.transpose(ctx, a, (1, 0, 2))
el2 = tcapi.get_elem(ctx, a, (0, 1, 0))
# el1 == el2

b = tcapi.transpose(ctx, a, (2, 1, 0))
sb = tcapi.shape(ctx, b)
# sb == (4, 3, 2)
\end{minted}

\bnoindentnl{cplx\_conj}\hfill\break

\noindent\textit{\texttt{C++}:}
\begin{minted}[bgcolor=LightGray]{cpp}
template <typename TenT>
void cplx_conj(
    tcapi::context_handle_t<TenT> &ctx,
    TenT &inout
);                                      // (1)

template <typename TenT>
void cplx_conj(
    tcapi::context_handle_t<TenT> &ctx,
    const TenT &in,
    TenT &out
);                                      // (2)
\end{minted}
\noindent\textit{\texttt{Python}:}
\begin{minted}[bgcolor=LightGray]{python}
tcapi.cplx_conj(ctx, a)
\end{minted}

Apply elementwise complex conjugation.
In the \texttt{C++} interface, (1) conjugates \texttt{inout} in place and (2) conjugates \texttt{in} and writes the result into \texttt{out}.
If \texttt{tcapi::elem\_t<TenT>} is real, (1) performs no change and (2) produces a deep copy.
In the \texttt{Python} interface, the returned value is a raw backend tensor; for a real-valued input, the result has the same shape and element values as \texttt{a}.

\example\hfill\break

\noindent\textit{\texttt{C++}:}
\begin{minted}[bgcolor=LightGray]{cpp}
auto a = tcapi::random<CplxTen>(
    ctx, {3, 2, 4}, gen
);
CplxTen a_conj;
tcapi::cplx_conj(ctx, a, a_conj);
auto el1 = tcapi::get_elem(ctx, a, {0, 0, 0});
auto el2 = tcapi::get_elem(
  ctx, a_conj, {0, 0, 0}
);
// std::conj(el1) == el2
\end{minted}
\noindent\textit{\texttt{Python}:}
\begin{minted}[bgcolor=LightGray]{python}
a = tcapi.random(ctx, CplxTen, (3, 2, 4), gen)
a_conj = tcapi.cplx_conj(ctx, a)
el1 = tcapi.get_elem(ctx, a, (0, 0, 0))
el2 = tcapi.get_elem(ctx, a_conj, (0, 0, 0))
# conjugate(el1) == el2
\end{minted}

\bnoindentnl{to\_cplx}\hfill\break

\noindent\textit{\texttt{C++}:}
\begin{minted}[bgcolor=LightGray]{cpp}
template <typename TenT>
tcapi::cplx_ten_t<TenT> to_cplx(
    tcapi::context_handle_t<TenT> &ctx,
    const TenT &in
);
\end{minted}
\noindent\textit{\texttt{Python}:}
\begin{minted}[bgcolor=LightGray]{python}
tcapi.to_cplx(ctx, a)
\end{minted}

Return a complex-valued tensor with the same shape as \texttt{in}.
Real-valued elements are embedded as complex numbers with zero imaginary parts.
If \texttt{tcapi::elem\_t<TenT>} is already complex, the function returns a deep copy of \texttt{in}.
In the \texttt{Python} interface, the returned value is a raw backend tensor whose \texttt{TenKind} is the complex counterpart of \texttt{tcapi.ten\_kind(ctx, a)}.

\example\hfill\break

\noindent\textit{\texttt{C++}:}
\begin{minted}[bgcolor=LightGray]{cpp}
auto a = tcapi::eye<Ten>(ctx, 3);
auto cplx_a = tcapi::to_cplx(ctx, a);
auto el = tcapi::get_elem(ctx, cplx_a, {1, 1});
// el == 1.0 + 0.0j
\end{minted}
\noindent\textit{\texttt{Python}:}
\begin{minted}[bgcolor=LightGray]{python}
a = tcapi.eye(ctx, Ten, 3)
cplx_a = tcapi.to_cplx(ctx, a)
el = tcapi.get_elem(ctx, cplx_a, (1, 1))
# el == 1.0 + 0.0j
\end{minted}

\bnoindentnl{real}\hfill\break

\noindent\textit{\texttt{C++}:}
\begin{minted}[bgcolor=LightGray]{cpp}
template <typename TenT>
tcapi::real_ten_t<TenT> real(
    tcapi::context_handle_t<TenT> &ctx,
    const TenT &in
);
\end{minted}
\noindent\textit{\texttt{Python}:}
\begin{minted}[bgcolor=LightGray]{python}
tcapi.real(ctx, a)
\end{minted}

Extract the real part of \texttt{in} elementwise and return a tensor with a real-valued element type.
If \texttt{tcapi::elem\_t<TenT>} is already real, this function returns a deep copy of \texttt{in}.
In the \texttt{Python} interface, the returned value is a raw backend tensor whose \texttt{TenKind} is the real counterpart of \texttt{tcapi.ten\_kind(ctx, a)}.

\example\hfill\break

\noindent\textit{\texttt{C++}:}
\begin{minted}[bgcolor=LightGray]{cpp}
auto a = tcapi::random<CplxTen>(
    ctx, {3, 2, 4}, gen
);
auto a_real = tcapi::real(ctx, a);
auto el1 = tcapi::get_elem(ctx, a, {0, 0, 0});
auto el2 = tcapi::get_elem(
  ctx, a_real, {0, 0, 0}
);
// std::real(el1) == el2
\end{minted}
\noindent\textit{\texttt{Python}:}
\begin{minted}[bgcolor=LightGray]{python}
a = tcapi.random(ctx, CplxTen, (3, 2, 4), gen)
a_real = tcapi.real(ctx, a)
el1 = tcapi.get_elem(ctx, a, (0, 0, 0))
el2 = tcapi.get_elem(ctx, a_real, (0, 0, 0))
# real(el1) == el2
\end{minted}

\bnoindentnl{imag}\hfill\break

\noindent\textit{\texttt{C++}:}
\begin{minted}[bgcolor=LightGray]{cpp}
template <typename TenT>
tcapi::real_ten_t<TenT> imag(
    tcapi::context_handle_t<TenT> &ctx,
    const TenT &in
);
\end{minted}
\noindent\textit{\texttt{Python}:}
\begin{minted}[bgcolor=LightGray]{python}
tcapi.imag(ctx, a)
\end{minted}

Extract the imaginary part of \texttt{in} elementwise and return a tensor with a real-valued element type.
If \texttt{tcapi::elem\_t<TenT>} is already real, the function returns a real-valued tensor of the same shape whose elements are all zero.
In the \texttt{Python} interface, the returned value is a raw backend tensor whose \texttt{TenKind} is the real counterpart of \texttt{tcapi.ten\_kind(ctx, a)}.

\example\hfill\break

\noindent\textit{\texttt{C++}:}
\begin{minted}[bgcolor=LightGray]{cpp}
auto a = tcapi::random<CplxTen>(
    ctx, {3, 2, 4}, gen
);
auto a_imag = tcapi::imag(ctx, a);
auto el1 = tcapi::get_elem(ctx, a, {0, 0, 0});
auto el2 = tcapi::get_elem(
  ctx, a_imag, {0, 0, 0}
);
// std::imag(el1) == el2
\end{minted}
\noindent\textit{\texttt{Python}:}
\begin{minted}[bgcolor=LightGray]{python}
a = tcapi.random(ctx, CplxTen, (3, 2, 4), gen)
a_imag = tcapi.imag(ctx, a)
el1 = tcapi.get_elem(ctx, a, (0, 0, 0))
el2 = tcapi.get_elem(ctx, a_imag, (0, 0, 0))
# imag(el1) == el2
\end{minted}

\bnoindentnl{expand}\hfill\break

\noindent\textit{\texttt{C++}:}
\begin{minted}[bgcolor=LightGray]{cpp}
template <typename TenT>
void expand(
    tcapi::context_handle_t<TenT> &ctx,
    TenT &inout,
    const Map<
        tcapi::bond_idx_t<TenT>,
        tcapi::bond_dim_t<TenT>
        > &bond_idx_increment_map
);                                      // (1)

template <typename TenT>
void expand(
    tcapi::context_handle_t<TenT> &ctx,
    const TenT &in,
    const Map<
        tcapi::bond_idx_t<TenT>,
        tcapi::bond_dim_t<TenT>
        > &bond_idx_increment_map,
    TenT &out
);                                      // (2)
\end{minted}
\noindent\textit{\texttt{Python}:}
\begin{minted}[bgcolor=LightGray]{python}
tcapi.expand(ctx, a, bond_idx_increment_map)
\end{minted}

Expand selected bonds by increasing their sizes according to \texttt{bond\_idx\_increment\_map}.
The keys specify the bonds to expand, and the corresponding values specify the additional sizes to append along those bonds.
Elements whose coordinates lie within the original bounds are preserved.
Any element with at least one coordinate in an appended range is set to zero.
In the \texttt{C++} interface, (1) applies the expansion to \texttt{inout} in place and (2) reads from \texttt{in} and writes the expanded tensor into \texttt{out}.
In the \texttt{Python} interface, \texttt{bond\_idx\_increment\_map} is a mapping from bond indices to bond-dimension increments, and the returned value is a raw backend tensor.

\example\hfill\break

\noindent\textit{\texttt{C++}:}
\begin{minted}[bgcolor=LightGray]{cpp}
auto a = tcapi::zeros<Ten>(ctx, {2, 2, 2});
tcapi::expand(ctx, a, {{1, 2}, {0, 1}});
auto s = tcapi::shape(ctx, a);
// s == {3, 4, 2}
auto el = tcapi::get_elem(ctx, a, {2, 3, 0});
// el == 0.0
\end{minted}
\noindent\textit{\texttt{Python}:}
\begin{minted}[bgcolor=LightGray]{python}
a = tcapi.zeros(ctx, Ten, (2, 2, 2))
a = tcapi.expand(ctx, a, {1: 2, 0: 1})
s = tcapi.shape(ctx, a)
# s == (3, 4, 2)
el = tcapi.get_elem(ctx, a, (2, 3, 0))
# el == 0.0
\end{minted}

\bnoindentnl{shrink}\hfill\break

\noindent\textit{\texttt{C++}:}
\begin{minted}[bgcolor=LightGray]{cpp}
template <typename TenT>
void shrink(
    tcapi::context_handle_t<TenT> &ctx,
    TenT &inout,
    const bond_idx_elem_coor_pair_map<TenT> &
          bd_idx_el_coor_pair_map
);                                      // (1)
 
template <typename TenT>
void shrink(
    tcapi::context_handle_t<TenT> &ctx,
    const TenT &in,
    const bond_idx_elem_coor_pair_map<TenT> &
          bd_idx_el_coor_pair_map,
    TenT &out
);                                      // (2)
\end{minted}
\noindent\textit{\texttt{Python}:}
\begin{minted}[bgcolor=LightGray]{python}
tcapi.shrink(ctx, a, bd_idx_el_coor_pair_map)
\end{minted}

Shrink selected bonds by slicing them to half-open coordinate ranges $[coor\_first,~coor\_second)$ specified by \texttt{bd\_idx\_el\_coor\_pair\_map}.
Bonds not listed in \texttt{bd\_idx\_el\_coor\_pair\_map} retain their original ranges (and hence their original dimensions).
In the \texttt{C++} interface, (1) shrinks \texttt{inout} in place and (2) reads from \texttt{in} and writes the sliced tensor into \texttt{out}.
In the \texttt{Python} interface, \texttt{bd\_idx\_el\_coor\_pair\_map} is a mapping from bond indices to coordinate-pair sequences, and the returned value is a raw backend tensor.

\example\hfill\break

\noindent\textit{\texttt{C++}:}
\begin{minted}[bgcolor=LightGray]{cpp}
auto a = tcapi::random<Ten>(
  ctx, {3, 4, 2}, gen
);
auto el1 = tcapi::get_elem(ctx, a, {0, 1, 1});
tcapi::shrink(ctx,
  a,
  {{1, {1, 3}}, {0, {0, 2}}}
);
auto s = tcapi::shape(ctx, a);
// s == {2, 2, 2}
auto el2 = tcapi::get_elem(ctx, a, {0, 0, 1});
// el1 == el2
\end{minted}
\noindent\textit{\texttt{Python}:}
\begin{minted}[bgcolor=LightGray]{python}
a = tcapi.random(ctx, Ten, (3, 4, 2), gen)
el1 = tcapi.get_elem(ctx, a, (0, 1, 1))
a = tcapi.shrink(
    ctx, a, {1: (1, 3), 0: (0, 2)}
)
s = tcapi.shape(ctx, a)
# s == (2, 2, 2)
el2 = tcapi.get_elem(ctx, a, (0, 0, 1))
# el1 == el2
\end{minted}

\bnoindentnl{extract\_sub}\hfill\break

\noindent\textit{\texttt{C++}:}
\begin{minted}[bgcolor=LightGray]{cpp}
template <typename TenT>
void extract_sub(
    tcapi::context_handle_t<TenT> &ctx,
    TenT &inout,
    const List<
        Pair<
            tcapi::elem_coor_t<TenT>,
            tcapi::elem_coor_t<TenT>
        >> & coor_pairs
);                                      // (1)

template <typename TenT>
void extract_sub(
    tcapi::context_handle_t<TenT> &ctx,
    const TenT &in,
    const List<
        Pair<
            tcapi::elem_coor_t<TenT>,
            tcapi::elem_coor_t<TenT>
        >> & coor_pairs,
    TenT &out
);                                      // (2)
\end{minted}
\noindent\textit{\texttt{Python}:}
\begin{minted}[bgcolor=LightGray]{python}
tcapi.extract_sub(ctx, a, coor_pairs)
\end{minted}

Extract a subtensor specified by per-bond half-open coordinate ranges $[coor\_first,~coor\_second)$, provided in \texttt{coor\_pairs} in bond-index order.
In the \texttt{C++} interface, (1) overwrites \texttt{inout} with the extracted subtensor and (2) reads from \texttt{in} and writes the extracted subtensor into \texttt{out}.
In the \texttt{Python} interface, \texttt{coor\_pairs} is a \texttt{Python} sequence of coordinate-pair sequences, and the returned value is a raw backend tensor.

\example\hfill\break

\noindent\textit{\texttt{C++}:}
\begin{minted}[bgcolor=LightGray]{cpp}
auto a = tcapi::random<Ten>(
  ctx, {3, 4, 2}, gen
);
Ten sub;
tcapi::extract_sub(ctx,
  a,
  {{1, 3}, {0, 2}, {0, 2}},
  sub
);
auto el1 = tcapi::get_elem(ctx, a, {1, 0, 0});
auto el2 = tcapi::get_elem(
  ctx, sub, {0, 0, 0}
);
// el1 == el2
\end{minted}
\noindent\textit{\texttt{Python}:}
\begin{minted}[bgcolor=LightGray]{python}
a = tcapi.random(ctx, Ten, (3, 4, 2), gen)
sub = tcapi.extract_sub(
    ctx, a, ((1, 3), (0, 2), (0, 2))
)
el1 = tcapi.get_elem(ctx, a, (1, 0, 0))
el2 = tcapi.get_elem(ctx, sub, (0, 0, 0))
# el1 == el2
\end{minted}

\bnoindentnl{replace\_sub}\hfill\break

\noindent\textit{\texttt{C++}:}
\begin{minted}[bgcolor=LightGray]{cpp}
template <typename TenT>
void replace_sub(
    tcapi::context_handle_t<TenT> &ctx,
    TenT &inout,
    const TenT &sub,
    const tcapi::elem_coors_t<TenT> &begin_pt
);                                      // (1)

template <typename TenT>
void replace_sub(
    tcapi::context_handle_t<TenT> &ctx,
    const TenT &in,
    const TenT &sub,
    const tcapi::elem_coors_t<TenT> &begin_pt,
    TenT &out
);                                      // (2)
\end{minted}
\noindent\textit{\texttt{Python}:}
\begin{minted}[bgcolor=LightGray]{python}
tcapi.replace_sub(ctx, a, sub, begin_pt)
\end{minted}

Replace a subregion of a tensor with \texttt{sub}, placing the origin (all-zero coordinate) of \texttt{sub} at \texttt{begin\_pt} in the target tensor.
In the \texttt{C++} interface, (1) modifies \texttt{inout} in place and (2) reads from \texttt{in} and writes the updated tensor into \texttt{out}.
In the \texttt{Python} interface, \texttt{begin\_pt} is a \texttt{Python} sequence, and the returned value is a raw backend tensor with the specified subregion replaced.

\example\hfill\break

\noindent\textit{\texttt{C++}:}
\begin{minted}[bgcolor=LightGray]{cpp}
auto a = tcapi::zeros<Ten>(ctx, {3, 4, 2});
Ten sub = tcapi::random<Ten>(
    ctx, {2, 2, 2}, gen
);
tcapi::replace_sub(ctx,
  a,
  sub,
  {1, 2, 0}
);
auto el1 = tcapi::get_elem(ctx, a, {1, 2, 0});
auto el2 = tcapi::get_elem(
  ctx, sub, {0, 0, 0}
);
// el1 == el2
\end{minted}
\noindent\textit{\texttt{Python}:}
\begin{minted}[bgcolor=LightGray]{python}
a = tcapi.zeros(ctx, Ten, (3, 4, 2))
sub = tcapi.random(ctx, Ten, (2, 2, 2), gen)
a = tcapi.replace_sub(ctx, a, sub, (1, 2, 0))
el1 = tcapi.get_elem(ctx, a, (1, 2, 0))
el2 = tcapi.get_elem(ctx, sub, (0, 0, 0))
# el1 == el2
\end{minted}

\bnoindentnl{concatenate}\hfill\break

\noindent\textit{\texttt{C++}:}
\begin{minted}[bgcolor=LightGray]{cpp}
template <typename TenT>
TenT concatenate(
    tcapi::context_handle_t<TenT> &ctx,
    const List<CRef<TenT>> &ins,
    const tcapi::bond_idx_t<TenT> concat_bdidx
);
\end{minted}
\noindent\textit{\texttt{Python}:}
\begin{minted}[bgcolor=LightGray]{python}
tcapi.concatenate(ctx, ins, concat_bdidx)
\end{minted}

Concatenate the tensors in \texttt{ins} along bond \texttt{concat\_bdidx} and return the result.
In the \texttt{C++} interface, \texttt{ins} is a list of non-owning const references to existing tensors; the referenced tensor objects are not copied, consumed, or modified and must remain valid for the duration of the call.
All input tensors must have the same order and identical bond dimensions on every bond except \texttt{concat\_bdidx}.
The output dimension along \texttt{concat\_bdidx} equals the sum of the corresponding input dimensions.
In the \texttt{Python} interface, \texttt{ins} is a \texttt{Python} sequence of raw backend tensors, and the returned value is a raw backend tensor.

\example\hfill\break

\noindent\textit{\texttt{C++}:}
\begin{minted}[bgcolor=LightGray]{cpp}
auto a = tcapi::random<Ten>(
  ctx, {2, 3, 4}, gen
);
auto b = tcapi::random<Ten>(
  ctx, {2, 3, 4}, gen
);
auto c = tcapi::random<Ten>(
  ctx, {2, 3, 4}, gen
);
auto d = tcapi::concatenate<Ten>(
    ctx,
    {std::cref(a), std::cref(b), std::cref(c)},
    1
);
auto s = tcapi::shape(ctx, d);
// s == {2, 9, 4}
auto el1 = tcapi::get_elem(ctx, b, {0, 0, 0});
auto el2 = tcapi::get_elem(ctx, d, {0, 3, 0});
// el1 == el2
\end{minted}
\noindent\textit{\texttt{Python}:}
\begin{minted}[bgcolor=LightGray]{python}
a = tcapi.random(ctx, Ten, (2, 3, 4), gen)
b = tcapi.random(ctx, Ten, (2, 3, 4), gen)
c = tcapi.random(ctx, Ten, (2, 3, 4), gen)
d = tcapi.concatenate(ctx, (a, b, c), 1)
s = tcapi.shape(ctx, d)
# s == (2, 9, 4)
el1 = tcapi.get_elem(ctx, b, (0, 0, 0))
el2 = tcapi.get_elem(ctx, d, (0, 3, 0))
# el1 == el2
\end{minted}

\bnoindentnl{stack}\hfill\break

\noindent\textit{\texttt{C++}:}
\begin{minted}[bgcolor=LightGray]{cpp}
template <typename TenT>
TenT stack(
    tcapi::context_handle_t<TenT> &ctx,
    const List<CRef<TenT>> &ins,
    const tcapi::bond_idx_t<TenT> stack_bdidx
);
\end{minted}
\noindent\textit{\texttt{Python}:}
\begin{minted}[bgcolor=LightGray]{python}
tcapi.stack(ctx, ins, stack_bdidx)
\end{minted}

Stack the tensors in \texttt{ins}, which must all have identical shape, along a new bond inserted at index \texttt{stack\_bdidx}.
In the \texttt{C++} interface, \texttt{ins} is a list of non-owning const references to existing tensors; the referenced tensor objects are not copied, consumed, or modified and must remain valid for the duration of the call.
If the inputs are $r$th-order tensors, the result is an ($r$+1)th-order tensor.
The dimension of the newly inserted bond equals \texttt{ins.size()} in \texttt{C++} or \texttt{len(ins)} in \texttt{Python}.
In the \texttt{Python} interface, \texttt{ins} is a \texttt{Python} sequence of raw backend tensors, and the returned value is a raw backend tensor.

\example\hfill\break

\noindent\textit{\texttt{C++}:}
\begin{minted}[bgcolor=LightGray]{cpp}
auto a = tcapi::random<Ten>(
  ctx, {2, 3, 4}, gen
);
auto b = tcapi::random<Ten>(
  ctx, {2, 3, 4}, gen
);
auto c = tcapi::random<Ten>(
  ctx, {2, 3, 4}, gen
);
auto d = tcapi::stack(
  ctx,
  {std::cref(a), std::cref(b), std::cref(c)},
  1
);
auto s = tcapi::shape(ctx, d);
// s == {2, 3, 3, 4}
auto el1 = tcapi::get_elem(ctx, b, {0, 2, 3});
auto el2 = tcapi::get_elem(
  ctx, d, {0, 1, 2, 3}
);
// el1 == el2
\end{minted}
\noindent\textit{\texttt{Python}:}
\begin{minted}[bgcolor=LightGray]{python}
a = tcapi.random(ctx, Ten, (2, 3, 4), gen)
b = tcapi.random(ctx, Ten, (2, 3, 4), gen)
c = tcapi.random(ctx, Ten, (2, 3, 4), gen)
d = tcapi.stack(ctx, (a, b, c), 1)
s = tcapi.shape(ctx, d)
# s == (2, 3, 3, 4)
el1 = tcapi.get_elem(ctx, b, (0, 2, 3))
el2 = tcapi.get_elem(ctx, d, (0, 1, 2, 3))
# el1 == el2
\end{minted}

\bnoindentnl{for\_each}\hfill\break

\noindent\textit{\texttt{C++}:}
\begin{minted}[bgcolor=LightGray]{cpp}
template <typename TenT, typename Func>
void for_each(
    tcapi::context_handle_t<TenT> &ctx,
    TenT &inout,
    Func &&f
);                                      // (1)
    
template <typename TenT, typename Func>
void for_each(
    tcapi::context_handle_t<TenT> &ctx,
    const TenT &in,
    Func &&f
);                                      // (2)
\end{minted}
\noindent\textit{\texttt{Python}:}
\begin{minted}[bgcolor=LightGray]{python}
tcapi.for_each(ctx, a, f)
\end{minted}

Visit every element exactly once and apply \texttt{f}.
The traversal order is implementation-defined.
In the \texttt{C++} interface, \texttt{Func} must be invocable with the corresponding signature.
(1) In-place (mutating) traversal: \texttt{f} is invoked as \texttt{f(elem)}, where \texttt{elem} has type \texttt{tcapi::elem\_t<TenT>\&}.
(2) Read-only \emph{const} traversal: \texttt{f} is invoked as \texttt{f(elem)}, where \texttt{elem} has type \texttt{const tcapi::elem\_t<TenT>} (or an equivalent const-qualified form, such as \texttt{const tcapi::elem\_t<TenT>\&}).
In the \texttt{Python} interface, \texttt{f} is invoked as \texttt{f(el)}, where \texttt{el} is a scalar value consistent with the relevant \texttt{TenKind}.
If \texttt{f(el)} returns \texttt{None}, the current element is copied unchanged into the result.
Otherwise, the returned value is used as the replacement element value.
\texttt{tcapi.for\_each} returns a raw backend tensor with the resulting element values.
The input tensor \texttt{a} is not modified.

\example\hfill\break

\noindent\textit{\texttt{C++}:}
\begin{minted}[bgcolor=LightGray]{cpp}
auto a = tcapi::eye<Ten>(ctx, 3);
float tot = 0.0;
auto sum = [&tot](const float el) {
  tot += el; 
};
const Ten &ca = a;
tcapi::for_each(ctx, ca, sum);  // call (2)
// tot == 3.0
auto plus_one = [](float &el) {
 el += 1.0;
};
tcapi::for_each(ctx, a, plus_one);  // call (1)
auto el = tcapi::get_elem(ctx, a, {1, 1});
// el == 2.0
\end{minted}
\noindent\textit{\texttt{Python}:}
\begin{minted}[bgcolor=LightGray]{python}
a = tcapi.eye(ctx, Ten, 3)
tot = {"value": 0.0}

def sum_elem(el):
    tot["value"] += el
    return None

tcapi.for_each(ctx, a, sum_elem)
# tot["value"] == 3.0

def plus_one(el):
    return el + 1.0

a = tcapi.for_each(ctx, a, plus_one)
el = tcapi.get_elem(ctx, a, (1, 1))
# el == 2.0
\end{minted}

\bnoindentnl{for\_each\_with\_coors}\hfill\break

\noindent\textit{\texttt{C++}:}
\begin{minted}[bgcolor=LightGray]{cpp}
template <typename TenT, typename Func>
void for_each_with_coors(
    tcapi::context_handle_t<TenT> &ctx,
    TenT &inout,
    Func &&f
);                                      // (1)
 
template <typename TenT, typename Func>
void for_each_with_coors(
    tcapi::context_handle_t<TenT> &ctx,
    const TenT &in,
    Func &&f
);                                      // (2)
\end{minted}
\noindent\textit{\texttt{Python}:}
\begin{minted}[bgcolor=LightGray]{python}
tcapi.for_each_with_coors(ctx, a, f)
\end{minted}

This function is similar to \texttt{for\_each}, but invokes \texttt{f} with both the element and its coordinates.
For a zeroth-order tensor, the callable is invoked once with the empty coordinate sequence.
The traversal order is implementation-defined.
In the \texttt{C++} interface, the callable \texttt{f} is invoked as \texttt{f(elem, coors)}, where \texttt{coors} has type \texttt{const tcapi::elem\_coors\_t<TenT>\&}.
In the \texttt{Python} interface, \texttt{f} is invoked as \texttt{f(el, coors)}, where \texttt{el} is a scalar value consistent with the relevant \texttt{TenKind}, and \texttt{coors} is a \texttt{Python} coordinate sequence in bond-index order.
If \texttt{f(el, coors)} returns \texttt{None}, the current element is copied unchanged into the result.
Otherwise, the returned value is used as the replacement element value.
\texttt{tcapi.for\_each\_with\_coors} returns a raw backend tensor with the resulting element values.
The input tensor \texttt{a} is not modified.

\example\hfill\break

\noindent\textit{\texttt{C++}:}
\begin{minted}[bgcolor=LightGray]{cpp}
auto a = tcapi::eye<Ten>(ctx, 3);
auto plus_coor0 = [](
    float &el,
    const tcapi::elem_coors_t<Ten> &coors
) {
  el += coors[0];
};
tcapi::for_each_with_coors(
    ctx, a, plus_coor0
);  // call (1)
auto el = tcapi::get_elem(ctx, a, {1, 1});
// el == 2.0
\end{minted}
\noindent\textit{\texttt{Python}:}
\begin{minted}[bgcolor=LightGray]{python}
a = tcapi.eye(ctx, Ten, 3)

def plus_coor0(el, coors):
    return el + coors[0]

a = tcapi.for_each_with_coors(
    ctx, a, plus_coor0
)
el = tcapi.get_elem(ctx, a, (1, 1))
# el == 2.0
\end{minted}

\subsubsection{Tensor linear-algebra operations \label{sec:tensor_linear_algebra}}

This subsection specifies tensor linear-algebra operations provided by TCAPI.
These routines operate on higher-order tensors by temporarily matricizing them according to a user-specified bond partition.
They then apply standard linear-algebra kernels and fold the results back into tensor form.
The functionality includes norms, normalization, scalar scaling, matrix functions, factorizations, and eigensolvers.
Unless otherwise noted, dimensional compatibility conditions required by the underlying matrix operations must be satisfied.
In the \texttt{Python} interface, tensor results are returned as raw backend tensors.
Multi-output routines return tuples.

\bnoindentnl{diag}\hfill\break

\noindent\textit{\texttt{C++}:}
\begin{minted}[bgcolor=LightGray]{cpp}
template <typename TenT>
void diag(
    tcapi::context_handle_t<TenT> &ctx,
    TenT &inout
);                                      // (1)
 
template <typename TenT>
void diag(
    tcapi::context_handle_t<TenT> &ctx,
    const TenT &in,
    TenT &out
);                                      // (2)
\end{minted}
\noindent\textit{\texttt{Python}:}
\begin{minted}[bgcolor=LightGray]{python}
tcapi.diag(ctx, a)
\end{minted}

If the input is first-order with shape $\{N\}$, the routine returns a second-order diagonal tensor with shape $\{N,N\}$, placing the input elements on the main diagonal and setting all off-diagonal elements to zero.
Implementations may store the result compactly when supported by the underlying TCF.
If the input is second-order with shape $\{M,N\}$, the routine extracts the main diagonal as a first-order tensor of length $\min(M,N)$.
The input tensor must be either first- or second-order.
In the \texttt{C++} interface, (1) modifies \texttt{inout} in place, and (2) reads from \texttt{in} and writes the result into \texttt{out}.
In the \texttt{Python} interface, the returned value is a raw backend tensor.

\example\hfill\break

\noindent\textit{\texttt{C++}:}
\begin{minted}[bgcolor=LightGray]{cpp}
auto a = tcapi::eye<Ten>(ctx, 3);
auto a_cpy = tcapi::copy(ctx, a);
tcapi::diag(ctx, a);
auto r = tcapi::order(ctx, a);
// r == 1
tcapi::diag(ctx, a);
// a == a_cpy
\end{minted}
\noindent\textit{\texttt{Python}:}
\begin{minted}[bgcolor=LightGray]{python}
a = tcapi.eye(ctx, Ten, 3)
a_cpy = tcapi.copy(ctx, a)
a = tcapi.diag(ctx, a)
r = tcapi.order(ctx, a)
# r == 1
a = tcapi.diag(ctx, a)
# a == a_cpy
\end{minted}

\bnoindentnl{norm}\hfill\break

\noindent\textit{\texttt{C++}:}
\begin{minted}[bgcolor=LightGray]{cpp}
template <typename TenT>
tcapi::real_t<TenT> norm(
    tcapi::context_handle_t<TenT> &ctx,
    const TenT &a
);
\end{minted}
\noindent\textit{\texttt{Python}:}
\begin{minted}[bgcolor=LightGray]{python}
tcapi.norm(ctx, a)
\end{minted}

Return the Frobenius norm of \texttt{a}.
In the \texttt{C++} interface, the returned value has type \texttt{tcapi::real\_t<TenT>}.
In the \texttt{Python} interface, the returned value is a scalar with the real dtype corresponding to \texttt{tcapi.ten\_kind(ctx, a)}.
For an $r$th-order tensor $A \in \mathbb{K}^{d_0\times \cdots \times d_{r-1}}$, the Frobenius norm is
\begin{equation}
\label{eq:frob_norm}
\lVert A\rVert_F =
\sqrt{\sum_{i_0=0}^{d_0-1}\cdots\sum_{i_{r-1}=0}^{d_{r-1}-1}
\big|A_{i_0\cdots i_{r-1}}\big|^2},
\end{equation}
where $|A_{i_{0} \cdots i_{r-1}}|$ denotes the modulus when $A$ is complex-valued.

\example\hfill\break

\noindent\textit{\texttt{C++}:}
\begin{minted}[bgcolor=LightGray]{cpp}
auto a = tcapi::eye<Ten>(ctx, 3);
auto norm = tcapi::norm(ctx, a);
// norm == std::sqrt(3.0)
\end{minted}
\noindent\textit{\texttt{Python}:}
\begin{minted}[bgcolor=LightGray]{python}
a = tcapi.eye(ctx, Ten, 3)
norm = tcapi.norm(ctx, a)
# norm == sqrt(3.0)
\end{minted}

\bnoindentnl{normalize}\hfill\break

\noindent\textit{\texttt{C++}:}
\begin{minted}[bgcolor=LightGray]{cpp}
template <typename TenT>
tcapi::real_t<TenT> normalize(
    tcapi::context_handle_t<TenT> &ctx,
    TenT &inout
);                                      // (1)
 
template <typename TenT>
tcapi::real_t<TenT> normalize(
    tcapi::context_handle_t<TenT> &ctx,
    const TenT &in,
    TenT &out
);                                      // (2)
\end{minted}
\noindent\textit{\texttt{Python}:}
\begin{minted}[bgcolor=LightGray]{python}
tcapi.normalize(ctx, a)
\end{minted}

Scale \texttt{a} to unit Frobenius norm.
In the \texttt{C++} interface, (1) normalizes \texttt{inout} in place, and (2) reads from \texttt{in} and writes the normalized tensor into \texttt{out}; both overloads return the original norm as \texttt{tcapi::real\_t<TenT>}.
In the \texttt{Python} interface, the returned value is the tuple \texttt{(normalized\_a, norm)}, where \texttt{normalized\_a} is a raw backend tensor containing the normalized tensor and \texttt{norm} is the original Frobenius norm.

\example\hfill\break

\noindent\textit{\texttt{C++}:}
\begin{minted}[bgcolor=LightGray]{cpp}
auto a = tcapi::random<Ten>(
  ctx, {3, 4, 2}, gen
);
tcapi::normalize(ctx, a);
auto norm = tcapi::norm(ctx, a);
// norm == 1.0

a = tcapi::random<Ten>(ctx, {3, 4, 2}, gen);
Ten b;
auto m = tcapi::normalize(ctx, a, b);
auto m2 = tcapi::norm(ctx, b);
// m2 == 1.0
\end{minted}
\noindent\textit{\texttt{Python}:}
\begin{minted}[bgcolor=LightGray]{python}
a = tcapi.random(ctx, Ten, (3, 4, 2), gen)
a, orig_norm = tcapi.normalize(ctx, a)
norm = tcapi.norm(ctx, a)
# norm == 1.0
\end{minted}

\bnoindentnl{scale}\hfill\break

\noindent\textit{\texttt{C++}:}
\begin{minted}[bgcolor=LightGray]{cpp}
template <typename TenT>
void scale(
    tcapi::context_handle_t<TenT> &ctx,
    TenT &inout,
    const tcapi::elem_t<TenT> s
);                                      // (1)

template <typename TenT>
void scale(
    tcapi::context_handle_t<TenT> &ctx,
    const TenT &in,
    const tcapi::elem_t<TenT> s,
    TenT &out
);                                      // (2)
\end{minted}
\noindent\textit{\texttt{Python}:}
\begin{minted}[bgcolor=LightGray]{python}
tcapi.scale(ctx, a, s)
\end{minted}

Multiply a tensor by a scalar \texttt{s}.
In the \texttt{C++} interface, \texttt{s} has type \texttt{tcapi::elem\_t<TenT>}, (1) scales \texttt{inout} in place, and (2) reads from \texttt{in} and writes the scaled tensor into \texttt{out}.
In the \texttt{Python} interface, \texttt{s} is a scalar value consistent with the dtype associated with \texttt{tcapi.ten\_kind(ctx, a)}, and the returned value is a raw backend tensor.

\example\hfill\break

\noindent\textit{\texttt{C++}:}
\begin{minted}[bgcolor=LightGray]{cpp}
auto a = tcapi::eye<Ten>(ctx, 3);

tcapi::scale(ctx, a, 3.0);
auto el = tcapi::get_elem(ctx, a, {2, 2});
// el == 3.0

Ten c;
tcapi::scale(ctx, a, -2.0, c);
auto el2 = tcapi::get_elem(ctx, c, {2, 2});
// el2 == -6.0
\end{minted}
\noindent\textit{\texttt{Python}:}
\begin{minted}[bgcolor=LightGray]{python}
a = tcapi.eye(ctx, Ten, 3)

a = tcapi.scale(ctx, a, 3.0)
el = tcapi.get_elem(ctx, a, (2, 2))
# el == 3.0

c = tcapi.scale(ctx, a, -2.0)
el2 = tcapi.get_elem(ctx, c, (2, 2))
# el2 == -6.0
\end{minted}

\bnoindentnl{trace}\hfill\break

\noindent\textit{\texttt{C++}:}
\begin{minted}[bgcolor=LightGray]{cpp}
template <typename TenT>
void trace(
    tcapi::context_handle_t<TenT> &ctx,
    TenT &inout,
    const bond_idx_pairs_t<TenT> &bdidx_pairs
);                                      // (1)

template <typename TenT>
void trace(
    tcapi::context_handle_t<TenT> &ctx,
    const TenT &in,
    const bond_idx_pairs_t<TenT> &bdidx_pairs,
    TenT &out
);                                      // (2)
\end{minted}
\noindent\textit{\texttt{Python}:}
\begin{minted}[bgcolor=LightGray]{python}
tcapi.trace(ctx, a, bdidx_pairs)
\end{minted}

Perform a partial trace by summing over the bond pairs specified by \texttt{bdidx\_pairs}.
The type \texttt{bond\_idx\_pairs\_t} is defined in Sec.~\ref{sec:aux_types}.
In the \texttt{Python} interface, \texttt{bdidx\_pairs} is a \texttt{Python} sequence of bond-index pairs.

Each paired bond must have the same dimension.
If all bonds are paired, the result is a zeroth-order tensor.
The relative order and dimensions of any remaining (untraced) bonds are preserved.

In the \texttt{C++} interface, (1) applies the trace to \texttt{inout} in place, and (2) reads from \texttt{in} and writes the result into \texttt{out}.
In the \texttt{Python} interface, the returned value is a raw backend tensor.

\example\hfill\break

\noindent\textit{\texttt{C++}:}
\begin{minted}[bgcolor=LightGray]{cpp}
auto a = tcapi::random<Ten>(
  ctx, {3, 4, 2, 4, 2}, gen
);
tcapi::trace(ctx, a, {{1, 3}, {2, 4}});
auto s = tcapi::shape(ctx, a);
// s == {3}
\end{minted}
\noindent\textit{\texttt{Python}:}
\begin{minted}[bgcolor=LightGray]{python}
a = tcapi.random(
    ctx, Ten, (3, 4, 2, 4, 2), gen
)
a = tcapi.trace(ctx, a, ((1, 3), (2, 4)))
s = tcapi.shape(ctx, a)
# s == (3,)
\end{minted}

\bnoindentnl{exp}\hfill\break

\noindent\textit{\texttt{C++}:}
\begin{minted}[bgcolor=LightGray]{cpp}
template <typename TenT>
void exp(
    tcapi::context_handle_t<TenT> &ctx,
    TenT &inout,
    const tcapi::order_t<TenT>
        num_of_bds_as_row
);                                      // (1)

template <typename TenT>
void exp(
    tcapi::context_handle_t<TenT> &ctx,
    const TenT &in,
    const tcapi::order_t<TenT>
        num_of_bds_as_row,
    TenT &out
);                                      // (2)
\end{minted}
\noindent\textit{\texttt{Python}:}
\begin{minted}[bgcolor=LightGray]{python}
tcapi.exp(ctx, a, num_of_bds_as_row)
\end{minted}

Compute the matrix exponential by matricizing the tensor with the first \texttt{num\_of\_bds\_as\_row} bonds as row indices and the remaining bonds as column indices.
The routine applies the matrix exponential to the resulting matrix and refolds the result back to the original tensor shape.
The total row dimension must equal the total column dimension.
In the \texttt{C++} interface, (1) modifies \texttt{inout} in place, and (2) reads from \texttt{in} and writes the result into \texttt{out}.
In the \texttt{Python} interface, \texttt{num\_of\_bds\_as\_row} is an integer-like value, and the returned value is a raw backend tensor.

\example\hfill\break

\noindent\textit{\texttt{C++}:}
\begin{minted}[bgcolor=LightGray]{cpp}
auto a = tcapi::eye<Ten>(ctx, 3);
tcapi::exp(ctx, a, 1);
auto el = tcapi::get_elem(ctx, a, {1, 1});
// el == 2.71828...
\end{minted}
\noindent\textit{\texttt{Python}:}
\begin{minted}[bgcolor=LightGray]{python}
a = tcapi.eye(ctx, Ten, 3)
a = tcapi.exp(ctx, a, 1)
el = tcapi.get_elem(ctx, a, (1, 1))
# el == 2.71828...
\end{minted}

\bnoindentnl{inverse}\hfill\break

\noindent\textit{\texttt{C++}:}
\begin{minted}[bgcolor=LightGray]{cpp}
template <typename TenT>
void inverse(
    tcapi::context_handle_t<TenT> &ctx,
    TenT &inout,
    const tcapi::order_t<TenT>
        num_of_bds_as_row
);                                      // (1)

template <typename TenT>
void inverse(
    tcapi::context_handle_t<TenT> &ctx,
    const TenT &in,
    const tcapi::order_t<TenT>
        num_of_bds_as_row,
    TenT &out
);                                      // (2)
\end{minted}
\noindent\textit{\texttt{Python}:}
\begin{minted}[bgcolor=LightGray]{python}
tcapi.inverse(ctx, a, num_of_bds_as_row)
\end{minted}

Compute the matrix inverse by matricizing the tensor with the first \texttt{num\_of\_bds\_as\_row} bonds as row indices and the remaining bonds as column indices.
The routine forms the corresponding square matrix, inverts it, and refolds the result back to the original tensor shape.
The total row dimension must equal the total column dimension, and the resulting matrix must be invertible.
In the \texttt{C++} interface, (1) modifies \texttt{inout} in place, and (2) reads from \texttt{in} and writes the result into \texttt{out}.
In the \texttt{Python} interface, \texttt{num\_of\_bds\_as\_row} is an integer-like value, and the returned value is a raw backend tensor.

\example\hfill\break

\noindent\textit{\texttt{C++}:}
\begin{minted}[bgcolor=LightGray]{cpp}
auto a = tcapi::eye<Ten>(ctx, 3);
tcapi::inverse(ctx, a, 1);
auto el = tcapi::get_elem(ctx, a, {1, 1});
// el == 1.0
\end{minted}
\noindent\textit{\texttt{Python}:}
\begin{minted}[bgcolor=LightGray]{python}
a = tcapi.eye(ctx, Ten, 3)
a = tcapi.inverse(ctx, a, 1)
el = tcapi.get_elem(ctx, a, (1, 1))
# el == 1.0
\end{minted}

\bnoindentnl{contract}\hfill\break

\noindent\textit{\texttt{C++}:}
\begin{minted}[bgcolor=LightGray]{cpp}
template <typename TenT>
void contract(
    tcapi::context_handle_t<TenT> &ctx,
    const TenT &a,
    const List<
        tcapi::bond_label_t<TenT>
        > &bd_labs_a,
    const TenT &b,
    const List<
        tcapi::bond_label_t<TenT>
        > &bd_labs_b,
    TenT &c,
    const List<
        tcapi::bond_label_t<TenT>
        > &bd_labs_c
);                                      // (1)

template <typename TenT>
void contract(
    tcapi::context_handle_t<TenT> &ctx,
    const TenT &a,
    const std::string_view bd_labs_str_a,
    const TenT &b,
    const std::string_view bd_labs_str_b,
    TenT &c,
    const std::string_view bd_labs_str_c
);                                      // (2)
\end{minted}
\noindent\textit{\texttt{Python}:}
\begin{minted}[bgcolor=LightGray]{python}
tcapi.contract(
    ctx, a, bd_labs_a, b, bd_labs_b,
    bd_labs_c
)
\end{minted}

This routine performs a label-based Einstein contraction.
Let the label lists for \texttt{a}, \texttt{b}, and the result be $\alpha$, $\beta$, and $\gamma$, respectively.
All labels that appear in both $\alpha$ and $\beta$ but not in $\gamma$ are contracted (summed over).
The labels in $\gamma$ define both the free bonds of the result and their ordering.
Bond dimensions associated with identical labels must agree.
In the \texttt{C++} interface, (1) provides labels as integer-like values of type \texttt{tcapi::bond\_label\_t<TenT>}, and (2) provides labels as strings (e.g., \texttt{"ijk"}), where each character is interpreted as a label.
In the \texttt{Python} interface, labels may be provided as sequences of integers or, for the string-label API, as strings; the returned value is a raw backend tensor.
If $\gamma$ is empty, the result is a zeroth-order tensor.
This API guarantees correctness even when the \texttt{C++} output tensor \texttt{c} aliases \texttt{a}, \texttt{b}, or both.
Repeating a label within a single operand (i.e., an implicit trace within \texttt{a} or \texttt{b}) is not supported in the current specification; use \texttt{trace} first if needed.

\example\hfill\break

\noindent\textit{\texttt{C++}:}
\begin{minted}[bgcolor=LightGray]{cpp}
auto a = tcapi::random<Ten>(
  ctx, {3, 4, 2}, gen
);
auto b = tcapi::random<Ten>(
  ctx, {2, 4, 5}, gen
);
Ten c;
tcapi::contract(ctx,
  a, {1, -1, -2},
  b, {-2, -1, 0},
  c, {0, 1}
);
// which is equivalent to
tcapi::contract(ctx,
  a, "ijk", b, "kjl",
  c, "li"
);
auto s = tcapi::shape(ctx, c);
// s == {5, 3}
\end{minted}
\noindent\textit{\texttt{Python}:}
\begin{minted}[bgcolor=LightGray]{python}
a = tcapi.random(ctx, Ten, (3, 4, 2), gen)
b = tcapi.random(ctx, Ten, (2, 4, 5), gen)
c = tcapi.contract(
    ctx,
    a, (1, -1, -2),
    b, (-2, -1, 0),
    (0, 1),
)
# which is equivalent to
c = tcapi.contract(
    ctx, a, "ijk", b, "kjl", "li"
)
s = tcapi.shape(ctx, c)
# s == (5, 3)
\end{minted}

\bnoindentnl{linear\_combine}\hfill\break

\noindent\textit{\texttt{C++}:}
\begin{minted}[bgcolor=LightGray]{cpp}
template <typename TenT>
TenT linear_combine(
    tcapi::context_handle_t<TenT> &ctx,
    const List<CRef<TenT>> &ins
);                                      // (1)
 
template <typename TenT>
TenT linear_combine(
    tcapi::context_handle_t<TenT> &ctx,
    const List<CRef<TenT>> &ins,
    const List<tcapi::elem_t<TenT>> &coefs
);                                      // (2)
\end{minted}
\noindent\textit{\texttt{Python}:}
\begin{minted}[bgcolor=LightGray]{python}
tcapi.linear_combine(ctx, ins)
tcapi.linear_combine(ctx, ins, coefs)
\end{minted}

Form and return the linear combination $\sum_{i=0}^{m-1} s_i A_i$, where $\{A_i\} = \texttt{ins}$ and $\{s_i\} = \texttt{coefs}$.
In overload (1), all coefficients are implicitly set to $s_i=1$.
All input tensors $A_i$ must have identical shapes.
In the \texttt{C++} interface, \texttt{ins} is a list of non-owning const references to existing tensors; the referenced tensor objects are not copied, consumed, or modified and must remain valid for the duration of the call.
In the \texttt{Python} interface, \texttt{ins} is a \texttt{Python} sequence of raw backend tensors, \texttt{coefs} is a \texttt{Python} sequence of scalar values consistent with the input tensor kind, and the returned value is a raw backend tensor.

\example\hfill\break

\noindent\textit{\texttt{C++}:}
\begin{minted}[bgcolor=LightGray]{cpp}
auto a = tcapi::random<Ten>(
  ctx, {3, 4, 2}, gen
);
auto b = tcapi::random<Ten>(
  ctx, {3, 4, 2}, gen
);
auto c = tcapi::random<Ten>(
  ctx, {3, 4, 2}, gen
);
auto d = tcapi::linear_combine(ctx,
  {std::cref(a), std::cref(b), std::cref(c)},
  {1.0, 2.0, 3.0}
);
\end{minted}
\noindent\textit{\texttt{Python}:}
\begin{minted}[bgcolor=LightGray]{python}
a = tcapi.random(ctx, Ten, (3, 4, 2), gen)
b = tcapi.random(ctx, Ten, (3, 4, 2), gen)
c = tcapi.random(ctx, Ten, (3, 4, 2), gen)
d = tcapi.linear_combine(
    ctx,
    (a, b, c),
    (1.0, 2.0, 3.0),
)
\end{minted}

\bnoindentnl{svd}\hfill\break

\noindent\textit{\texttt{C++}:}
\begin{minted}[bgcolor=LightGray]{cpp}
template <typename TenT>
void svd(
    tcapi::context_handle_t<TenT> &ctx,
    const TenT &a,
    const tcapi::order_t<TenT>
        num_of_bds_as_row,
    TenT &u,
    tcapi::real_ten_t<TenT> &sigma,
    TenT &v_dag
);
\end{minted}
\noindent\textit{\texttt{Python}:}
\begin{minted}[bgcolor=LightGray]{python}
tcapi.svd(ctx, a, num_of_bds_as_row)
\end{minted}

Perform an SVD of an $r$th-order tensor \texttt{a} by first matricizing it into a matrix.
Let $\texttt{shape(a)} = \{d_0,\cdots,d_{r-1}\}$ and $k = \texttt{num\_of\_bds\_as\_row}$ with $1 \leq k < r$.
The tensor is matricized by grouping the first $k$ bonds into the row index and the remaining bonds into the column index:
\begin{equation}
    A' \in \mathbb{K}^{I\times J},\quad I=\prod_{b=0}^{k-1}d_b,\quad J = \prod_{b=k}^{r-1}d_b~.
\end{equation}
An SVD
\begin{equation}
    A' = U\Sigma V^{\dagger}~
\end{equation}
is then computed, where $\Sigma = \text{diag}(s_0,\cdots,s_{\kappa-1})$ with $s_0 \geq s_1 \geq \cdots \geq s_{\kappa-1}\geq 0$ and $\kappa = \min(I,J)$.
Finally, the factors are folded back into tensors: $\texttt{shape(u)} = \{d_0,\cdots, d_{k-1}, \kappa\}$, $\texttt{shape(sigma)} = \{\kappa,\kappa\}$, and $\texttt{shape(v\_dag)} = \{\kappa, d_k, \cdots, d_{r-1}\}$.
\texttt{sigma} is a real second-order diagonal tensor corresponding to $\Sigma$; its diagonal elements are the singular values in non-increasing order, and all off-diagonal elements are zero.
The first $k$ bonds of \texttt{u} match the first $k$ bonds of \texttt{a}, and the remaining $r-k$ bonds of \texttt{v\_dag} match the remaining bonds of \texttt{a}.
In the \texttt{Python} interface, \texttt{num\_of\_bds\_as\_row} is an integer-like value, and the returned value is the tuple \texttt{(u, sigma, v\_dag)} of raw backend tensors.

\example\hfill\break

\noindent\textit{\texttt{C++}:}
\begin{minted}[bgcolor=LightGray]{cpp}
auto a = tcapi::random<Ten>(
  ctx, {3, 4, 12}, gen
);
Ten u, sigma, v_t;
tcapi::svd(ctx, a, 2, u, sigma, v_t);
\end{minted}
\noindent\textit{\texttt{Python}:}
\begin{minted}[bgcolor=LightGray]{python}
a = tcapi.random(ctx, Ten, (3, 4, 12), gen)
u, sigma, v_t = tcapi.svd(ctx, a, 2)
\end{minted}

\bnoindentnl{trunc\_svd}\hfill\break

\noindent\textit{\texttt{C++}:}
\begin{minted}[bgcolor=LightGray]{cpp}
template <typename TenT>
void trunc_svd(
    tcapi::context_handle_t<TenT> &ctx,
    const TenT &a,
    const tcapi::order_t<TenT>
        num_of_bds_as_row,
    TenT &u,
    tcapi::real_ten_t<TenT> &sigma,
    TenT &v_dag,
    tcapi::real_t<TenT> &trunc_err,
    const tcapi::bond_dim_t<TenT> chi_max,
    const tcapi::real_t<TenT> s_min
);                                      // (1)
    
template <typename TenT>
void trunc_svd(
    tcapi::context_handle_t<TenT> &ctx,
    const TenT &a,
    const tcapi::order_t<TenT>
        num_of_bds_as_row,
    TenT &u,
    tcapi::real_ten_t<TenT> &sigma,
    TenT &v_dag,
    tcapi::real_t<TenT> &trunc_err,
    const tcapi::bond_dim_t<TenT> chi_max,
    const tcapi::real_t<TenT> s_min,
    const tcapi::bond_dim_t<TenT> chi_min,
    const tcapi::real_t<TenT> target_trunc_err
);                                      // (2)
\end{minted}
\noindent\textit{\texttt{Python}:}
\begin{minted}[bgcolor=LightGray]{python}
tcapi.trunc_svd(
    ctx, a, num_of_bds_as_row, chi_max, s_min
)
tcapi.trunc_svd(
    ctx, a, num_of_bds_as_row,
    chi_max, s_min, chi_min, target_trunc_err
)
\end{minted}

This routine follows the same matricization and SVD convention as \texttt{svd}, but additionally truncates the singular values according to a prescribed strategy.
For pre-truncation singular values $s_0 \geq \cdots \geq s_{\kappa-1}$, define the (relative) truncation error for keeping $\chi$ values as
\begin{equation}
    \epsilon = \frac{\sum_{i = \chi}^{\kappa-1} s_i^2}{\sum_{i = 0}^{\kappa-1} s_i^2}~,
\end{equation}
which is returned in \texttt{trunc\_err}.
(1) Fixed maximum bond dimension kept: keep at most $\texttt{chi\_max}$ singular values after discarding those below $\texttt{s\_min}$.
Equivalently, (2) with $\texttt{chi\_min} = 1$ and $\texttt{target\_trunc\_err} = 0$.
(2) General truncation strategy:
\begin{itemize}
    \item Discard all $s_i < \texttt{s\_min}$.
    \item Among the singular values that survive the \texttt{s\_min} cutoff, retain at least \texttt{chi\_min} values when possible. If fewer than \texttt{chi\_min} values survive the cutoff, retain all surviving values; values below \texttt{s\_min} are not restored to satisfy \texttt{chi\_min}.
    \item Increase $\chi$ (retaining additional values in descending order) until either $\epsilon \leq \texttt{target\_trunc\_err}$ or $\chi = \texttt{chi\_max}$.
\end{itemize}
The output \texttt{sigma} is a second-order diagonal tensor of retained singular values with shape $\{\chi,\chi\}$, where $\chi$ is the retained bond dimension.
In the \texttt{Python} interface, \texttt{num\_of\_bds\_as\_row}, \texttt{chi\_min}, and \texttt{chi\_max} are integer-like values, \texttt{target\_trunc\_err} and \texttt{s\_min} are real scalar values, and the returned value is the tuple \texttt{(u, sigma, v\_dag, trunc\_err)}.

\example\hfill\break

\noindent\textit{\texttt{C++}:}
\begin{minted}[bgcolor=LightGray]{cpp}
Ten u, sigma, v_t;
auto a = tcapi::random<Ten>(
  ctx, {3, 4, 12}, gen
);
float trunc_err = 0.0;
tcapi::trunc_svd(ctx,
  a, 2,
  u, sigma, v_t,
  trunc_err,
  6, 1e-12, 3, 1e-2
);
// trunc_err != 0.0
\end{minted}
\noindent\textit{\texttt{Python}:}
\begin{minted}[bgcolor=LightGray]{python}
a = tcapi.random(ctx, Ten, (3, 4, 12), gen)
u, sigma, v_t, trunc_err = tcapi.trunc_svd(
    ctx,
    a, 2,
    6, 1e-12, 3, 1e-2,
)
# trunc_err != 0.0
\end{minted}

\bnoindentnl{qr}\hfill\break

\noindent\textit{\texttt{C++}:}
\begin{minted}[bgcolor=LightGray]{cpp}
template <typename TenT>
void qr(
    tcapi::context_handle_t<TenT> &ctx,
    const TenT &a,
    const tcapi::order_t<TenT>
        num_of_bds_as_row,
    TenT &q,
    TenT &r
);
\end{minted}
\noindent\textit{\texttt{Python}:}
\begin{minted}[bgcolor=LightGray]{python}
tcapi.qr(ctx, a, num_of_bds_as_row)
\end{minted}

Perform an economy (thin) QR decomposition of an $r$th-order tensor \texttt{a} using the same matricization convention as \texttt{svd}.
Let $\texttt{shape(a)} = \{d_0,\cdots,d_{r-1}\}$ and $k = \texttt{num\_of\_bds\_as\_row}$ with $1 \leq k < r$.
The tensor is first matricized by grouping the first $k$ bonds into the row index and the remaining bonds into the column index:
\begin{equation}
    A' \in \mathbb{K}^{I\times J},\quad I=\prod_{b=0}^{k-1}d_b,\quad J = \prod_{b=k}^{r-1}d_b~.
\end{equation}
A QR decomposition
\begin{equation}
    A' = QR~
\end{equation}
is then computed, where $Q \in \mathbb{K}^{I\times \rho}$ has orthonormal columns and $R \in \mathbb{K}^{\rho \times J}$ is upper triangular (upper trapezoidal if $I < J$) with $\rho = \min(I,J)$.
Finally, the factors are folded back into tensors: $\texttt{shape(q)} = \{d_0,\cdots, d_{k-1}, \rho\}$ and $\texttt{shape(r)} = \{\rho, d_k, \cdots, d_{r-1}\}$.
In the \texttt{Python} interface, \texttt{num\_of\_bds\_as\_row} is an integer-like value, and the returned value is the tuple \texttt{(q, r)} of raw backend tensors.

\example\hfill\break

\noindent\textit{\texttt{C++}:}
\begin{minted}[bgcolor=LightGray]{cpp}
auto a = tcapi::random<Ten>(
  ctx, {3, 4, 12}, gen
);
Ten q, r;
tcapi::qr(ctx, a, 2, q, r);
\end{minted}
\noindent\textit{\texttt{Python}:}
\begin{minted}[bgcolor=LightGray]{python}
a = tcapi.random(ctx, Ten, (3, 4, 12), gen)
q, r = tcapi.qr(ctx, a, 2)
\end{minted}

\bnoindentnl{lq}\hfill\break

\noindent\textit{\texttt{C++}:}
\begin{minted}[bgcolor=LightGray]{cpp}
template <typename TenT>
void lq(
    tcapi::context_handle_t<TenT> &ctx,
    const TenT &a,
    const tcapi::order_t<TenT>
        num_of_bds_as_row,
    TenT &l,
    TenT &q
);
\end{minted}
\noindent\textit{\texttt{Python}:}
\begin{minted}[bgcolor=LightGray]{python}
tcapi.lq(ctx, a, num_of_bds_as_row)
\end{minted}

Perform an economy (thin) LQ decomposition of an $r$th-order tensor \texttt{a} using the same matricization convention as \texttt{svd}.
Let $\texttt{shape(a)} = \{d_0,\cdots,d_{r-1}\}$ and $k = \texttt{num\_of\_bds\_as\_row}$ with $1 \leq k < r$.
The tensor is first matricized by grouping the first $k$ bonds into the row index and the remaining bonds into the column index:
\begin{equation}
    A' \in \mathbb{K}^{I\times J},\quad I=\prod_{b=0}^{k-1}d_b,\quad J = \prod_{b=k}^{r-1}d_b~.
\end{equation}
An LQ decomposition
\begin{equation}
    A' = LQ~,
\end{equation}
is then computed, where $L \in \mathbb{K}^{I\times \rho}$ is lower triangular (lower trapezoidal if $I > J$) and $Q \in \mathbb{K}^{\rho \times J}$ has orthonormal rows with $\rho = \min(I,J)$.
Finally, the factors are folded back into tensors: $\texttt{shape(l)} = \{d_0,\cdots, d_{k-1}, \rho\}$ and $\texttt{shape(q)} = \{\rho, d_k, \cdots, d_{r-1}\}$.
In the \texttt{Python} interface, \texttt{num\_of\_bds\_as\_row} is an integer-like value, and the returned value is the tuple \texttt{(l, q)} of raw backend tensors.

\example\hfill\break

\noindent\textit{\texttt{C++}:}
\begin{minted}[bgcolor=LightGray]{cpp}
auto a = tcapi::random<Ten>(
  ctx, {3, 4, 12}, gen
);
Ten l, q;
tcapi::lq(ctx, a, 2, l, q);
\end{minted}
\noindent\textit{\texttt{Python}:}
\begin{minted}[bgcolor=LightGray]{python}
a = tcapi.random(ctx, Ten, (3, 4, 12), gen)
l, q = tcapi.lq(ctx, a, 2)
\end{minted}

\bnoindentnl{eigvals}\hfill\break

\noindent\textit{\texttt{C++}:}
\begin{minted}[bgcolor=LightGray]{cpp}
template <typename TenT>
void eigvals(
    tcapi::context_handle_t<TenT> &ctx,
    const TenT &a,
    const tcapi::order_t<TenT>
        num_of_bds_as_row,
    tcapi::cplx_ten_t<TenT> &w
);
\end{minted}
\noindent\textit{\texttt{Python}:}
\begin{minted}[bgcolor=LightGray]{python}
tcapi.eigvals(ctx, a, num_of_bds_as_row)
\end{minted}

Compute the eigenvalues of an $r$th-order tensor \texttt{a} by first reshaping it into a matrix.
Let $\texttt{shape(a)} = \{d_0,\cdots,d_{r-1}\}$ and $k = \texttt{num\_of\_bds\_as\_row}$ with $1 \leq k < r$.
The tensor is matricized by grouping the first $k$ bonds into the row index and the remaining bonds into the column index:
\begin{equation}
    A' \in \mathbb{K}^{I\times J},\quad I=\prod_{b=0}^{k-1}d_b,\quad J = \prod_{b=k}^{r-1}d_b~,
\end{equation}
where $I=J$ is required.
The spectrum $\{ \lambda_i \}_{i=0}^{I-1}$ of $A'$ is then computed (e.g., via a dense eigensolver) and is returned as a first-order complex tensor \texttt{w}.
Unlike \texttt{eig}, which returns the eigenvalue matrix as a second-order diagonal tensor for use in decompositions, this routine returns only the eigenvalue list as a first-order tensor.
In the \texttt{Python} interface, \texttt{num\_of\_bds\_as\_row} is an integer-like value, and the returned value is a raw backend tensor whose kind is the complex counterpart of the input tensor kind.

\example\hfill\break

\noindent\textit{\texttt{C++}:}
\begin{minted}[bgcolor=LightGray]{cpp}
auto a = tcapi::random<Ten>(
  ctx, {3, 4, 12}, gen
);
CplxTen w;
tcapi::eigvals(ctx, a, 2, w);
\end{minted}
\noindent\textit{\texttt{Python}:}
\begin{minted}[bgcolor=LightGray]{python}
a = tcapi.random(ctx, Ten, (3, 4, 12), gen)
w = tcapi.eigvals(ctx, a, 2)
\end{minted}

\bnoindentnl{eigvalsh}\hfill\break

\noindent\textit{\texttt{C++}:}
\begin{minted}[bgcolor=LightGray]{cpp}
template <typename TenT>
void eigvalsh(
    tcapi::context_handle_t<TenT> &ctx,
    const TenT &a,
    const tcapi::order_t<TenT>
        num_of_bds_as_row,
    tcapi::real_ten_t<TenT> &w
);
\end{minted}
\noindent\textit{\texttt{Python}:}
\begin{minted}[bgcolor=LightGray]{python}
tcapi.eigvalsh(ctx, a, num_of_bds_as_row)
\end{minted}

Hermitian/symmetric variant of \texttt{eigvals}.
The input is interpreted as a real symmetric or complex Hermitian matrix after matricization, and the eigenvalues are returned as a real first-order tensor \texttt{w} in ascending order.
Unlike \texttt{eigh}, which returns the eigenvalue matrix as a second-order diagonal tensor for use in decompositions, this routine returns only the eigenvalue list as a first-order tensor.
In the \texttt{Python} interface, \texttt{num\_of\_bds\_as\_row} is an integer-like value, and the returned value is a raw backend tensor whose kind is the real counterpart of the input tensor kind.

\example\hfill\break

\noindent\textit{\texttt{C++}:}
\begin{minted}[bgcolor=LightGray]{cpp}
auto a = tcapi::random<Ten>(
  ctx, {3, 4, 3, 4}, gen
);
Ten a_t, w;
tcapi::transpose(ctx, a, {2, 3, 0, 1}, a_t);
auto a_sym = tcapi::linear_combine(
  ctx, {std::cref(a), std::cref(a_t)}
);
tcapi::eigvalsh(ctx, a_sym, 2, w);
\end{minted}
\noindent\textit{\texttt{Python}:}
\begin{minted}[bgcolor=LightGray]{python}
a = tcapi.random(ctx, Ten, (3, 4, 3, 4), gen)
a_t = tcapi.transpose(ctx, a, (2, 3, 0, 1))
a_sym = tcapi.linear_combine(ctx, (a, a_t))
w = tcapi.eigvalsh(ctx, a_sym, 2)
\end{minted}

\bnoindentnl{eig}\hfill\break

\noindent\textit{\texttt{C++}:}
\begin{minted}[bgcolor=LightGray]{cpp}
template <typename TenT>
void eig(
    tcapi::context_handle_t<TenT> &ctx,
    const TenT &a,
    const tcapi::order_t<TenT>
        num_of_bds_as_row,
    tcapi::cplx_ten_t<TenT> &lambda_mat,
    tcapi::cplx_ten_t<TenT> &v
);
\end{minted}
\noindent\textit{\texttt{Python}:}
\begin{minted}[bgcolor=LightGray]{python}
tcapi.eig(ctx, a, num_of_bds_as_row)
\end{minted}

Perform a right eigendecomposition of an $r$th-order tensor \texttt{a} by first matricizing it into a matrix.
Let $\texttt{shape(a)} = \{d_0,\cdots,d_{r-1}\}$ and $k = \texttt{num\_of\_bds\_as\_row}$ with $1 \leq k < r$.
The tensor is matricized by grouping the first $k$ bonds into the row index and the remaining bonds into the column index:
\begin{equation}
    A' \in \mathbb{K}^{I\times J},\quad I=\prod_{b=0}^{k-1}d_b,\quad J = \prod_{b=k}^{r-1}d_b~,
\end{equation}
where $I=J$ is required.
A right-eigendecomposition
\begin{equation}
    A'V=V\Lambda~,
\end{equation}
is then computed, where the diagonal elements of $\Lambda = \text{diag}(\lambda_0,\cdots,\lambda_{I-1})$ are the eigenvalues and the columns of $V$ are the corresponding right eigenvectors.
The routine returns the eigenvalues as \texttt{lambda\_mat}, a second-order diagonal tensor representing $\Lambda$ with shape $\{I,I\}$, and returns the eigenvectors in \texttt{v}, obtained by folding $V$ back to a tensor with $\texttt{shape(v)} = \{d_0,\cdots,d_{k-1},I\}$, so that the last bond indexes eigenvectors and the first $k$ bonds match those of \texttt{a}.
In the \texttt{Python} interface, \texttt{num\_of\_bds\_as\_row} is an integer-like value, and the returned value is the tuple \texttt{(lambda\_mat, v)} of raw backend tensors whose kinds are the complex counterpart of the input tensor kind.

\example\hfill\break

\noindent\textit{\texttt{C++}:}
\begin{minted}[bgcolor=LightGray]{cpp}
auto a = tcapi::random<Ten>(
  ctx, {3, 4, 12}, gen
);
CplxTen lambda_mat, v;
tcapi::eig(ctx, a, 2, lambda_mat, v);
auto s = tcapi::shape(ctx, v);
// s == {3, 4, 12}
\end{minted}
\noindent\textit{\texttt{Python}:}
\begin{minted}[bgcolor=LightGray]{python}
a = tcapi.random(ctx, Ten, (3, 4, 12), gen)
lambda_mat, v = tcapi.eig(ctx, a, 2)
s = tcapi.shape(ctx, v)
# s == (3, 4, 12)
\end{minted}

\bnoindentnl{eigh}\hfill\break

\noindent\textit{\texttt{C++}:}
\begin{minted}[bgcolor=LightGray]{cpp}
template <typename TenT>
void eigh(
    tcapi::context_handle_t<TenT> &ctx,
    const TenT &a,
    const tcapi::order_t<TenT>
        num_of_bds_as_row,
    tcapi::real_ten_t<TenT> &lambda_mat,
    TenT &v
);
\end{minted}
\noindent\textit{\texttt{Python}:}
\begin{minted}[bgcolor=LightGray]{python}
tcapi.eigh(ctx, a, num_of_bds_as_row)
\end{minted}

Hermitian/symmetric variant of \texttt{eig}.
The input is interpreted as a real symmetric or complex Hermitian matrix after matricization.
The eigenvalues are returned as \texttt{lambda\_mat}, a real second-order diagonal tensor whose diagonal elements are the eigenvalues in ascending order, and the corresponding eigenvectors are returned in \texttt{v}.
In the \texttt{Python} interface, \texttt{num\_of\_bds\_as\_row} is an integer-like value, and the returned value is the tuple \texttt{(lambda\_mat, v)} of raw backend tensors; \texttt{lambda\_mat} has the tensor kind obtained by taking the real counterpart of the input tensor kind, while \texttt{v} has the same kind as the input tensor.

\example\hfill\break

\noindent\textit{\texttt{C++}:}
\begin{minted}[bgcolor=LightGray]{cpp}
auto a = tcapi::random<Ten>(
  ctx, {3, 4, 3, 4}, gen
);
Ten a_t, lambda_mat, v;
tcapi::transpose(ctx, a, {2, 3, 0, 1}, a_t);
auto a_sym = tcapi::linear_combine(
  ctx, {std::cref(a), std::cref(a_t)}
);
tcapi::eigh(ctx, a_sym, 2, lambda_mat, v);
auto s = tcapi::shape(ctx, v);
// s == {3, 4, 12}
\end{minted}
\noindent\textit{\texttt{Python}:}
\begin{minted}[bgcolor=LightGray]{python}
a = tcapi.random(ctx, Ten, (3, 4, 3, 4), gen)
a_t = tcapi.transpose(ctx, a, (2, 3, 0, 1))
a_sym = tcapi.linear_combine(ctx, (a, a_t))
lambda_mat, v = tcapi.eigh(ctx, a_sym, 2)
s = tcapi.shape(ctx, v)
# s == (3, 4, 12)
\end{minted}

\subsubsection{Miscellaneous routines}

This subsection specifies auxiliary routines that support context management, data movement, debugging, and interoperability.
These routines are not performance-critical.

\bnoindentnl{create\_context}\hfill\break

\noindent\textit{\texttt{C++}:}
\begin{minted}[bgcolor=LightGray]{cpp}
template <typename ContextHandleT,
          /* implementation-defined */ >
void create_context(
    ContextHandleT &ctx,
    /* implementation-defined */
);
\end{minted}
\noindent\textit{\texttt{Python}:}
\begin{minted}[bgcolor=LightGray]{python}
tcapi.create_context(Ten, **options)
\end{minted}

Create and initialize a backend execution context.
In the \texttt{C++} interface, the concrete type \texttt{ContextHandleT} selects the target TCF/backend.
The handle \texttt{ctx} is default-constructible and is initially not bound to a live execution context; on success, it becomes a valid context handle.
In the \texttt{Python} interface, \texttt{Ten} is a \texttt{TenKind} that selects the implementation family for the context, and the returned value is an implementation-defined context handle.
A context may support multiple \texttt{C++} tensor types \texttt{TenT} or \texttt{Python} \texttt{TenKind} objects provided by the same TCAPI implementation.
Implementations may accept additional parameters, such as device IDs, streams/queues, allocators, or logging/debug flags.

\example\hfill\break

\noindent\textit{\texttt{C++}:}
\begin{minted}[bgcolor=LightGray]{cpp}
// Pseudocode
// For a tensor on CPU with the type CPUTen
tcapi::context_handle_t<CPUTen> ctx_cpu;
// default threads, default allocator
tcapi::create_context(ctx_cpu);

// For a tensor on GPU with the type GPUTen
tcapi::context_handle_t<GPUTen> ctx_gpu;
int device_id = 0;
GpuStream stream = /* ... */;
// select device and stream/queue
tcapi::create_context(
  ctx_gpu, device_id, stream
);
\end{minted}
\noindent\textit{\texttt{Python}:}
\begin{minted}[bgcolor=LightGray]{python}
# Pseudocode
# For a tensor on CPU with the TenKind CPUTen
ctx_cpu = tcapi.create_context(CPUTen)

# For a tensor on GPU with the TenKind GPUTen
ctx_gpu = tcapi.create_context(
    GPUTen,
    device_id=0,
    stream=stream,
)
\end{minted}

\bnoindentnl{destroy\_context}\hfill\break

\noindent\textit{\texttt{C++}:}
\begin{minted}[bgcolor=LightGray]{cpp}
template <typename ContextHandleT>
void destroy_context(ContextHandleT &ctx);
\end{minted}
\noindent\textit{\texttt{Python}:}
\begin{minted}[bgcolor=LightGray]{python}
tcapi.destroy_context(ctx)
\end{minted}

Destroy the backend execution context managed by \texttt{ctx} and release any owned resources (e.g., devices, streams/queues, library handles, memory pools).
After this call, \texttt{ctx} is invalidated and must not be passed to further TCAPI routines unless it is re-initialized via \texttt{create\_context}.

\example\hfill\break

\noindent\textit{\texttt{C++}:}
\begin{minted}[bgcolor=LightGray]{cpp}
tcapi::destroy_context(ctx);
\end{minted}
\noindent\textit{\texttt{Python}:}
\begin{minted}[bgcolor=LightGray]{python}
tcapi.destroy_context(ctx)
\end{minted}

\bnoindentnl{to\_range}\hfill\break

\noindent\textit{\texttt{C++}:}
\begin{minted}[bgcolor=LightGray]{cpp}
template <typename TenT,
          typename RandomIt,
          typename Func>
void to_range(
    tcapi::context_handle_t<TenT> &ctx,
    const TenT &a,
    RandomIt first,
    Func &&coors2idx
);
\end{minted}
\noindent\textit{\texttt{Python}:}
\begin{minted}[bgcolor=LightGray]{python}
tcapi.to_range(ctx, a, coors2idx)
\end{minted}

Copy all elements of \texttt{a} into a one-dimensional container using a user-supplied index map.
In the \texttt{C++} interface, elements are copied into a random-access output range beginning at \texttt{first}; for each coordinate tuple \texttt{coors}, the destination is written as \texttt{*(first + coors2idx(coors))}.
The callable \texttt{coors2idx} must have the following signature:
\begin{minted}[bgcolor=LightGray]{cpp}
std::iterator_traits<
        RandomIt
>::difference_type coors2idx(
    const tcapi::elem_coors_t<TenT> &coors
);
\end{minted}
It must map every valid coordinate of \texttt{a} to a valid index in $[0,\texttt{tcapi::size(ctx,a)})$.
In the \texttt{Python} interface, the returned value is a \texttt{Python} \texttt{list} with length \texttt{tcapi.size(ctx, a)}.
The callable \texttt{coors2idx} is invoked as \texttt{coors2idx(coors)}, where \texttt{coors} is a \texttt{Python} coordinate sequence in bond-index order, and must return an integer-like index in $[0,\texttt{tcapi.size(ctx,a)})$.
For a zeroth-order tensor, the output range or list has length 1; \texttt{coors2idx} is invoked once with the empty coordinate sequence and must return 0.
Each list element is a scalar value consistent with the relevant \texttt{TenKind}.
For a diagonal tensor, this routine writes all tensor elements, including off-diagonal zeros.
To extract only diagonal elements, use \texttt{diag}.

\example\hfill\break

\noindent\textit{\texttt{C++}:}
\begin{minted}[bgcolor=LightGray]{cpp}
using Elem = typename tcapi::elem_t<Ten>;
using ElemVec = std::vector<Elem>;
std::iterator_traits<
    typename ElemVec::iterator
>::difference_type coors2idx(
    const tcapi::elem_coors_t<Ten> &coors
) {
  return 3 * coors[0] + coors[1]; 
}
ElemVec els {1.0, 2.0, 3.0, 4.0, 5.0, 6.0};
auto a = tcapi::assign_from_range<Ten>(
    ctx, {2, 3}, els.begin(), coors2idx
);
ElemVec els2(6);
tcapi::to_range(ctx,
  a,
  els2.begin(),
  coors2idx
);
// els2 == els
\end{minted}
\noindent\textit{\texttt{Python}:}
\begin{minted}[bgcolor=LightGray]{python}
def coors2idx(coors):
    return 3 * coors[0] + coors[1]

els = [1.0, 2.0, 3.0, 4.0, 5.0, 6.0]
a = tcapi.assign_from_range(
    ctx, Ten, (2, 3), els, coors2idx
)
els2 = tcapi.to_range(ctx, a, coors2idx)
# els2 == els
\end{minted}

\bnoindentnl{show}\hfill\break

\noindent\textit{\texttt{C++}:}
\begin{minted}[bgcolor=LightGray]{cpp}
template <typename TenT>
void show(
    tcapi::context_handle_t<TenT> &ctx,
    const TenT &a
);
\end{minted}
\noindent\textit{\texttt{Python}:}
\begin{minted}[bgcolor=LightGray]{python}
tcapi.show(ctx, a)
\end{minted}

Print the contents of \texttt{a} to standard output in a human-readable form.
The exact formatting (layout, precision, ordering, etc.) is implementation-defined.

\example\hfill\break

\noindent\textit{\texttt{C++}:}
\begin{minted}[bgcolor=LightGray]{cpp}
tcapi::show(ctx, a);
\end{minted}
\noindent\textit{\texttt{Python}:}
\begin{minted}[bgcolor=LightGray]{python}
tcapi.show(ctx, a)
\end{minted}

\bnoindentnl{close}\hfill\break

\noindent\textit{\texttt{C++}:}
\begin{minted}[bgcolor=LightGray]{cpp}
template <typename TenT>
bool close(
    tcapi::context_handle_t<TenT> &ctx,
    const TenT &a,
    const TenT &b,
    const tcapi::real_t<TenT> epsilon
);
\end{minted}
\noindent\textit{\texttt{Python}:}
\begin{minted}[bgcolor=LightGray]{python}
tcapi.close(ctx, a, b, epsilon)
\end{minted}

This routine performs an elementwise absolute-tolerance comparison.
The comparison succeeds iff $\texttt{shape(a)} = \texttt{shape(b)} = \{d_0,\cdots, d_{r-1}\}$, and
\begin{equation}
    \text{Close}(A,B;\epsilon) \equiv \max_{0\leq i_b < d_b} |A_{i_0\cdots i_{r-1}} - B_{i_0\cdots i_{r-1}}|\leq \epsilon~,
\end{equation}
where $|\cdot|$ denotes absolute value, or modulus for complex values.
The parameter $\texttt{epsilon}$ must satisfy $\texttt{epsilon} \geq 0$.
In the \texttt{C++} interface, the returned value is \texttt{true} when the comparison succeeds and \texttt{false} otherwise.
In the \texttt{Python} interface, \texttt{epsilon} is a real scalar value and the returned value is a Boolean-like value.

\example\hfill\break

\noindent\textit{\texttt{C++}:}
\begin{minted}[bgcolor=LightGray]{cpp}
auto A = tcapi::eye<Ten>(ctx, 3);
auto B = tcapi::copy(ctx, A);
tcapi::set_elem(ctx, B, {2, 0}, 1e-5);

bool ok_abs = tcapi::close(ctx, A, B, 1e-6);
// false
ok_abs = tcapi::close(ctx, A, B, 1e-4f);
// true
\end{minted}
\noindent\textit{\texttt{Python}:}
\begin{minted}[bgcolor=LightGray]{python}
A = tcapi.eye(ctx, Ten, 3)
B = tcapi.copy(ctx, A)
B = tcapi.set_elem(ctx, B, (2, 0), 1e-5)

ok_abs = tcapi.close(ctx, A, B, 1e-6)
# False
ok_abs = tcapi.close(ctx, A, B, 1e-4)
# True
\end{minted}

\bnoindentnl{convert}\hfill\break

\noindent\textit{\texttt{C++}:}
\begin{minted}[bgcolor=LightGray]{cpp}
template <typename Ten1T, typename Ten2T>
void convert(
    tcapi::context_handle_t<Ten1T> &ctx1,
    const Ten1T &t1,
    tcapi::context_handle_t<Ten2T> &ctx2,
    Ten2T &t2
);
\end{minted}
\noindent\textit{\texttt{Python}:}
\begin{minted}[bgcolor=LightGray]{python}
tcapi.convert(ctx1, t1, ctx2, Ten2)
\end{minted}

Convert \texttt{t1} into a destination tensor, possibly changing the element type, the underlying TCF/backend, or both.
This includes conversions between different floating-point precisions within the same TCF.
Memory movement between \texttt{ctx1} (source TCF) and \texttt{ctx2} (destination TCF) is implementation-defined.
In the \texttt{C++} interface, the result is written into \texttt{t2}; on success, \texttt{t2} must be fully constructed and ready for use.
In the \texttt{Python} interface, \texttt{Ten2} is the destination \texttt{TenKind}, and the returned value is a raw backend tensor associated with \texttt{ctx2}.
If the source and destination have the same tensor type or tensor kind, this routine is equivalent to a deep copy.

In the precision-conversion examples below, \texttt{Real64Ten} denotes an implementation-provided \texttt{C++} tensor type or \texttt{Python} \texttt{TenKind} with 64-bit real floating-point elements.
Both \texttt{Ten} and \texttt{Real64Ten} are assumed to be provided by the same TCF and supported by \texttt{ctx\_cpu}.

\example\hfill\break

\noindent\textit{\texttt{C++}:}
\begin{minted}[bgcolor=LightGray]{cpp}
// CPU float -> CPU complex (promotion)
auto A = tcapi::random<Ten>(
  ctx_cpu, {4, 4}, gen
);
CplxTen Ac;
tcapi::convert(ctx_cpu, A, ctx_cpu, Ac);

// CPU float -> GPU float (device transfer)
GPUTen Ag;
tcapi::convert(ctx_cpu, A, ctx_gpu, Ag);

// CPU float32 -> CPU float64 (precision conversion)
Real64Ten A64;
tcapi::convert(ctx_cpu, A, ctx_cpu, A64);
\end{minted}
\noindent\textit{\texttt{Python}:}
\begin{minted}[bgcolor=LightGray]{python}
# CPU float -> CPU complex (promotion)
A = tcapi.random(ctx_cpu, Ten, (4, 4), gen)
Ac = tcapi.convert(
    ctx_cpu, A, ctx_cpu, CplxTen
)

# CPU float -> GPU float (device transfer)
Ag = tcapi.convert(ctx_cpu, A, ctx_gpu, GPUTen)

# CPU float32 -> CPU float64 (precision conversion)
A64 = tcapi.convert(ctx_cpu, A, ctx_cpu, Real64Ten)
\end{minted}

\bnoindentnl{version}\hfill\break

\noindent\textit{\texttt{C++}:}
\begin{minted}[bgcolor=LightGray]{cpp}
template <typename TenT>
std::string version();
\end{minted}
\noindent\textit{\texttt{Python}:}
\begin{minted}[bgcolor=LightGray]{python}
tcapi.version(Ten)
\end{minted}

Return the TCAPI specification version implemented for the tensor type \texttt{TenT} in \texttt{C++} or for the \texttt{TenKind} \texttt{Ten} in \texttt{Python}.
The version string has the form \texttt{"M.m"}, where \texttt{M} is the major version and \texttt{m} is the minor version, both nonnegative integers.

\example\hfill\break

\noindent\textit{\texttt{C++}:}
\begin{minted}[bgcolor=LightGray]{cpp}
auto version = tcapi::version<Ten>();
// may equal "1.0"
\end{minted}
\noindent\textit{\texttt{Python}:}
\begin{minted}[bgcolor=LightGray]{python}
version = tcapi.version(Ten)
# may equal "1.0"
\end{minted}

\subsection{Environment variables}

The final component of TCAPI is a set of environment variables that specify optional runtime behavior of TCAPI implementations.
These variables provide a lightweight mechanism to enable diagnostics and performance-related reporting without modifying application code.

\bnoindentnl{TCAPI\_VERBOSE}

Control diagnostic output and optional performance reporting for TCAPI function calls.
Similar in spirit to the verbose mode in Intel Math Kernel Library (MKL), a TCAPI implementation may emit implementation-defined messages whenever a TCAPI routine is invoked.
Such messages may include the function name or signature, selected properties of the input tensors (e.g., order, shape, and element type), and, at higher verbosity levels, measured wall-clock execution time.
The behavior of \texttt{TCAPI\_VERBOSE} is defined as follows:
\begin{itemize}
    \item \texttt{TCAPI\_VERBOSE = 0} (default): no diagnostic output is produced.
    \item \texttt{TCAPI\_VERBOSE = 1}: print a single-line message per TCAPI call containing the function name and salient input information; the exact format is implementation-defined.
    \item \texttt{TCAPI\_VERBOSE = 2}: in addition to the output at \texttt{TCAPI\_VERBOSE = 1}, include the measured execution time of each call.
\end{itemize}

\clearpage

% Required submission file: retain references.bbl when cleaning build artifacts.

\end{document}